\documentclass{acmsiggraph}
\usepackage{amsmath,bm}
\usepackage[charter,cal=cmcal]{mathdesign}
\usepackage[activate={true,nocompatibility},final,tracking=false,kerning=true,factor=1100,stretch=10,shrink=10]{microtype}
\usepackage{afterpage}
\usepackage{graphicx}
\usepackage{comment}
\usepackage{rotating}
\usepackage{adjustbox}
\usepackage{empheq} 
\usepackage[usenames,dvipsnames]{color}
\usepackage{algorithm,algorithmicx,algpseudocode}
\usepackage{enumitem}
\usepackage{pbox,array,multirow}
\setlist[itemize]{noitemsep,topsep=0pt,parsep=4pt,partopsep=0pt,leftmargin=*}
\setlist[enumerate]{noitemsep,topsep=0pt,parsep=4pt,partopsep=0pt,leftmargin=*}
\usepackage[compact]{titlesec}
\titlespacing{\section}{0pt}{*0}{*0}
\titlespacing{\subsection}{0pt}{*0}{*0}
\titlespacing{\subsubsection}{0pt}{*0}{*0}
\titlespacing{\paragraph}{0pt}{*0}{10pt}

\pdfpageattr {/Group << /S /Transparency /I true /CS /DeviceRGB>>}

\newlength\savedwidth
\newcommand\whline[1]{\noalign{\global\savedwidth\arrayrulewidth
                               \global\arrayrulewidth #1} %
                      \hline
                      \noalign{\global\arrayrulewidth\savedwidth}}

\newcommand{\sLNM}{}
\newcommand{\tLNM}{}

\newcommand{\xx}{\bm{x}}

\newcommand{\vv}{\bm{v}}

\newcommand{\BO}{\mathbb{B}_0}
\newcommand{\CC}{\mathcal{C}}
\newcommand{\DO}{\mathbb{D}_0}
\newcommand{\intd}{\mathrm{d}}

\newcommand{\FC}{Fr\'{e}chet }

\let\originalparagraph\paragraph
\renewcommand{\paragraph}[2][.]{\originalparagraph{#2#1}}

\definecolor{mygray}{rgb}{.92, .92, .94}
\newlength\mytemplen
\newsavebox\mytempbox

\makeatletter
\newcommand\mybox{%
    \@ifnextchar[
       {\@mybox}%
       {\@mybox[0pt]}}

\def\@mybox[#1]{%
    \@ifnextchar[
       {\@@mybox[#1]}%
       {\@@mybox[#1][0pt]}}

\def\@@mybox[#1][#2]#3{
    \sbox\mytempbox{#3}%
    \mytemplen\ht\mytempbox
    \advance\mytemplen #1\relax
    \ht\mytempbox\mytemplen
    \mytemplen\dp\mytempbox
    \advance\mytemplen #2\relax
    \dp\mytempbox\mytemplen
    \colorbox{mygray}{\hspace{1em}\usebox{\mytempbox}\hspace{1em}}}

\makeatother

\title{Inverse Diffusion Curves using Shape Optimization}

\author{\begin{tabular*}{0.75\textwidth}{@{\extracolsep{\fill}}ccc}
		Shuang Zhao & Fr\'edo Durand & Changxi Zheng\\
		University of California, Irvine & MIT & Columbia University
	\end{tabular*}}
\pdfauthor{Shuang Zhao}


\begin{document}
\teaser{
	\includegraphics[width=\textwidth]{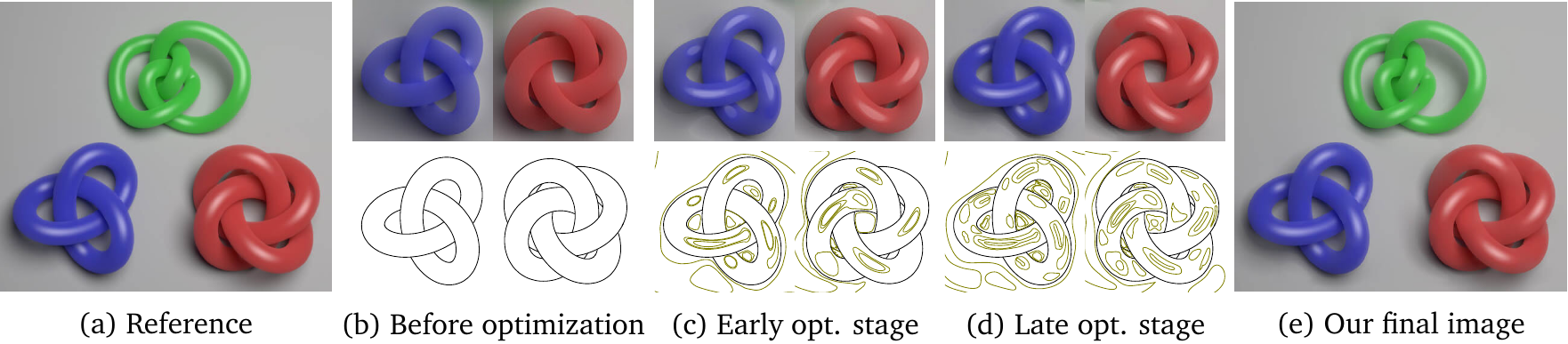}
	\caption{\label{fig:teaser} We present a new approach to automatically build diffusion curve images approximating provided color fields (a). Starting from a set of boundary curves (black strokes) indicating color jump discontinuities (b), our method iteratively adds curves (dark yellow strokes) and refines their shapes in an optimized manner (c, d). The resulting image (e) accurately matches the input.}
}

\maketitle

\begin{abstract}
The inverse diffusion curve problem focuses on automatic creation of diffusion curve images that resemble user provided color fields.
This problem is challenging since the 1D curves have a nonlinear and global impact on resulting color fields via a partial differential equation (PDE).
We introduce a new approach complementary to previous methods by optimizing curve geometry.
In particular, we propose a novel iterative algorithm based on the theory of shape derivatives.
The resulting diffusion curves are clean and well-shaped, and the final image closely approximates the input. 
Our method provides a user-controlled parameter to regularize curve complexity, and generalizes to handle input color fields represented in a variety of formats.
\end{abstract}

\keywordlist

\section{Introduction}
Vector graphic images remain invaluable for a broad range of 2D applications because of their resolution independence, compactness of representation, and powerful editability.
Recently, diffusion curve images~\cite{Orzan:2008} further improve the expressiveness of vector graphics by providing flexible and easy-to-manipulate smooth gradients, 
and since then inspired a variety of novel applications~\cite{Jeschke:2009,TSNI10,Sun:2012:DCT,Ilbery:2013:BDC,Sun:2014:FMR}.
Defined along the curves, colors are diffused across the image 
by a Poisson or Laplace reconstruction, 
and their smoothness can be further controlled by the curve's blurriness through post-processing.
Although efficient rendering of diffusion curve images has been well explored, 
its inverse problem 
of creating diffusion curves automatically given desired target images remains 
challenging.

The \emph{inverse diffusion curve problem} is 
difficult because even though the curves themselves are 1D, their impact on the final image is nonlinear and global over a 2D domain through a partial differential equation
(PDE), Laplace's equation.
The geometry of curves largely determines the reconstruction quality.
Previous methods have used {\em local} heuristics to obtain curve geometry. 
They place curves at locations indicated by edge detectors applied to the target
image~\cite{Orzan:2008} and its Laplacian or bi-Laplacian~\cite{Xie:2014:HDC}.
While these heuristics work well around sharp edges such as object boundaries,
they have difficulty handling color variations in regions that are smooth yet
visually rich.

We introduce a new approach to the inverse diffusion curve problem. 
Complementary to existing methods, our approach solves for curve geometry through a \emph{global optimization} that takes into account the curves' full impact on the PDE-based color field.
To achieve this, we characterize how modifications to a diffusion curve can reduce a global cost
function determined by the solution of a PDE with the curves acting as
boundary conditions. 

Our method is grounded on the theory of \emph{shape optimization}~\cite{sokolowski1992introduction}.
Given a color field, it computes curve geometry by minimizing a measure of the color reconstruction residual.
Starting from an initial set of curves, it iteratively evolves their shapes toward an optimal configuration.
Mathematically, the curves are treated as continuous functionals.
This allows them to deform arbitrarily, enabling full exploration of possible curve configurations.
This iterative process is similar to a surface normal flow: our curve evolution at every iteration is guided by the \FC derivative of the residual function with respect to the curve's boundary velocity, leading to an efficient ``gradient descent'' of the residual.
This method is mathematically clean and easy to implement: all computations at each iteration boil down to solving a Laplace equation and a Poisson equation.%

Building on our curve placement algorithm, we introduce a complete pipeline for solving the inverse diffusion curve problem.
Our method generates curves in a clean and concise way, and the resulting images can accurately capture complex color variations of input color fields (see Figures~\ref{fig:teaser}, \ref{fig:comparisons}, \ref{fig:3d_render}, and \ref{fig:grad_mesh} {as well as the supplementary images}).

We demonstrate that our method promises practical applications beyond pixel image vectorization.
For instance, it enables automatic rendering of vector graphic images from 3D geometries, analogous to the traditional pixel image rendering.
Further, using our algorithm, one can directly transform other formats of vector graphics, such as gradient meshes, into diffusion curve images without rasterizing the input.

\section{Related Work}
\label{sec:related}
Diffusion curves~\cite{Orzan:2008} represent a color field by diffusing the colors defined along control curves over the entire image plane.
The diffusion process is described by Laplace's equation solved using a finite volume method.
Later, solving Laplace's equation was improved using a multigrid method~\cite{Jeschke:2009}, triangle mesh interpolation~\cite{Pang:12}, Boundary Element
method~\cite{Sun:2012:DCT}, 2D ray tracing~\cite{prevost2014vectorial}, and
Fast Multipole method~\cite{Sun:2014:FMR}.

\paragraph{Inverse diffusion curve problem}
Our work focuses on the \emph{inverse problem} of diffusion curves.
Previously, Orzan et al.~\shortcite{Orzan:2008} proposed to place diffusion curves along edges extracted from input images using the Canny detector~\cite{canny1986computational}.
Jeschke et al.~\shortcite{jeschke2011estimating} introduced a technique to improve curve colorings.
Xie et al.~\shortcite{Xie:2014:HDC} further improved this method by detecting edges in a Laplacian (and/or bi-Laplacian) domain and constructing curves hierarchically.
They solve the Laplacian and bi-Laplacian weights using least-squares fitting.
In all methods, diffusion curves are placed along the detected edges, and never moved or added in continuous color regions.
These methods then rely on optimizing curve coloring for better accuracy.

We introduce a fundamentally complementary solution to the inverse diffusion curve problem.
Instead of predetermining curve geometry and optimizing their coloring, we propose doing the opposite by first optimizing the geometry and then determining the coloring accordingly.
We demonstrate that with a very simple coloring scheme, our method outperforms prior methods under many situations (\S\ref{sec:main_res}).
Furthermore, our approach accepts input color fields beyond pixel images.

\paragraph{Extensions of diffusion curves}
Several methods have been proposed to extend the expressiveness of diffusion curves.
Sun et al.~\shortcite{Sun:2014:FMR} enabled fast diffusion curve cloning and multi-layer composition.
Finch et al.~\shortcite{Finch:2011} introduced a higher-order notion of smoothness: the colors are defined using a 4th-order linear elliptic PDE rather than a Laplace equation.
To accelerate the color evaluation, Boy\'{e} et al.~\shortcite{Boye:2012:VSF} developed a vectorial solver using the Finite Element Method, and Sun et al.~\shortcite{Sun:2012:DCT} proposed a boundary element based formula, which was later improved in~\cite{Ilbery:2013:BDC} to handle both Laplacian and bi-Laplacian curves in a unified framework.
Higher-order curves offer greater flexibility than the standard diffusion curves, but their inverse problems are more difficult and remain unsolved.
In this paper, we focus on the inverse problem for \emph{original diffusion curves} and discuss potential extension to higher-order domains in \S\ref{sec:addtl_res}.

\paragraph{Theory and applications of shape optimization}
We build our curve optimization on the theoretical foundation of shape optimization~\cite{sokolowski1992introduction,haslinger2003introduction}, a subfield of optimal control theory.
Mathematically, it solves the problem of finding a bounded set $\Omega$ to minimize a continuous functional on $\Omega$.
The core idea of shape optimization has been used for image segmentation since the seminal work of~\cite{kass1988snakes,mumford1989optimal}.
It is also related to surface gradient flow widely studied in geometry processing~\cite{schneider2001geometric,Crane:2013:RFC}.
In areas outside of computer graphics, shape optimization has been used to 
enhance mechanical structures such as airfoils~\cite{mohammadi2001applied} and photonic crystals~\cite{burger2004inverse}.
It has also been used in computer vision for image segmentation (e.g., \cite{herbulot2006segmentation,jung2012nonlocal}).
To our knowledge, shape optimization has not yet been applied in vector graphics.
In this paper, we solve a shape optimization problem with a PDE constraint (\S\ref{sec:opt_prob}), which is significantly more challenging than a conventional shape optimization problem.

\begin{figure*}[t]
	\includegraphics[width=\textwidth]{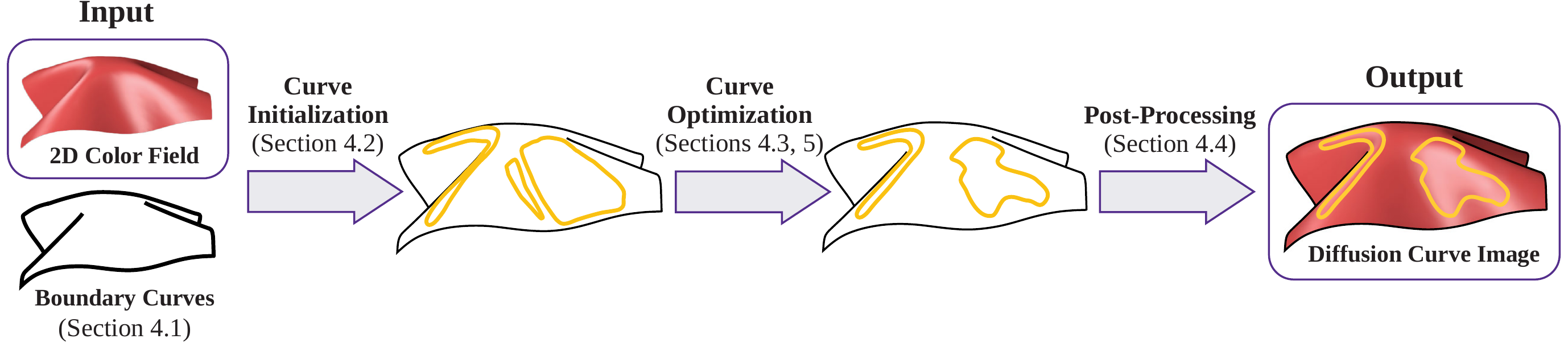}
	\caption{\label{fig:pipeline}
		{\bf Our pipeline.} The input to our method is a 2D color field.
		After obtaining a set of boundary curves indicating color jump discontinuities (\S\ref{subsec:curve_bound}), our method constructs a set of initial curves (\S\ref{subsec:curve_init}) and optimizes their shapes and trajectories (\S\ref{subsec:curve_place} and \S\ref{sec:curve_opt}).
		Finally, we post-process the optimized curves and obtain the resulting diffusion curve image (\S\ref{subsec:curve_post}).}
\end{figure*}

\section{Background and Overview}\label{sec:background}
We start by briefly revisiting the mathematical formulation of diffusion curve images.
We then present the main focus of this work,
the inverse diffusion curve problem, and overview our proposed solution.

\paragraph{Diffusion curve images}
In a diffusion curve image, as originally formulated in~\cite{Orzan:2008,Jeschke:2009}, the color field $u$ is a harmonic function, satisfying a Laplace equation with a Dirichlet boundary condition:
\begin{equation}\label{eq:laplace}
\begin{aligned}
    u(\xx) &= \left\{ C_{\ell}(\xx), C_r(\xx) \right\}, & \xx\in\mathbb{B} \\
    \Delta u(\xx) &= 0, & \mathrm{otherwise},
\end{aligned}
\end{equation}
where the boundary $\mathbb{B}$ consists of the entire set of diffusion curves; $C_{\ell}$ and $C_r$ specify the colors on the \emph{left} and \emph{right} side of each curve, respectively.
Typically, both the shapes of the curves and their left- and right-side colors are specified by the user, and the entire color field is uniquely determined by solving the Laplace equation~\eqref{eq:laplace}. 

Since its invention, diffusion curves have been augmented.
Orzan et al.~\shortcite{Orzan:2008} proposed to apply per-pixel blurring to the rasterized image of $u$, the solution of \eqref{eq:laplace}.
Finch et al.~\shortcite{Finch:2011} further extended to diffuse colors using higher-order elliptic PDEs such as the biharmonic equations.

\paragraph{Inverse diffusion curve problem}
While plenty of extensions of the \emph{forward} diffusion curve problem have been proposed, largely underexplored is the \emph{inverse problem},
one that computes a set of diffusion curves such that the resulting vector image
closely resembles a user-provided 2D color field. In this paper, we address this inverse problem. 
In particular, we note that the inverse problem involves two subproblems:
\begin{itemize}
\item \textbf{Curve geometry.} To build a diffusion curve image, one needs to decide where to place the curves (namely, to determine $\mathbb{B}$).
\item \textbf{Curve coloring.} Given the curve geometry, the colors on both sides of each curve (namely $C_{\ell}$ and $C_r$) need to be specified.
\end{itemize}
As discussed in \S\ref{sec:related}, recent work~\cite{jeschke2011estimating,Xie:2014:HDC} has largely focused on optimizing curve coloring with their geometry predetermined (using edge detection).
In contrast, we focus on a complementary problem, the problem of directly optimizing curve geometry.
We demonstrate (in \S\ref{sec:results}) that curves with 
optimized geometries generally yield higher-quality of reconstructions, regardless of the curve coloring schemes.

\subsection{Overview}
\begin{figure}[b]
	\addtolength{\tabcolsep}{-3pt}
	\centering
	\begin{tabular}{ccc}
		\hspace{-7pt}
		\includegraphics[height=1in]{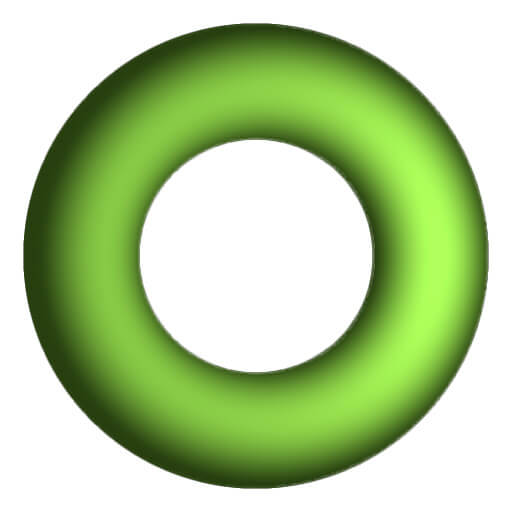}
		\hspace{-6pt}
		&
		\hspace{-6pt}        
		\includegraphics[height=1in]{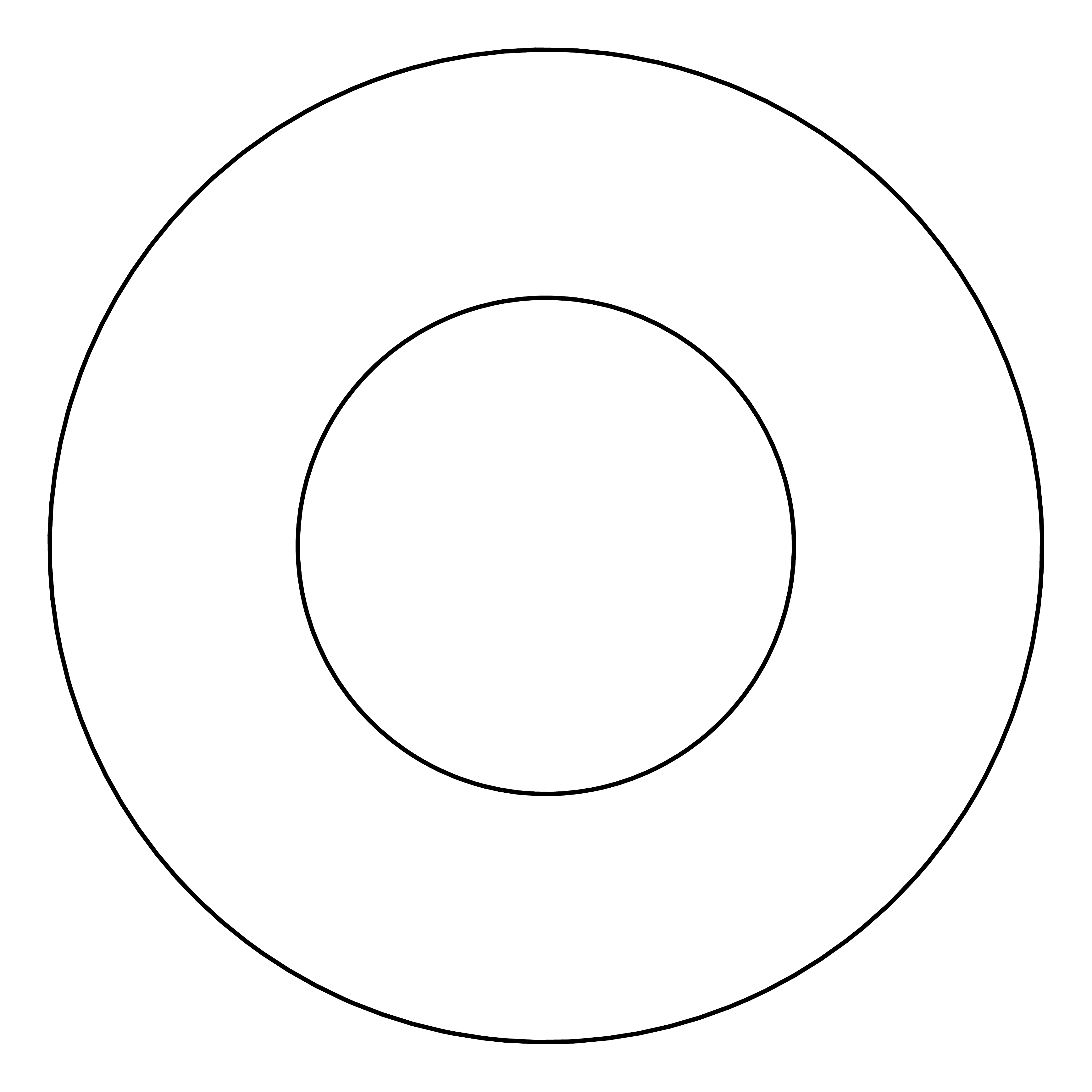}
		\hspace{-6pt}
		&
		\includegraphics[height=0.95in]{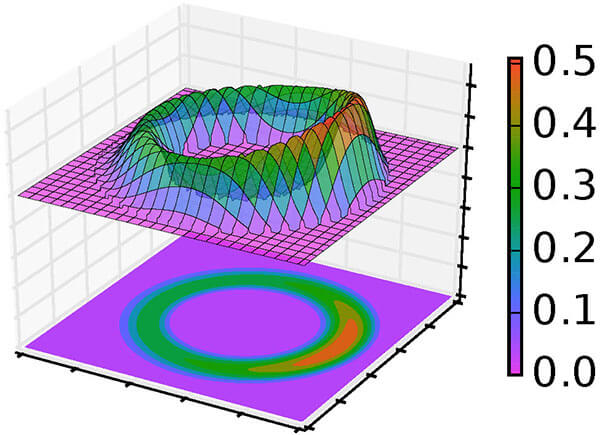}
		\\
		(a) & (b) & (c)
	\end{tabular}
	\caption{\label{fig:torusA_ref} {\bf A sample color field:} (a) the color
		field representing a smoothly shaded torus viewed from the top; (b) the corresponding boundary curves; (c) a visualization of the color field.
	}
\end{figure}

\paragraph{Pipeline} 
We develop a complete pipeline for automatic creation of diffusion curve images (\S\ref{sec:curvegen}).
Figure~\ref{fig:pipeline} shows an overview of our pipeline.
We take as input a color field $I$ allowing to query for color values at for all $\xx \in \Omega$ (where $\Omega$ denotes the image domain).
Starting with extracting a set of boundary curves (\S\ref{subsec:curve_bound}) indicating jump discontinuities in $I$, our method generates a set of curves as ``initial guesses'' (\S\ref{subsec:curve_init}) which are then deformed by our core curve optimization algorithm (\S\ref{sec:curve_opt}) to minimize reconstruction error (\S\ref{subsec:curve_place}).
Lastly, we post-process the deformed curves (\S\ref{subsec:curve_post}) to generate final diffusion curve images.

\paragraph{Curve optimization}
As one of our main contributions, the key component of our pipeline is a \emph{curve optimization algorithm} that deforms diffusion curves to minimize the reconstruction error.
In our algorithm, curves are 1D continuous geometries discretized as polylines, while the cost functional includes an integral over a 2D image region.
The optimization problem is constrained by a Laplace equation for color diffusion. 
Our general idea of solving this optimization problem is using iteration: given a set of curves $\mathbb{B}_0$, we construct a velocity field $\vv$ so that after deforming $\mathbb{B}_0$ according to $\vv$ for a small time step $t$, the updated set of curves $\mathbb{B}_t$ leads to improved reconstruction accuracy.
This process is repeated until convergence.
Conceptually, this routine is similar to gradient descent---in each iteration, a local adjustment is made to obtain a better solution.
Our main challenge is to derive a proper form for the derivative of the cost function with respect to the shapes of the curves and use it to construct the velocity field $\vv$.

In what follows, we will first provide a complete exposition of our end-to-end pipeline (in \S\ref{sec:curvegen} and outlined in Figure~\ref{fig:pipeline}), and then describe in details the algorithms and mathematical derivations for the curve optimization (in \S\ref{sec:curve_opt}).

%
\section{Our Pipeline}\label{sec:curvegen}

\subsection{Boundary Curves}\label{subsec:curve_bound}
Provided an input color field $I$, we start the pipeline by obtaining a set of \emph{boundary curves} $\partial\Omega$ indicating the outer boundary and jump discontinuities of $I$.%
\footnote{The necessity of boundary curves is explained in \S\ref{sec:curve_opt}.}
An example color field and corresponding boundary curves are shown in Figure~\ref{fig:torusA_ref}-ab.
In practice, we obtain the boundary curves $\partial\Omega$ depending on specific representation of the input color field $I$:
\begin{itemize}
	\item \textbf{Pixel images.} A common way to represent color fields is using standard pixel images.
	The boundary curves, however, are not uniquely defined in this case.
	To obtain these curves in practice, we use Canny edge detection similar to \cite{Orzan:2008}.
	\item \textbf{3D renderings.} If the color field is defined by the rendering of a 3D scene, the boundary curves can be obtained by extracting object contours.
	\item \textbf{Other vector formats.} For input color fields represented in other vector formats (e.g., gradient mesh), $\partial\Omega$ can be determined directly based on the underlying vector representation (e.g., triangle edges).
\end{itemize}
Please refer to \S\ref{sec:main_res} for more details and experimental results on  boundary curve computation.

\subsection{Curve Initialization}\label{subsec:curve_init}
\begin{algorithm}[b]
	\caption{Diffusion curve initialization}
	\begin{algorithmic}[1]
		\Require Color field $I$ (defined on $\Omega$) and boundary curves $\partial\Omega$
		\Procedure{CurveInit}{scheme, $\partial\Omega$, $I$, $\Omega$}
		\State generate uniform point samples in $\Omega$
		\State form ${\bf r}_0$ by evaluating $R_0(\partial\Omega;\xx)$ on the sampled points
		\If {scheme = `global'} \Comment{Global scheme}
		\State fit a piecewise function $f$ to ${\bf r}_0$
		\State let $A$ to be the (internal) piece boundaries of $f$
		\Else \Comment{Local scheme}
		\State $A \gets \{ 0.9 \max({\bf r}_0) \}$
		\EndIf
		\State \Return iso-contours with iso-values specified in $A$
		\EndProcedure
	\end{algorithmic}
	\label{alg:curve_init}
\end{algorithm}
\paragraph{Desired properties}
Similar to gradient descent methods, our curve optimization algorithm takes an initial guess to start with.
For ensuring high-quality optimization results, there are a few properties required of the initial curves:
\begin{enumerate}
\item \textbf{Easy to compute.} The curve initialization step should not require intense computation: we rely on the optimization step to refine the shapes of these curves.
\item \textbf{Good coverage.} The initial curves should provide a good coverage to the full image domain $\Omega$, so that the optimization is less prone to local optima.
\item \textbf{Being well-shaped.} The initial curves need to be well-shaped. For example, they should have low complexities and not self-intersect or collide with the boundary curves.
\end{enumerate}
To achieve these properties, we use iso-contours of the residual field as the initial curves:
\begin{equation} \label{eq:residual0}
R_0(\partial\Omega; \xx) = (u_0(\xx) - I(\xx))^2, \quad \forall \xx \in \Omega.
\end{equation}
In \eqref{eq:residual0}, $u_0$ is given by the diffusion curve image using only the input boundary curves $\partial\Omega$.
These curves can be computed easily from a set of iso-values (Property 1).
In addition, as long as the iso-values are distributed properly, the resulting iso-contours will provide a good coverage to the image domain while being well shaped (Properties 2 and 3).
For example, to make the initial curves never intersecting with $\partial\Omega$, we can simply pick strictly positive iso-values as $R_0(\partial\Omega; \xx) = 0$ for all $x \in \partial\Omega$.

\begin{figure}[t]
	\centering
	\includegraphics[width=0.85\columnwidth]{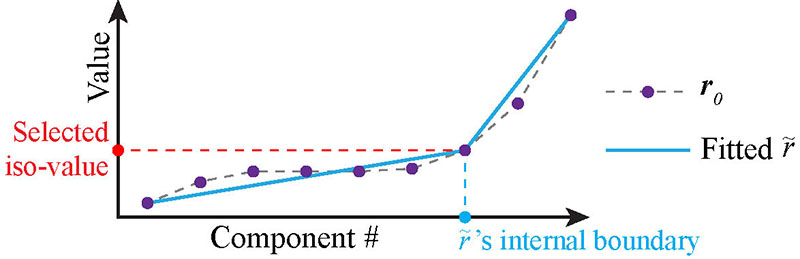}
	\caption{\label{fig:linearfit} {\bf Fitting} $\bm{r}_0$ with 9 components (indicated with purple dots) using a piecewise linear function $\tilde{r}$ with 2 pieces ($m = 1$). The value of $\tilde{r}$ at its internal boundary is selected as iso-value.}
\end{figure}

\paragraph{Our approach}
To choose a set of properly distributed iso-values, we start with sampling a set of points in $\Omega$ and stacking the residual values~\eqref{eq:residual0} at these points into a vector $\bm{r}_0$ in ascending order (lines 2 and 3 in Algorithm~\ref{alg:curve_init}).
The resulting vector $\bm{r}_0$ provides a picture on the distribution of residuals. 
We adopt two complementary schemes, \emph{global} and \emph{local}, to set iso-values using
$\bm{r}_0$, and thereby obtain the iso-contours:
\begin{itemize}
    \item {\bf Global.}
The global scheme constructs a relatively large set of initial curves over the entire
domain $\Omega$.
Assume that the number of iso-values $m$ is given.
Ideally, we would like to find $m$ values such that the consequent iso-contours optimally capture the structure of 2D residual field $R_0$.
In practice, we solve this problem approximately and rely on our curve optimization algorithm to refine the curves.
Particularly, we solve a well-studied 1D problem~\cite{ramer1972iterative}: to fit a piecewise linear function $\tilde{r}$ with $m + 1$ pieces that closely describes $\bm{r}_0$ (interpreted as a polyline).
Then, the values of $\tilde{r}$ at its $m$ internal piece boundaries are used as iso-values (see Figure~\ref{fig:linearfit}).
    \item {\bf Local.}
The local scheme, in contrast to the global one, adds curves locally in regions with high approximation error.
In this case, we use only one iso-value determined based on the maximal sampled residual (line~8 of Algorithm~\ref{alg:curve_init}).
\end{itemize}

In our curve placement algorithm (detailed in \S\ref{subsec:curve_place}), we use the global scheme at the beginning to ensure that the initial curves provide a good coverage to the domain $\Omega$ (Property 2).
Then, the local scheme is applied iteratively to add small sets of curves in high-residual areas.
The combination of both schemes offers sufficient approximation accuracy without introducing unnecessarily complex curves (Property 3).
We find that $m = 2$ works well in our experiments.

\subsection{Curve Placement}
\label{subsec:curve_place}
\begin{figure}[b]
	\vspace{5pt}
	\centering
	\includegraphics[width=\columnwidth]{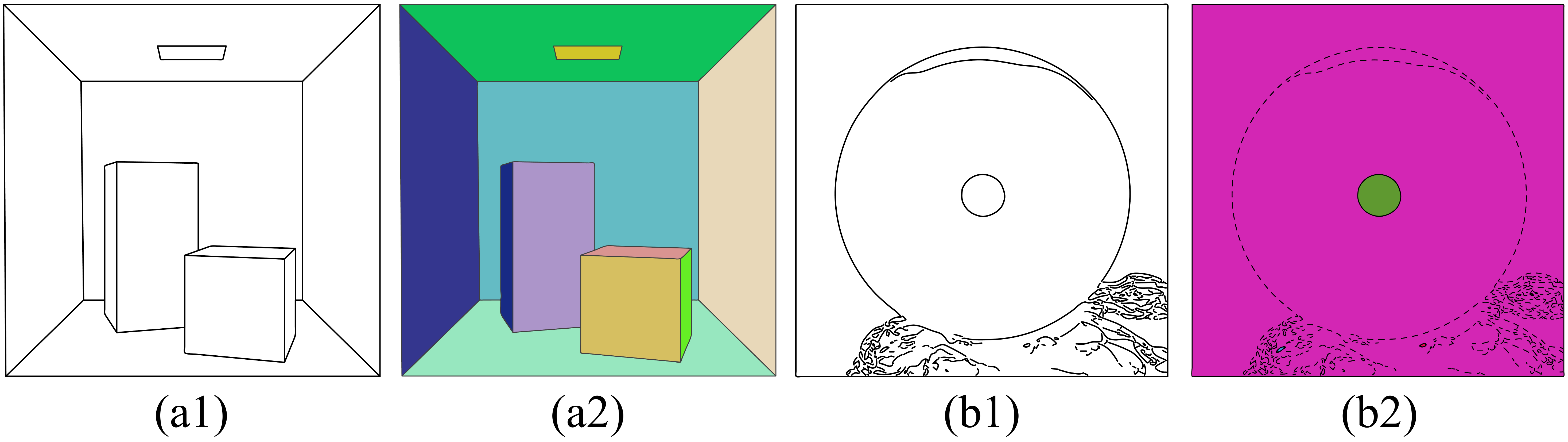}
	\caption{\label{fig:partition}
		Two examples of \textbf{boundary curves} and corresponding \textbf{partitioning} of domain $\Omega$.
		For clean and well defined boundaries~(a1), $\Omega$ can be divided into many well shaped components~(a2); for messier boundaries often resulting from edge detections~(b1), there are normally fewer components with more complex shapes~(b2).
		Our approach works well for both cases.}
	\vspace{-5pt}	
\end{figure}
\begin{algorithm}[b]
	\caption{Diffusion curve placement}
	\begin{algorithmic}[1]
		\Require Color field $I$ (defined on $\Omega$) and boundary curves $\partial\Omega$
		\Procedure{CurvePlacement}{$\partial\Omega,\ I,\ \Omega,\ \epsilon_0$}
		\State partition $\Omega$ into connected components
		\State $\mathbb{B} \gets \partial\Omega$ 
		\For {each component $\CC$}
		\State $\DO \gets$ \textsc{CurveInit}$(\text{`global'},\ \partial\CC,\ I,\ \CC)$
		\Comment{Alg.~\ref{alg:curve_init}}
		\State $\mathbb{D} \gets$ \textsc{CurveOpt}$(\DO,\ \partial\CC,\ I,\ \CC)$
		\Comment{Alg.~\ref{alg:gd}}
		\While {$R(\CC; \partial\CC \cup \mathbb{D}) > \epsilon_0$}
		\State $\DO' \gets$ \textsc{CurveInit}$(\text{`local'},\ \partial\CC \cup \mathbb{D},\ I,\ \CC)$
		\Comment{Alg.~\ref{alg:curve_init}}
		\State $\mathbb{D}' \gets$ \textsc{CurveOpt}$(\DO',\ \partial\CC \cup \mathbb{D},\ I,\ \CC)$
		\Comment{Alg.~\ref{alg:gd}}
		\State $\mathbb{D} \gets \mathbb{D} \cup \mathbb{D}'$
		\EndWhile
		\State $\mathbb{D} \gets$ \textsc{CurveOpt}$(\mathbb{D},\ \partial\CC,\ I,\ \CC)$
		\Comment{Alg.~\ref{alg:gd}}
		\State post-process $\mathbb{D}$ \Comment{\S\ref{subsec:curve_post}}
		\State $\mathbb{B} \gets \mathbb{B} \cup \mathbb{D}$        
		\EndFor
		\State \Return $\mathbb{B}$
		\EndProcedure
	\end{algorithmic}
	\label{alg:full_pipeline}
\end{algorithm}
Given the initial curves generated by Algorithm~\ref{alg:curve_init}, our
curve optimization algorithm iteratively refines their trajectories to reduce
reconstruction errors and finalize curve geometry.
We postpone the details of this algorithm (Algorithm~\ref{alg:gd}) and its derivations until \S\ref{sec:curve_opt}, 
but present here the complete curve placement steps (Algorithm~\ref{alg:full_pipeline}).

The curve placement process is built on the algorithms of curve initialization and
optimization schemes.
It takes as input the target color field $I$ defined on domain $\Omega$, the previously obtained boundary curves $\partial\Omega$, and a tolerance $\epsilon_0$ on reconstruction error.
Based on $\partial\Omega$, we partition the domain $\Omega$ into a number of connected components and process them individually in parallel (line~2 of Algorithm~\ref{alg:full_pipeline}).
Figure~\ref{fig:partition} illustrates example boundaries and resulting partitionings.
Notice that our approach allows boundary curves to exist inside individual components (e.g., Figure~\ref{fig:partition}-b)---these curves will remain fixed throughout the entire pipeline.

\begin{figure}[t]
	\centering
	\addtolength{\tabcolsep}{-5pt}	
	\begin{tabular}{cccccc}
		\raisebox{23pt}{\rotatebox[origin=c]{90}{Initial}~} &
		\includegraphics[width=0.21\columnwidth]{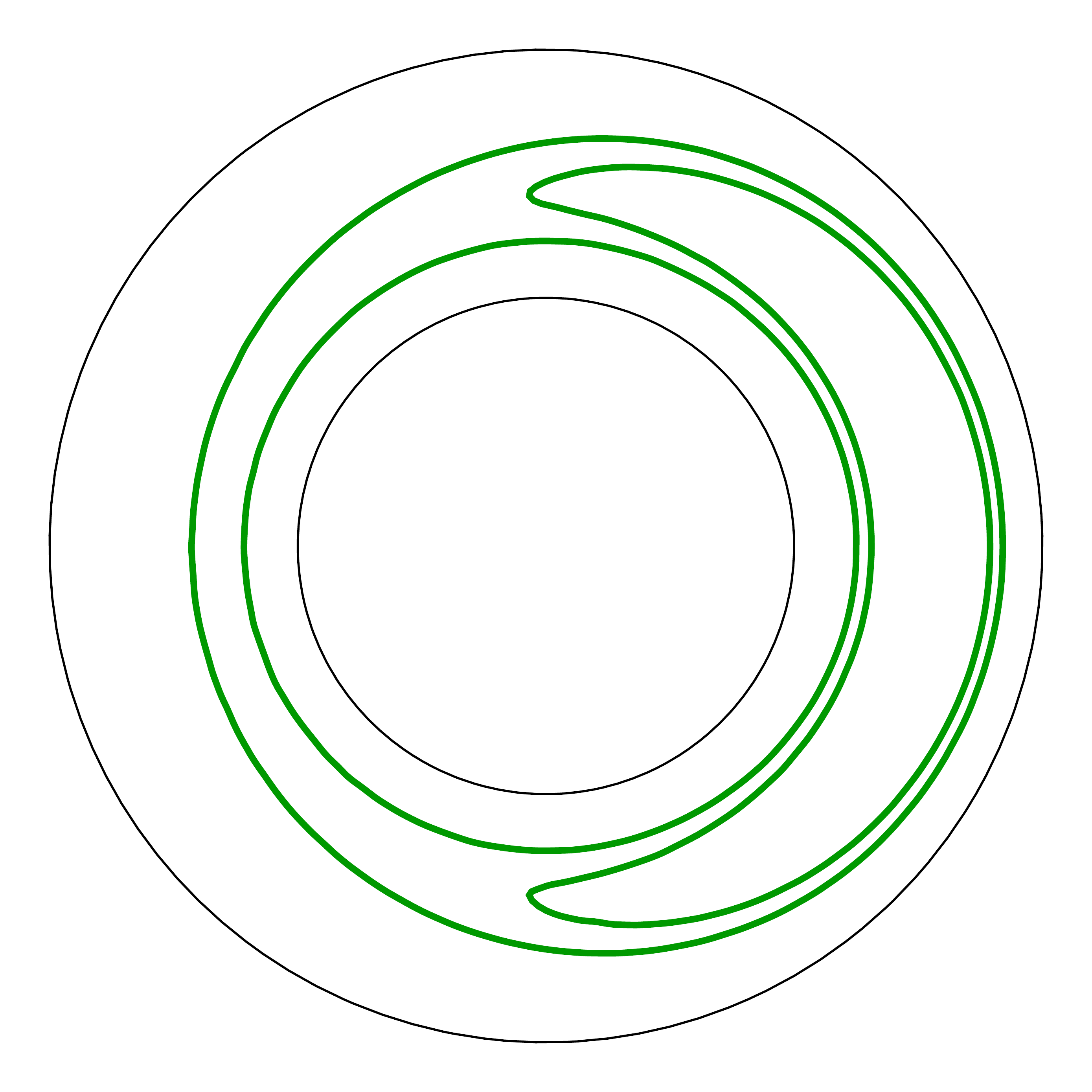} &
		\includegraphics[width=0.21\columnwidth]{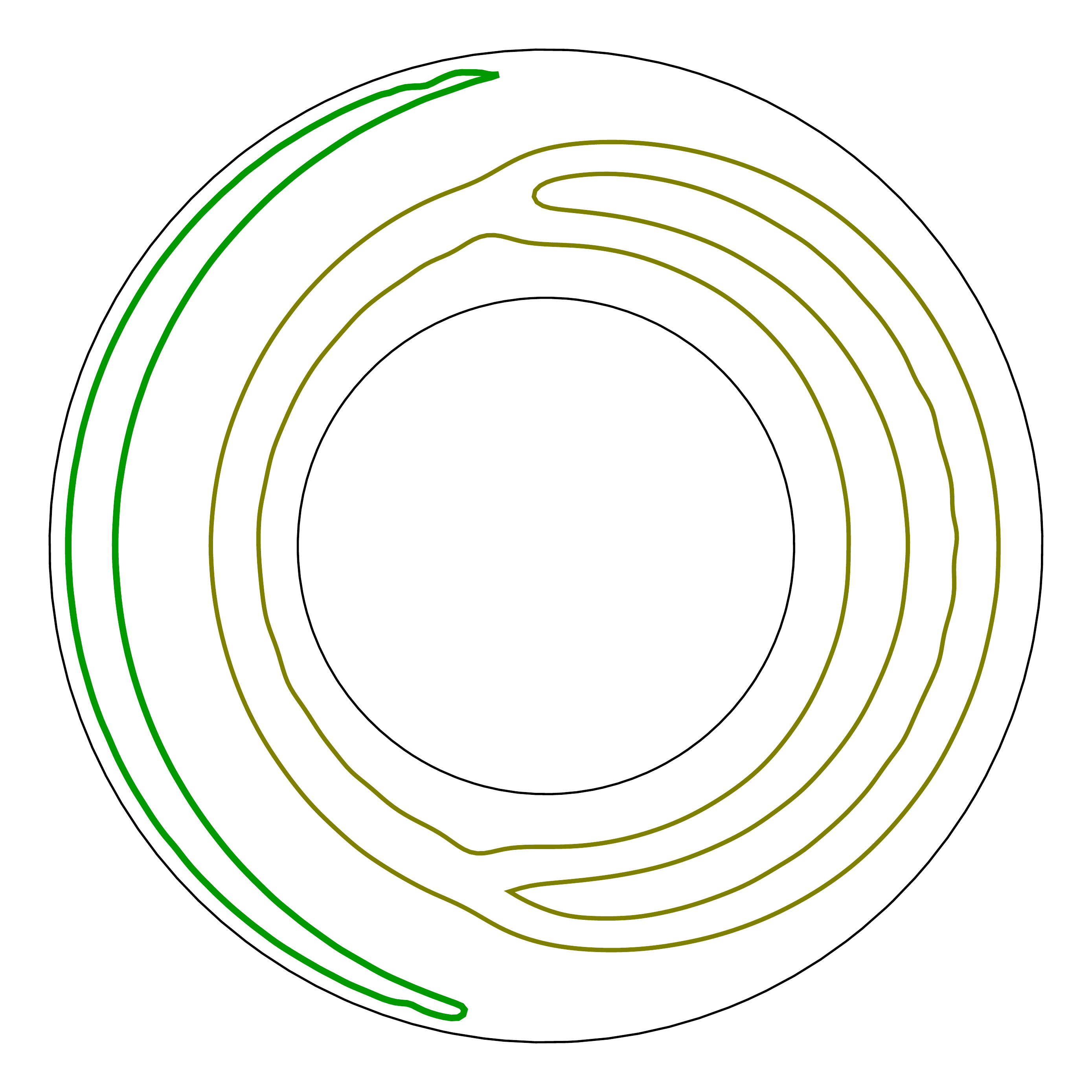} &
		\includegraphics[width=0.21\columnwidth]{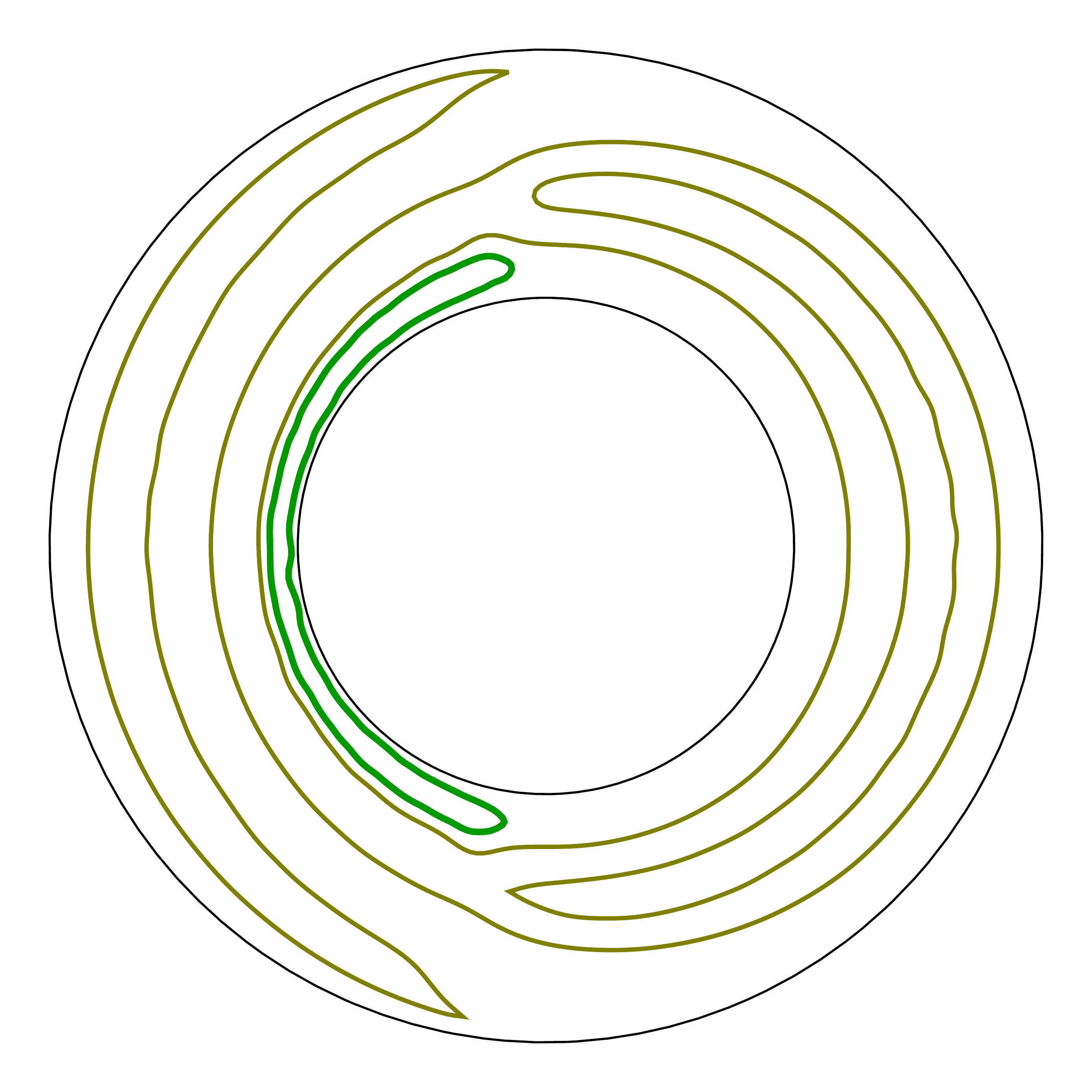} &
		\raisebox{23pt}{~$\ldots$~} &
		\includegraphics[width=0.21\columnwidth]{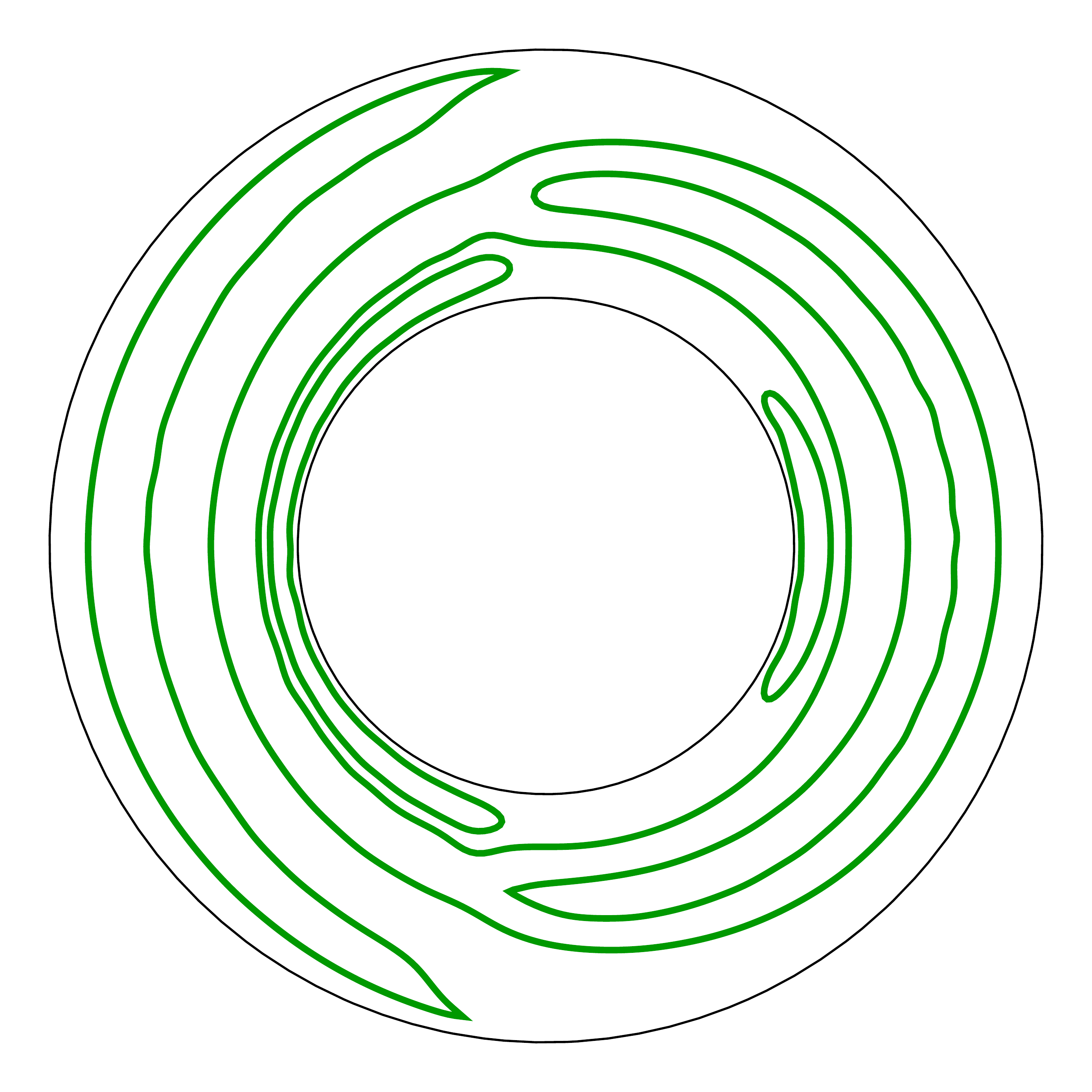}\\[-2pt]
		\raisebox{23pt}{\rotatebox[origin=c]{90}{Optimized}~} &
		\includegraphics[width=0.21\columnwidth]{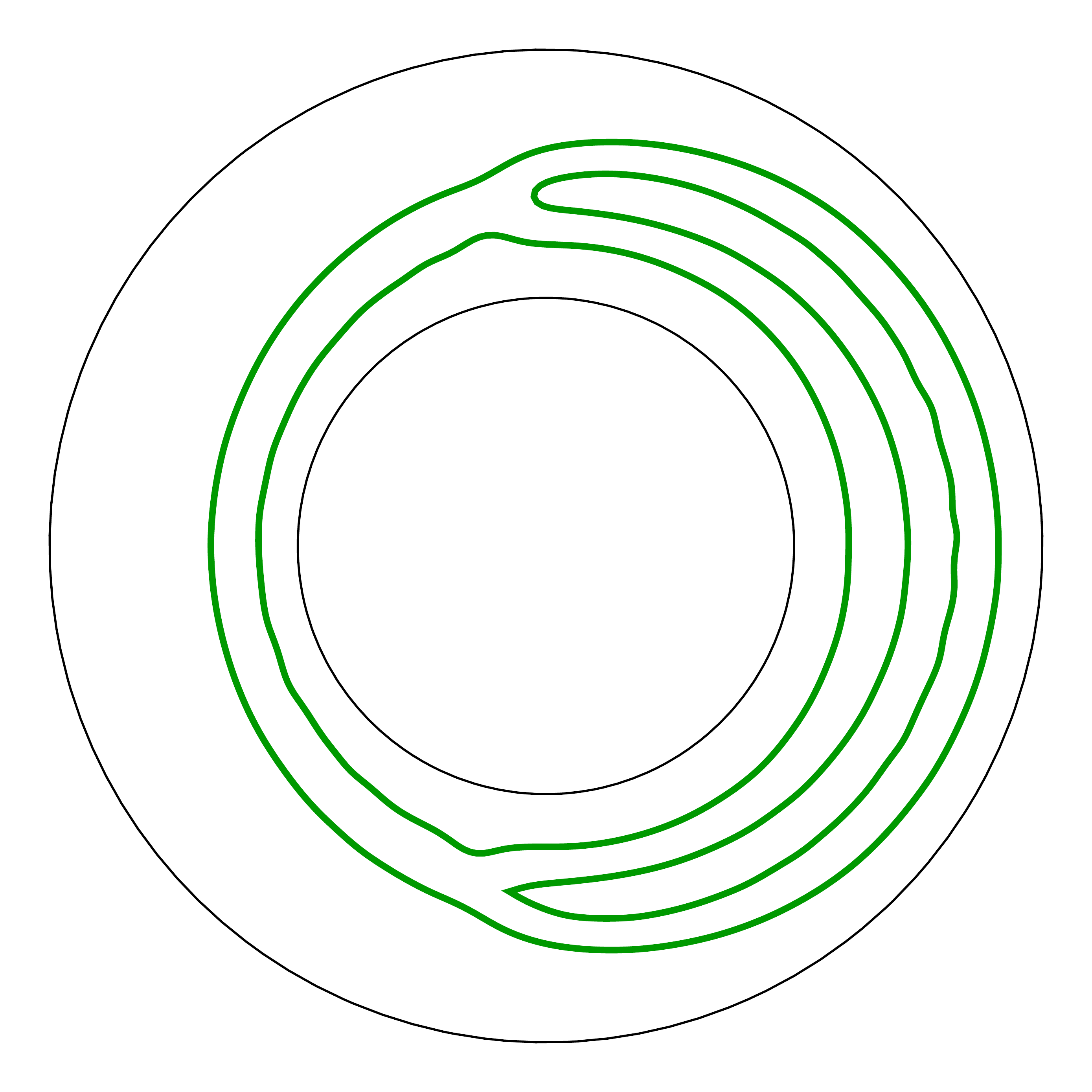} &
		\includegraphics[width=0.21\columnwidth]{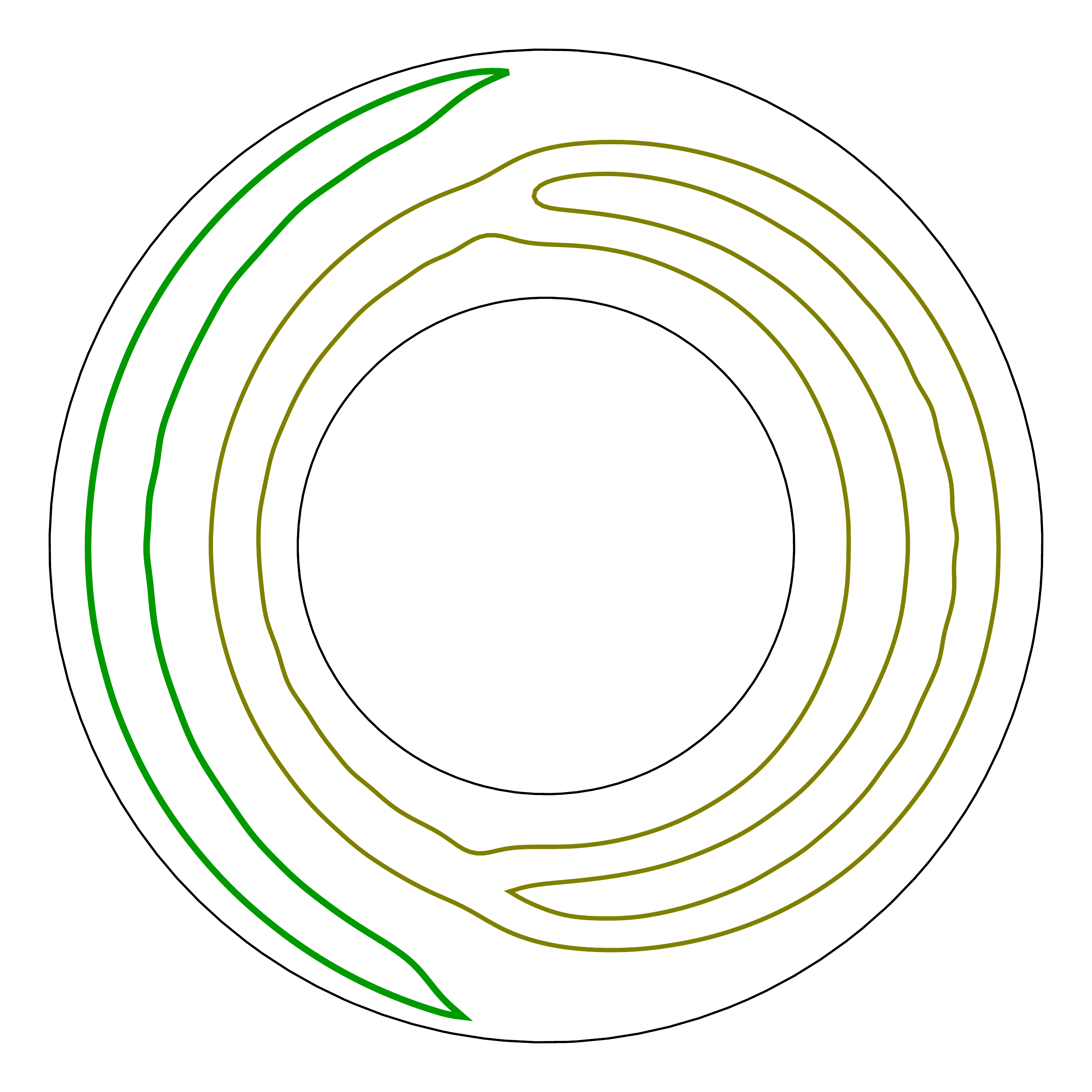} &
		\includegraphics[width=0.21\columnwidth]{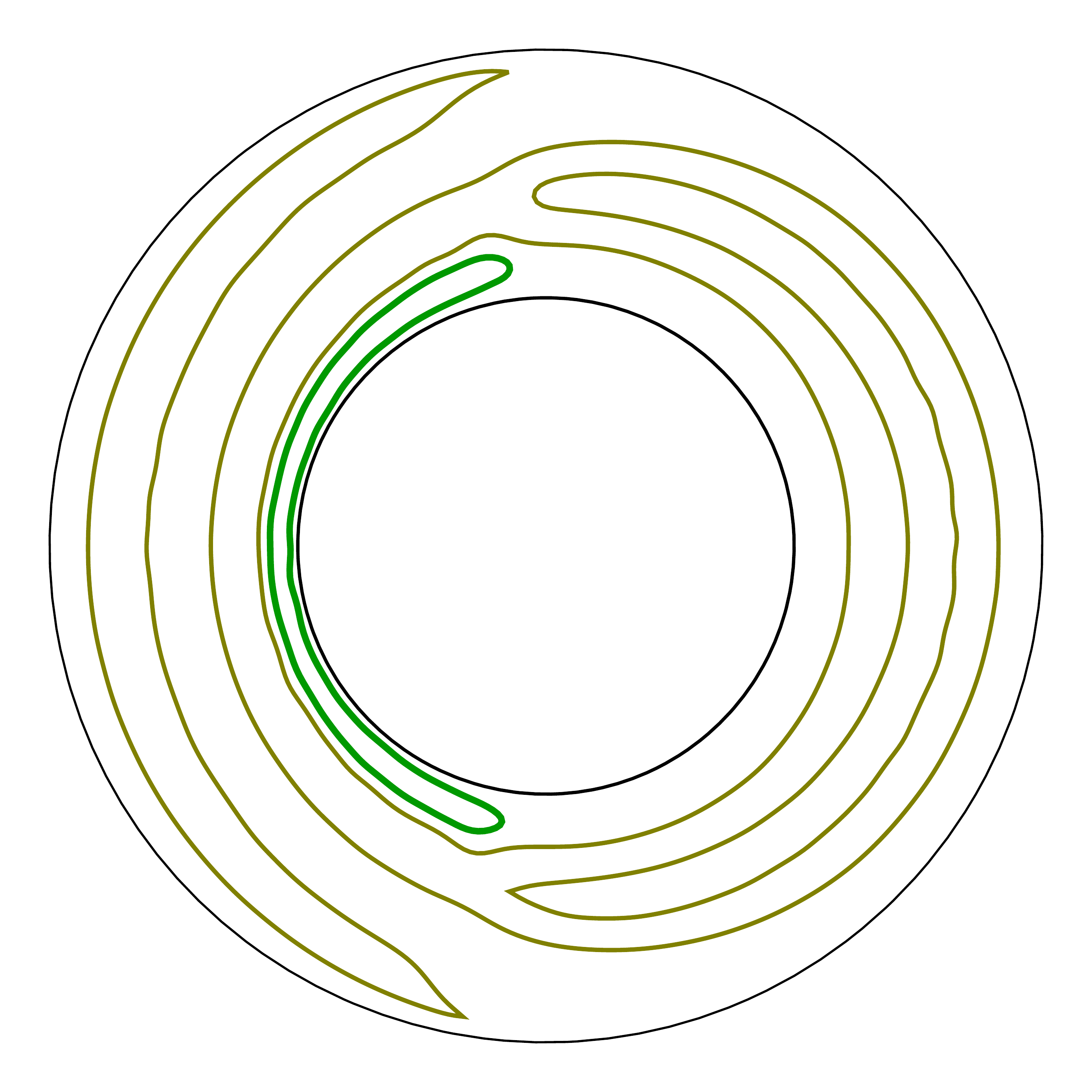} &
		\raisebox{23pt}{~$\ldots$~} &
		\includegraphics[width=0.22\columnwidth]{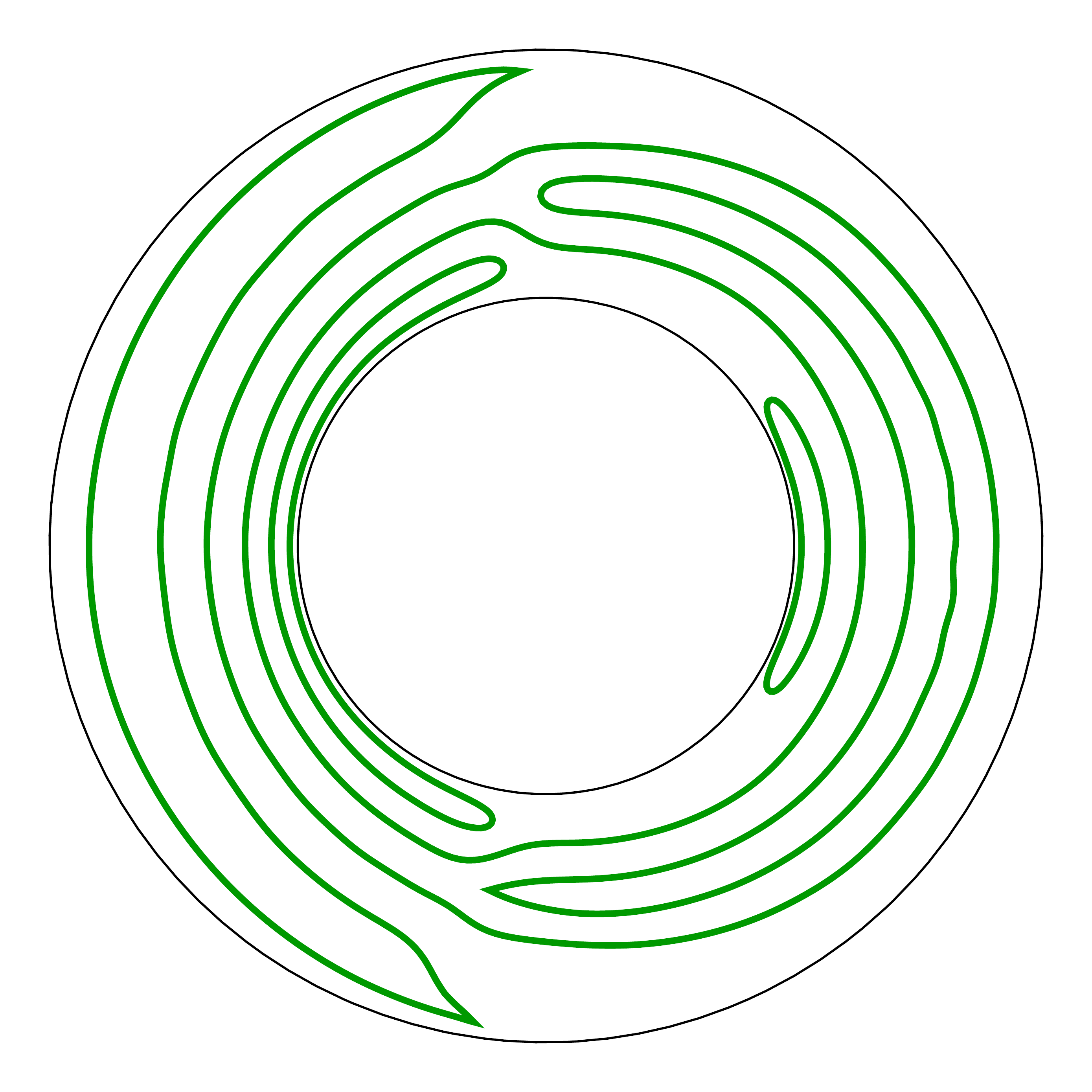}\\
		& Pass 1 & Pass 2 & Pass 3 & & Final pass
	\end{tabular}
	\caption{\label{fig:pipeline_example} An example of our {\bf curve placement} process (Algorithm~\protect\ref{alg:full_pipeline}) using Figure~\protect\ref{fig:torusA_ref}-ab as input. New curves are constructed using the global scheme in Pass 1 and the local one in the following passes. The final pass (Pass 5) generates no new curves. Instead, it starts with those created in all previous passes. In each pass, active curves (those being added and/or optimized) and previously generated ones are drawn as green and dark yellow strokes, respectively.}
\end{figure}

For each connected component $\mathcal{C}$, our approach generates diffusion curves via several \emph{passes}, each of which involves initializing a set of curves (Algorithm~\ref{alg:curve_init}) and optimizing their shapes (Algorithm~\ref{alg:gd} in \S\ref{sec:curve_opt}).
In the first pass, we start with initial curves constructed using the \emph{global} scheme (lines~5 and 6).
After this pass, if the approximation error remains beyond a tolerance $\epsilon_0$, additional passes are used in which new curves are initialized using the \emph{local} scheme (lines~8 to 10).
After the error drops below the threshold, we perform a final pass (line~12) 
in which all curves created in previous passes are optimized together.
Finally, we post-process the resulting curves to remove redundant curve segments (line~13 and \S\ref{subsec:curve_post}).
An example of this curve placement process is illustrated in Figure~\ref{fig:pipeline_example}.

\subsection{Curve Post-Processing}
\label{subsec:curve_post}
Lastly, we post-process the curves $\mathbb{B}$ returned by Algorithm~\ref{alg:full_pipeline} and generate the final diffusion curve image.

\paragraph{Curve Coloring}
Notice that Algorithm~\ref{alg:full_pipeline} returns optimized \emph{curve geometry} instead of actual diffusion curves.
Thus, to turn $\mathbb{B}$ into a set of diffusion curves, their \emph{coloring}, namely colors on both sides of each curve, needs to be provided.
This corresponds to specifying the values of $C_{\ell}$ and $C_r$ in (\ref{eq:laplace}).

As aforementioned, this curve coloring step is completely orthogonal and complementary to our core technique (Algorithm~\ref{alg:full_pipeline}).
Thus, in the rest of this paper, we use a simple scheme which directly sample color values on both sides of each curve from the input color field $I$.
That is, for any $\xx \in \mathbb{B}$, we set
\begin{equation}
	\label{eq:coloring}
	C_{\ell}(\xx) = I(\xx + \delta n_{\ell}), \quad
	C_r(\xx) = I(\xx + \delta n_r)
\end{equation}
where $n_{\ell}$ and $n_r$ respectively denote normal directions pointing left and right side of a point $\xx$ on a curve (thus, $n_l = -n_r$) and $\delta$ is a small positive number that can be set to the size of one pixel when $I$ is represented as a pixel image.
Our experiments demonstrate that this simple scheme can yield high-quality results thanks to our optimized curve geometry (\S\ref{sec:main_res}).
In \S\ref{sec:addtl_res}, we show that more advanced coloring techniques can further improve reconstruction accuracy. 

\paragraph{Removing redundant curve segments}
As mentioned in \S\ref{sec:num}, we represent diffusion curves as polylines
consisting of a number of line segments.  Some of these segments, however, may
be unnecessary. 
Note that the colors across a line segment are continuous because of the
boundary condition~\eqref{eq:bc} on $\mathbb{B}$.
If the color gradient normal to a segment is also continuous across, 
then the segment as a boundary has no influence on the 
solution color field $u$.
A mathematical explanation is in \S3 of the supplementary document.
Precisely, a normal gradient is continuous when
\begin{equation} \label{eq:redundant} 
d_{\bm{n}}(\xx) = \frac{\partial u(\xx)}{\partial n_{\ell}} + \frac{\partial u(\xx)}{\partial n_r}, \quad \xx \in \mathbb{B},
\end{equation}
is zero.
In practice, we solve $u$ using the Finite Element Method (\S\ref{sec:num}) and check if $|d_{\bm{n}}(\xx)|$ at the center point $\xx$ of each segment is below a threshold.
If so, we mark the segment as unnecessary.
Lastly, for each curve output by Algorithm~\ref{alg:full_pipeline}, we remove a largest set of connected redundant segments to avoid breaking the curve into many small disconnected components.

To transform the final polyline into a standard diffusion curve made from end-to-end connected B\'{e}zier curves, we adopt the Potrace algorithm~\cite{selinger2003potrace}, which was also used in~\cite{Orzan:2008}.

\paragraph{Per-pixel blurring (optional)}
The curves placed in a smooth color region have continuous color values across the curves.
However, since these curves serve as boundaries in the Laplace solve, color gradients may not necessarily remain continuous across curves generated by Algorithm~\ref{alg:full_pipeline}.
Such gradient discontinuities can sometimes lead to noticeable artifacts~\cite{Finch:2011}.
Thus, our pipeline includes an optional step following the original framework of diffusion curves~\cite{Orzan:2008} to perform per-pixel blurring on the rasterized image.
The size of blur kernel at each pixel is determined by another Laplace equation:
\sLNM \begin{align*}
K(\xx) &= K_0(\xx), & x \in \Gamma\\
\Delta K(\xx) &= 0, & \text{otherwise},
\end{align*} \tLNM
where $K_0(\xx)$ gives the desired kernel size along the curves.
In particular, we set $K_0(\xx) = 0$ for all $\xx \in \partial\Omega$ since the boundaries and discontinuities in the input color field should never be blurred.
For $\xx \in \mathbb{B}$, the value $d_{\bm{n}}(\xx)$ indicates the magnitude of the gradient domain discontinuity. 
Thus, we set $K_0(\xx) = b\,|d_{\bm{n}}(\xx)|^a$ for all $\xx\in\mathbb{B}$, where $a$ and $b$ are two global parameters.
In our implementation, we set $a = 0.2$ and $b$ to 5\% of the longest axis of $\Omega$'s bounding box.

Notice that more advanced curve coloring techniques, such as \cite{Xie:2014:HDC}, may optimize color gradients across the curves, largely removing gradient discontinuity artifacts.
In this case, per-pixel blurring is unnecessary (see \S\ref{sec:addtl_res}).

\section{Diffusion Curve Optimization}
\label{sec:curve_opt}
We now detail the core of our pipeline, the optimization of diffusion curve
geometries to approximate a given 2D color field.  We first describe an
algorithm minimizing the approximation error of diffusion curve images
(\S\ref{sec:opt_prob}-\ref{sec:gds}), and then extend it to balance accuracy
against curve length (\S\ref{sec:balance}).
Lastly, we provide implementation details (\S\ref{sec:num}), followed by the discussions of further extensions (\S\ref{sec:curveopt_discuss}).

We introduce \emph{Shape Optimization}~\cite{sokolowski1992introduction} to formulate the inverse diffusion curve problem.
While building our approach on existing shape optimization concepts and theories (\S\ref{sec:gds}), we also develop a new formula for regularizing curve length (\S\ref{sec:balance}).
Please refer to the supplementary document for complete derivations and a review of related background.

\begin{figure}[t]
	\centering
	\includegraphics[height=0.97in]{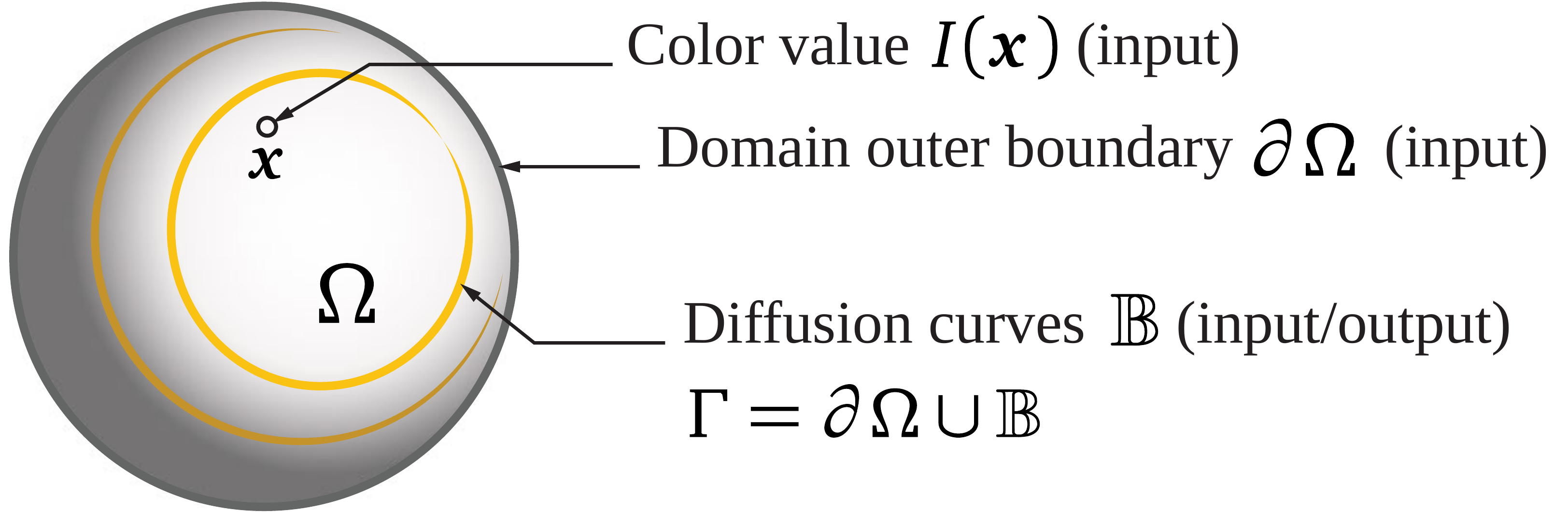}
	\caption{\label{fig:input_output}
		{\bf Input} and {\bf output} of our curve optimization algorithm. 
		{\bf Input:} color field $I$, domain $\Omega$ and its outer boundary $\partial\Omega$, initial diffusion curves $\mathbb{B}$; {\bf output:} refined curves $\mathbb{B}$.}
\end{figure}

\paragraph{Curve Optimization in a Nutshell}
The major steps of our approach are outlined in Algorithm~\ref{alg:gd}.
Its input includes the color field $I$, a 2D closed domain
$\Omega$ over which $I$ is defined, and a set of initial curves $\mathbb{B}$ in
$\Omega$ (Figure~\ref{fig:input_output}).
In this section, the color field $I$ is treated as a black box, allowing $I(\xx)$ 
and $\nabla I(\xx)$ to be evaluated for any $\xx \in \Omega$.
Our curve optimization algorithm then iteratively refines the curves by changing their shapes (i.e., the trajectories) and topologies to obtain better approximation. 
The resulting diffusion curve image consists of the optimized curves $\mathbb{B}$ and the domain boundary $\partial\Omega$.
During the optimization process, the colors along both sides of these curves
(i.e., $C_{\ell}$ and $C_r$ of the Laplace equation~\eqref{eq:laplace}) are sampled from the given color filed $I$, and the approximated color value $u(\xx)$ for all $\xx \in \Omega$ is determined
according to the equation~\eqref{eq:laplace} with the Dirichlet boundary condition
\begin{equation}\label{eq:bc}
u(\xx) = I(\xx), \;\;\forall \xx\in\mathbb{B}\cup\partial\Omega.
\end{equation}
We note that rather than sampling color values along the curves, 
prior methods~\cite{jeschke2011estimating,Xie:2014:HDC} post-optimize color values 
after the curves are determined. We will discuss the extension of our method to incorporate
their post-optimization later (\S\ref{sec:curveopt_discuss}) and examine it in our 
experiments (\S\ref{sec:addtl_res}).

\subsection{PDE-Constrained Optimization Problem}\label{sec:opt_prob}
Formally, our iterative curve optimization process minimizes a \emph{cost
functional} defined as the $L_2$ residual of the color approximation,
\begin{equation}\label{eq:residual}
R(\Omega;\mathbb{B}) = \frac{1}{2}\int_\Omega (u(\xx) - I(\xx))^2\,\intd\Omega,
\end{equation}
Here $u$ is the color field determined by diffusion curves.
We write $\mathbb{B}$ as a parameter of $R$ to emphasize the dependence of the
residual on $\mathbb{B}$ through the Dirichlet boundary condition~\eqref{eq:bc}.
Since $u$ is the solution of the Laplace equation~\eqref{eq:laplace}, we are concerned with an
optimization problem with a PDE constraint,
\sLNM \begin{empheq} [box={\mybox[1.8pt][1.5pt]}]{equation}\label{eq:curveopt} 
\min_{\mathbb{B}} R(\Omega;\mathbb{B})\;\;\text{s.t. $u$ satisfies the Laplace eqn.~\eqref{eq:laplace}}.
\end{empheq} \tLNM
\emph{PDE-constrained optimization problems} are known to be challenging in general~\cite{pinnau2008optimization}.
In our problem~\eqref{eq:curveopt}, the optimization variables are the shapes of diffusion curves, that is, the spatial trajectories and topologies of the curves. 
Ideally, a curve can have an arbitrarily continuous trajectory, and therefore needs to be represented using a continuous functional rather than using individual and discrete parameters. 
More importantly, the error residual $R(\Omega;\mathbb{B})$ depends on the optimization variables (the curves) through the Laplace equation~\eqref{eq:laplace} in a complex manner:
any local change to the curves $\mathbb{B}$ has a global impact, one that changes $u$ over the
entire domain $\Omega$, which further affects the residual via
\eqref{eq:residual}.

\begin{figure}[t]
	\centering
	\includegraphics[height=1.05in]{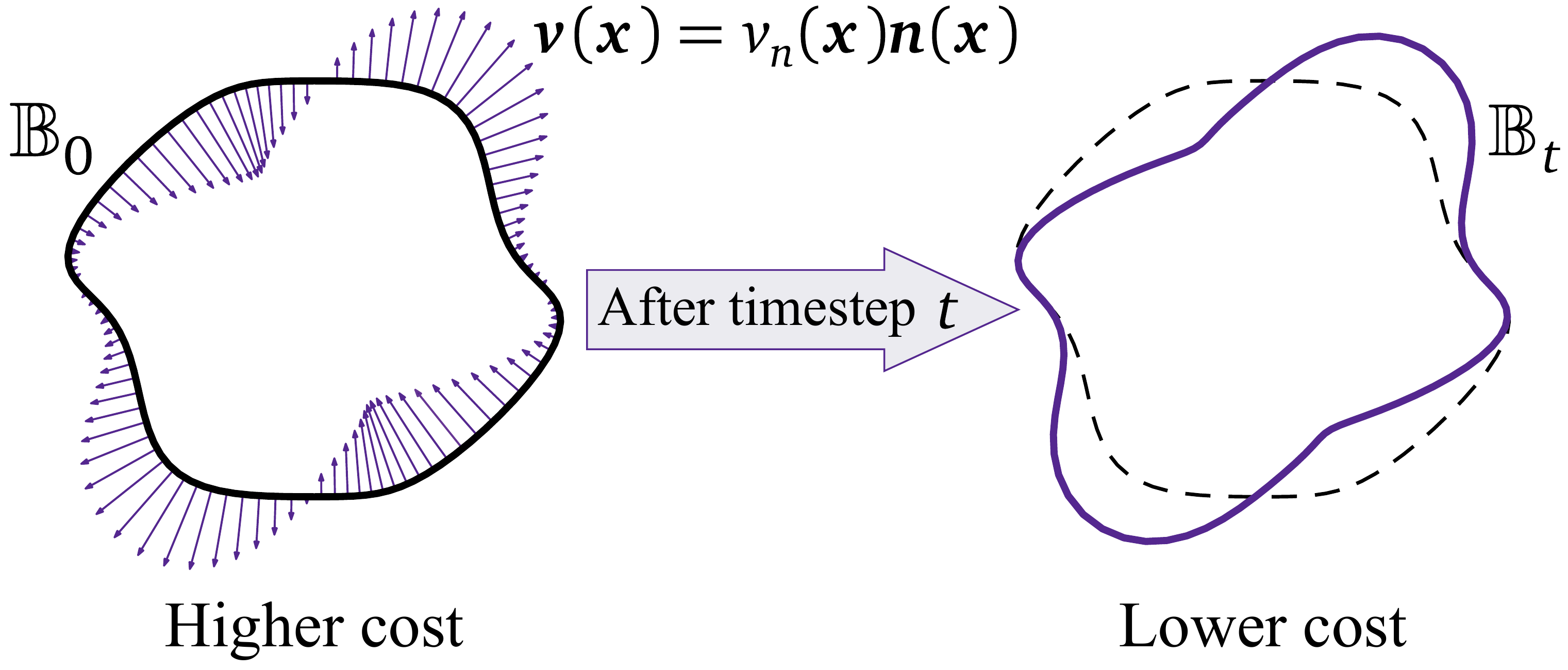}
	\caption{\label{fig:deform}
		{\bf Curve optimization.} Given a set of curves $\mathbb{B}_0$, we
		construct a velocity field $\vv$ so that if one deforms $\mathbb{B}_0$ according to $\vv$, the resulting curves $\mathbb{B}_t$ provide a lower cost.}
\end{figure}

\begin{algorithm}[b]
	\caption{Gradient-descent diffusion curve optimization}
	\begin{algorithmic}[1]
		\Require initial curves $\mathbb{B}_0$, color field $I$ on $\Omega$ with boundary $\partial\Omega$
		\Procedure{CurveOpt}{$\BO$, $\partial\Omega$, $I$, $\Omega$}
		\State $\Delta R\gets\infty;\;\;\mathbb{B}\gets\BO$\Comment{$\Delta R$ tracks residual change}
		\While{$\Delta R > \epsilon$}
		\State triangulate $\Omega$ using $\partial\Omega\cup\mathbb{B}$ as boundaries \Comment{\S\ref{sec:num}}
		\State solve the Laplace equation~\eqref{eq:laplace} for $u(\xx)$
		\State solve the Poisson equation~\eqref{eq:adj_laplace} for $p(\xx)$ \Comment{\S\ref{sec:gds}}
		\State compute $v_n(\xx)=-B_R(\xx)$ using~\eqref{eq:dJ1}
		\State forward-Euler curve advancement, $\mathbb{B}\gets\mathbb{B} + v_n t$
		\State evaluate $R$ using~\eqref{eq:residual}, and update its change $\Delta R$
		\EndWhile
		\State \Return current curves $\mathbb{B}$
		\EndProcedure
	\end{algorithmic}
	\label{alg:gd}
\end{algorithm}
\subsection{Gradient-Descent Solver}\label{sec:gds}
We propose a new approach for solving the curve optimization problem~\eqref{eq:curveopt}, following the general spirit of gradient descent.
Starting from a set of initial curves, our approach iteratively decreases the
residual~\eqref{eq:residual} by adjusting their shapes.
Throughout, a fundamental difficulty we need to address is the computation of the residual's ``gradient'' with respect to the shapes of the curves, as the conventional gradient in terms of continuous curves is undefined.

We develop our method from the perspective of functional analysis: in each
gradient-descent step, we first construct a velocity field $\bm{v}$ on the
curves, specifying $\bm{v}(\xx)$ for all $\xx\in\mathbb{B}$ (Figure~\ref{fig:deform}). 
We then use $\bm{v}(\xx)$ to deform the curves, analogous to a (2D) surface flow in geometry
processing~\cite{sethian2003level,brakke1992surface}.
In other words, we evolve the curves via a single step of the forward Euler method of integrating $\dot{\xx} = \vv(\xx),\,\forall\xx\in\mathbb{B}$. 

In this subsection, we present the details of computing such a $\vv$
that after deforming the curves accordingly, the residual is guaranteed to
decrease (Lines 5--7 of Algorithm~\ref{alg:gd}).
Briefly speaking, we will first assume that $\bm{v}(\xx)$ is known and analytically express
how much the residual would change if the curve is deformed according to $\bm{v}$.
This analytical expression allows us to formulate the condition of $\bm{v}$ resulting in a decrease of 
the residual, and thereby provides us a recipe for computing $\bm{v}$.

\paragraph{\FC derivative as a linear form} 
Given a domain $\Omega$ and a set of initial curves $\mathbb{B}_0$, we consider a general cost functional,
\sLNM \begin{equation}\label{eq:gcf}
C(\Omega; \BO) = \int_{\Omega_0} y(\xx;\BO)\,\intd\Omega,
\end{equation} \tLNM
where $y$ is continuous on $\Omega$ and may depend on the choice of $\BO$.
Our residual~\eqref{eq:residual} takes the form $y(\xx;\mathbb{B}_0) = \frac{1}{2}(u(\xx)-I(\xx))^2$ 
and depends on $\mathbb{B}_0$ via the Laplace solution $u$.
Assuming a known $\vv$, 
we introduce the \emph{\FC derivative}~\cite{coleman2012calculus}
of $C$ with respect to $\vv$.
Let $\mathbb{B}_t$ denote the curves evolved according to $\vv$ after an infinitesimal time period of 
$t$, that is, $\xx \mapsto \xx + \vv(\xx)\,t$ for all $\xx\in\mathbb{B}_0$ (Figure~\ref{fig:deform}).
The \FC derivative of $C$ is a linear form of $\vv$ satisfying that
\sLNM $$
\intd C(\Omega;\BO) = \lim_{t\downarrow 0}\frac{1}{t} (C(\Omega;\mathbb{B}_t) - C(\Omega;\BO)).
$$ \tLNM
Conceptually, this derivative measures how quickly the cost functional $C$
changes as we deform the curves using $\vv$ infinitesimally.
According to \emph{Hadamard-Zelo\'{e}sio Structure Theorem}~\cite{delfour2011shapes}, such a linear form always exists when $\Omega$, $\BO$ and $\vv$ are sufficiently regular, which is usually the case in practice.
For our cost functional~\eqref{eq:residual}, we further reduce the \FC derivative into a linear form expressed as a boundary integral
\begin{equation}\label{eq:surf_int}
\intd C(\Omega;\BO) = L[\vv(\xx)] := \int_{\Gamma_0} B(\xx) v_n(\xx)\,\intd\Gamma,
\end{equation}
where $v_n(\xx) := \bm{v}(\xx)\cdot\bm{n}(\xx)$ denotes the normal velocity on the curves (Figure~\ref{fig:deform}), $\Gamma_0 = \partial\Omega \cup \BO$ includes both the domain boundary $\partial\Omega$ and all the inner curves $\BO$ (see Figure~\ref{fig:input_output}), and $B$ is another function independent from $\vv$ but related to the specific integrand $y$.
In the rest of this subsection, we aim to derive a formula to evaluate $B(\xx)$ for any $\xx \in \BO$.

Once $B$ is known, setting
\sLNM \begin{equation}\label{eq:vn}
v_n(\xx) = \begin{cases}
    -B(\xx) & \text{if }\xx\in\BO,\\
    0       & \text{if }\xx\in\partial\Omega
\end{cases}
\end{equation} \tLNM
guarantees a negative derivative value in~\eqref{eq:surf_int} (assuming that $B$ does not vanish everywhere on $\BO$).
This provides a formula of constructing $v_n$, which
we then apply to deform the curves $\BO$.
With a sufficiently small timestep size $t$, the deformed curves
$\mathbb{B}_t$, computed by $\xx+v_n(\xx)\,t$ for all $\xx \in \BO$, is guaranteed by construction to yield a
smaller residual value and thus a better approximation of $I$.

\paragraph{Computational Recipe}
Shape Optimization Theory has provided a simple recipe of computing $B$ for our particular cost functional~\eqref{eq:residual}. 
Here we simply present the formulas.
Please see Appendix~\ref{app:deriv} for an outline of the derivation and \S3 of
the supplementary document for more details.

We first solve the Laplace equation~\eqref{eq:laplace} to compute $u(\xx)$, which in turn allows us to construct a Poisson equation with a Dirichlet boundary condition,
\begin{equation}\label{eq:adj_laplace}
\begin{aligned}
    \Delta p(\xx) &= u(\xx) - I(\xx) \\
    p(\xx) &= 0, \qquad\qquad\;\; \forall\xx\in\Gamma_0. 
\end{aligned}
\end{equation}
Next, the solution $p$ of this equation, together with $u$, allows the computation of $B(\xx)$ in a simple form
\sLNM
\begin{empheq} [box={\mybox[1.8pt][1.5pt]}]{equation}\label{eq:dJ1} 
B(\xx)=\frac{\partial p(\xx)}{\partial n} \left(\frac{\partial I(\xx)}{\partial n}-\frac{\partial u(\xx)}{\partial n}\right).
\end{empheq}
\tLNM
Combining \eqref{eq:dJ1} and \eqref{eq:vn} computes the normal velocity $v_n$,
the velocity that can deform the curves $\BO$ and lead to a decrease of the approximation
residual~\eqref{eq:residual}.
This computation is performed at each gradient-decent step, and the optimization
process stops when the residual change drops below a threshold $\epsilon$.
Figure~\ref{fig:syn_res} illustrates the optimization process with synthetic examples.

\subsection{Regularizing Curve Complexity}
\label{sec:balance}
\begin{figure}[t]
	\addtolength{\tabcolsep}{-5.5pt}
	\centering
	\begin{tabular}{ccccc}
		\hspace{-7pt}
		\includegraphics[height=1.08in]{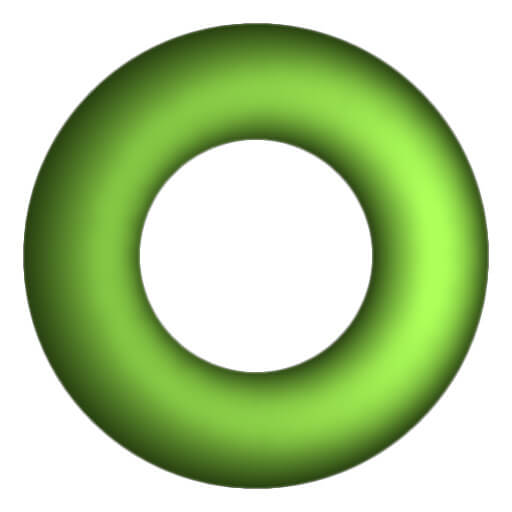}
		\hspace{-5pt}
		&
		\begin{adjustbox}{valign=t, raise=1.15in}
			\begin{tabular}{c}
				\hspace{-4pt}
				\includegraphics[height=0.6in]{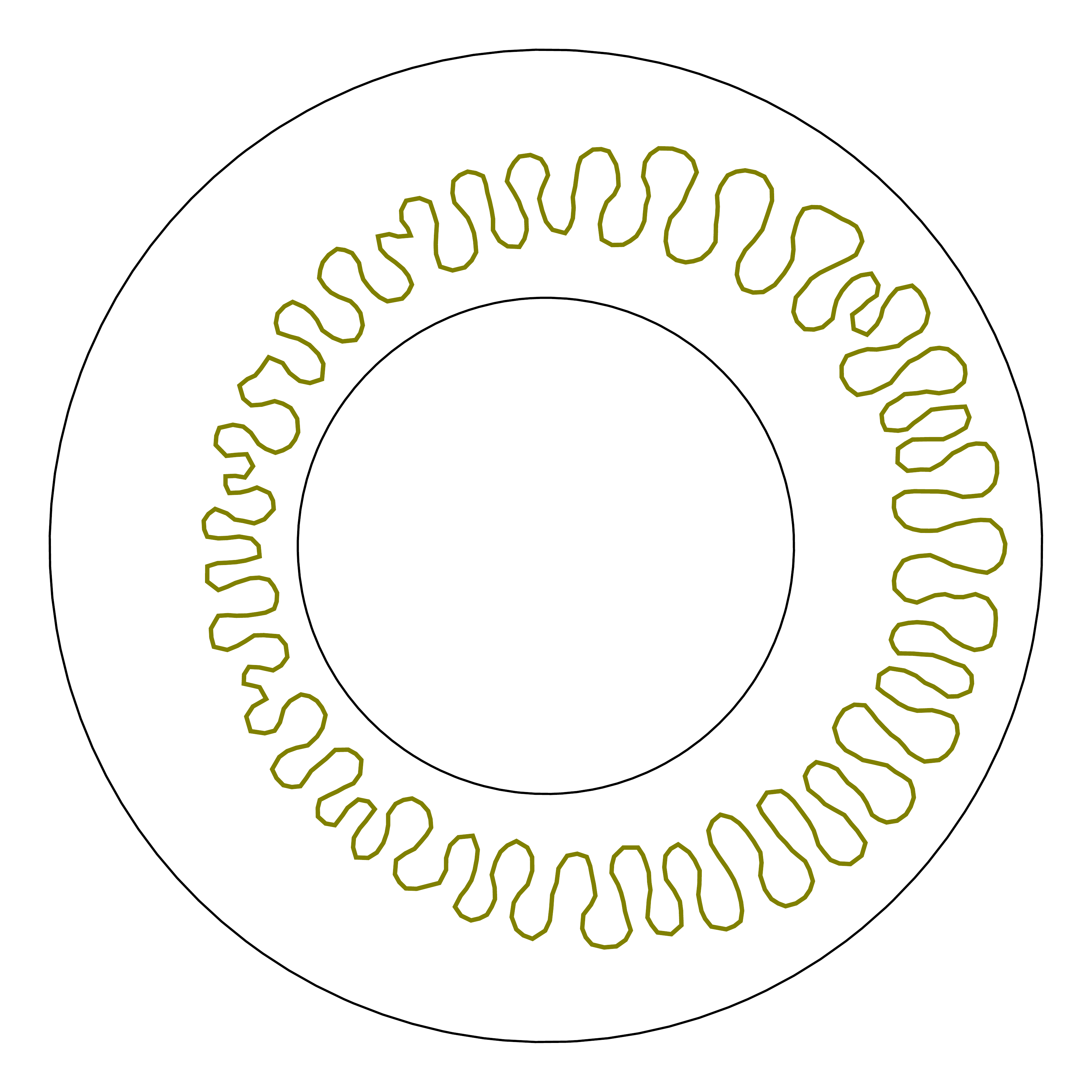}
				\hspace{-2pt}\\[-3pt]
				\hspace{-3pt}Curves\\
				\hspace{-4pt}
				\includegraphics[height=0.6in]{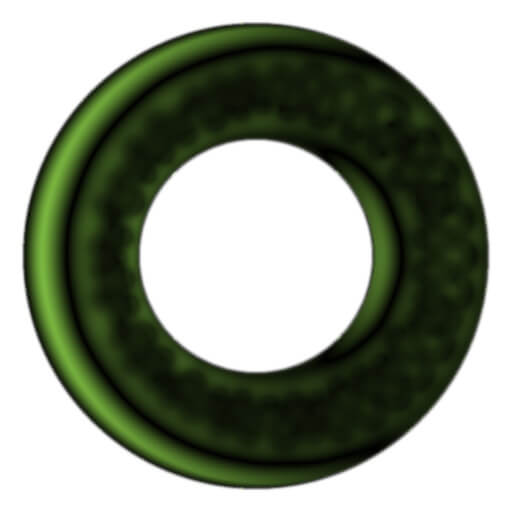}
				\hspace{-2pt}\\[-3pt]
				Error ($8\times$)
			\end{tabular}
		\end{adjustbox}
		&
		\hspace{10pt}
		&
		\hspace{-10pt}
		\includegraphics[height=1.08in]{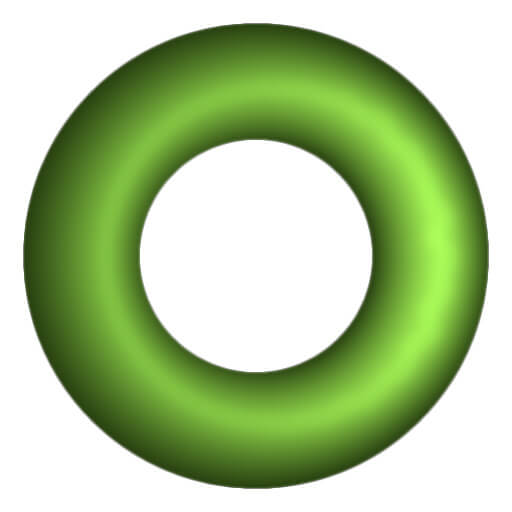}
		\hspace{-5pt}
		&
		\begin{adjustbox}{valign=t, raise=1.15in}
			\begin{tabular}{c}
				\hspace{-4pt}
				\includegraphics[height=0.6in]{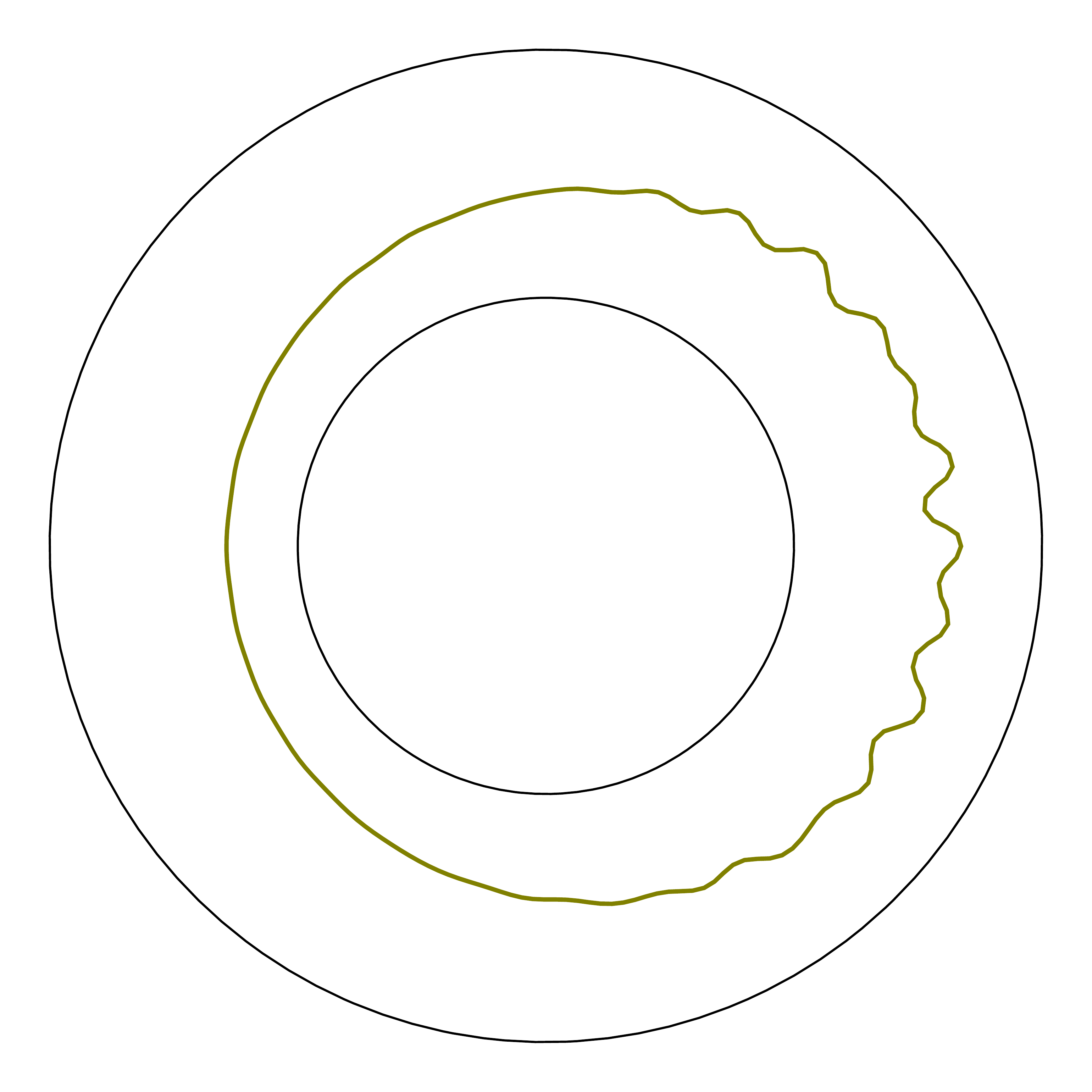}
				\hspace{-2pt}\\[-3pt]
				\hspace{-3pt}Curves\\
				\hspace{-4pt}
				\includegraphics[height=0.6in]{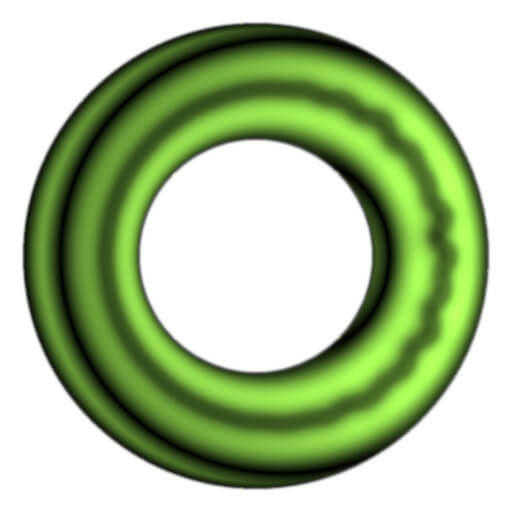}
				\hspace{-2pt}\\[-3pt]
				Error ($8\times$)
			\end{tabular}
		\end{adjustbox}	 	
		\\~\\[-5pt]
		\multicolumn{2}{c}{(a) Weak regularization} &
		&
		\multicolumn{2}{c}{(b) Strong regularization}
	\end{tabular}
	\caption{\label{fig:balance} Our method allows the user to {\bf regularize} curve complexity.
	Subfigures (a) and (b) show two optimization results using the color field illustrated in Figure~\protect\ref{fig:torusA_ref} as input.
	Both results are generated using Algorithm~\protect\ref{alg:gd} with identical initial configurations (a circle) but varying $\alpha$ values.
	The resulting curves and error images (scaled by $8 \times$) are shown to the right of the final images. See Figures~\protect\ref{fig:pipeline_example} and \protect\ref{fig:comparisons} for results created using our full pipeline (Algorithm~\protect\ref{alg:full_pipeline}).}
\end{figure}

So far, our optimization problem~\eqref{eq:curveopt} focuses solely on minimizing the $L_2$ residual~\eqref{eq:residual}.
However, because the $L_2$ error along a curve is always zero due to the boundary condition~\eqref{eq:bc}, one simple way to yield a very low residual is to use space-filling curves.
Indeed, if we start with one curve in a complex color region, it becomes zigzag after running the optimization for many iterations (Figure~\ref{fig:balance}-a).
While the numerical residual is low for such curves, their largely increased geometric complexity may be undesirable for certain applications (such as vector graphics editing).
Thus, we propose an extension to the cost functional~\eqref{eq:residual},
providing users the flexibility to trade approximation accuracy for simpler
curves.
To this end, we add a regularization term to \eqref{eq:residual} to penalize the total length of the curves:
\begin{equation}\label{eq:objfunc2}
\tilde{R}(\Omega;\mathbb{B}) = \frac{1}{2}\int_\Omega (u(\xx) - I(\xx))^2\,\intd\Omega +
\alpha\int_{\mathbb{B}} \intd\Gamma,
\end{equation}
where $\alpha$ is a user-specified scalar controlling the strength of regularization.
It can be shown that similar to~\eqref{eq:surf_int},  the \FC derivative of the second term is also a linear form of $v_n$.
Let $R_L(\mathbb{B}_0)=\int_{\mathbb{B}_0} d\Gamma$.
Then its derivative is
\begin{equation}
\intd R_L(\mathbb{B}_0) = \int_{\mathbb{B}_0} \kappa(\xx) v_n(\xx)\,\intd\Gamma,
\end{equation}
where $\kappa(\xx)$ measures the curvature of a point $\xx$ on the curves.
This formula has been used to derive the mean curvature
flow~\cite{mantegazza2011lecture} in geometry processing.
It is also a special case of the \FC derivative of a general boundary integral (see \S1.3 of the supplementary document).
Following the derivation of $B$ in \S\ref{sec:gds}, we obtain the normal velocity for decreasing $\tilde{R}(\Omega;\mathbb{B})$, that is, $v_n(\xx) = -B_R(\xx) - \alpha \kappa(\xx)$. 
With this slightly different velocity formula, the entire optimization algorithm remains the same as before. 
In addition, the user is able to control the complexity of resulting curves by adjusting 
the strength of regularization (Figure~\ref{fig:balance}).

\subsection{Implementation Details}
\label{sec:num}
We now present implementation details of Algorithm~\ref{alg:gd}, wherein two
major steps are solving the Laplace equation~\eqref{eq:laplace} and the
Poisson's equation~\eqref{eq:adj_laplace}.
Both PDEs have Dirichlet boundary conditions defined on the boundary of
$\Omega$ and the optimized curves $\mathbb{B}$ (recall~\eqref{eq:bc}).
Since we also need to evaluate the domain integral over $\Omega$ during the
iterations (Line 9 of Algorithm~\ref{alg:gd}), we triangulate the entire domain
of $\Omega$ and use the Finite Element Method~\cite{zienkiewicz1971finite} for
both solves, while other numerical solvers (e.g., the Boundary Element
Method) could also be applied.

\paragraph{Finite element discretization}
We discretize the boundary and optimized curves into piecewise linear segments,
and represent them using polylines. 
The velocity $v_n$ is discretized and stored at 
every vertex along the polylines.
We use the package \textsf{Triangle}~\cite{shewchuk1996triangle} to
triangulate the domain $\Omega$ (Line 4 of Algorithm~\ref{alg:gd}).
The resulting triangle mesh is then used in the finite element solves.
The computation of curves' normal velocity in~\eqref{eq:dJ1} involves boundary normal derivatives of the finite element solutions (i.e., $\partial p/\partial n$ and $\partial u/\partial n$).
We choose the second-order finite element basis, as it offers higher accuracy especially near the boundary (see \S4 of the supplementary document).

\paragraph{Curve tracking}
Advancing the curves using the computed normal velocity (i.e., computing $\mathbb{B}_t$ given $\BO$ and $v_n$ in Figure~\ref{fig:deform}) is a typical yet nontrivial surface tracking problem. 
We use a recently developed explicit tracking approach~\cite{brochu2009robust},
which advances the vertices on curve polylines using explicit forward Euler
method, and then carefully remeshes the polylines to ensure 
correct topology changes and a collision-free state.

\paragraph{Timestep size $t$}
To ensure robust curve tracking, we dynamically set the timestep size $t$ for the forward-Euler curve advancement (Line~8 in Algoirthm~\ref{alg:gd}).
We start with choosing a $t$ value such that the vertex displacement $v_n(\xx)\,t,\forall\xx\in\mathbb{B}$ would not collapse any polyline segment on $\mathbb{B}$.
This ensures that possible topology changes can be robustly processed.
From this starting value, we iteratively halve $t$ until the residual value (after a step of curve deformation) decreases.

\subsection{Discussions} \label{sec:curveopt_discuss}
\paragraph{Measuring geometric complexities}
In Equation~\eqref{eq:objfunc2} and the rest of this paper, we use the
\emph{total length} of all diffusion curves to measure their geometric complexity.
Depending on specific applications, there may exist other metrics more suitable
to user needs. 
As an example, in \S2 of the supplementary document, we discuss another
possible measure which can also be incorporated in our curve optimization
framework.

\paragraph{Coloring schemes}
As described at the beginning of this section, given the curve geometry
$\mathbb{B}$, we specify colors on both sides of each curve by directly
sampling color values from the input color field $I$. 
Alternatively,
prior work~\cite{jeschke2011estimating,Xie:2014:HDC} propose to post-optimize
curve colors for better reconstruction accuracy.
Our method can easily adopt this approach, post-optimizing the colors after the curves
are optimized. We implemented this approach and present the results in \S\ref{sec:addtl_res}.

\paragraph{Higher-order domains and curves}
While our approach focuses on solving the inverse problem of the standard (first-order)
diffusion curves, it can be also applied to higher-order domains.
For instance, as demonstrated in \S\ref{sec:addtl_res}, we can feed $\nabla I$
instead of $I$ to Algorithm~\ref{alg:gd} to compute curves offering a higher
order of smoothness.

In principle, it is possible to generalize \eqref{eq:residual} with
$u(\xx)$ directly given by higher-order (e.g., biharmonic diffusion) curves.
However, this would dramatically complicate the form of the \FC
derivate~\eqref{eq:surf_int} and thus that of the velocity field~\eqref{eq:vn},
causing them significantly more difficult to evaluate numerically.  Therefore,
we leave the extension of \eqref{eq:residual} to higher-order curves as future
work.

%
\begin{figure}[t]
	\centering
	\begingroup
	\addtolength{\tabcolsep}{-4pt}	
	\begin{tabular}{ccccc}
		\multicolumn{5}{c}{	$\overbrace{\hphantom{\hspace*{0.43\textwidth}}}^{\textbf{\normalsize Closed Curve}}$}\\
		\begin{adjustbox}{valign=t}
			\begin{tabular}{c}
				\includegraphics[height=0.9in]{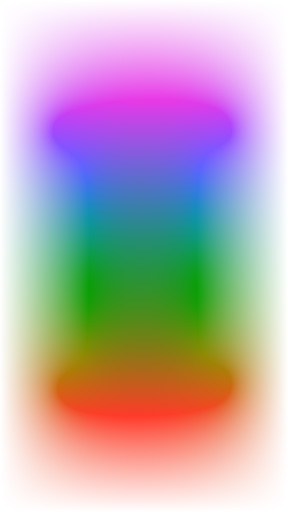}\\[-5pt]
				\includegraphics[height=0.9in]{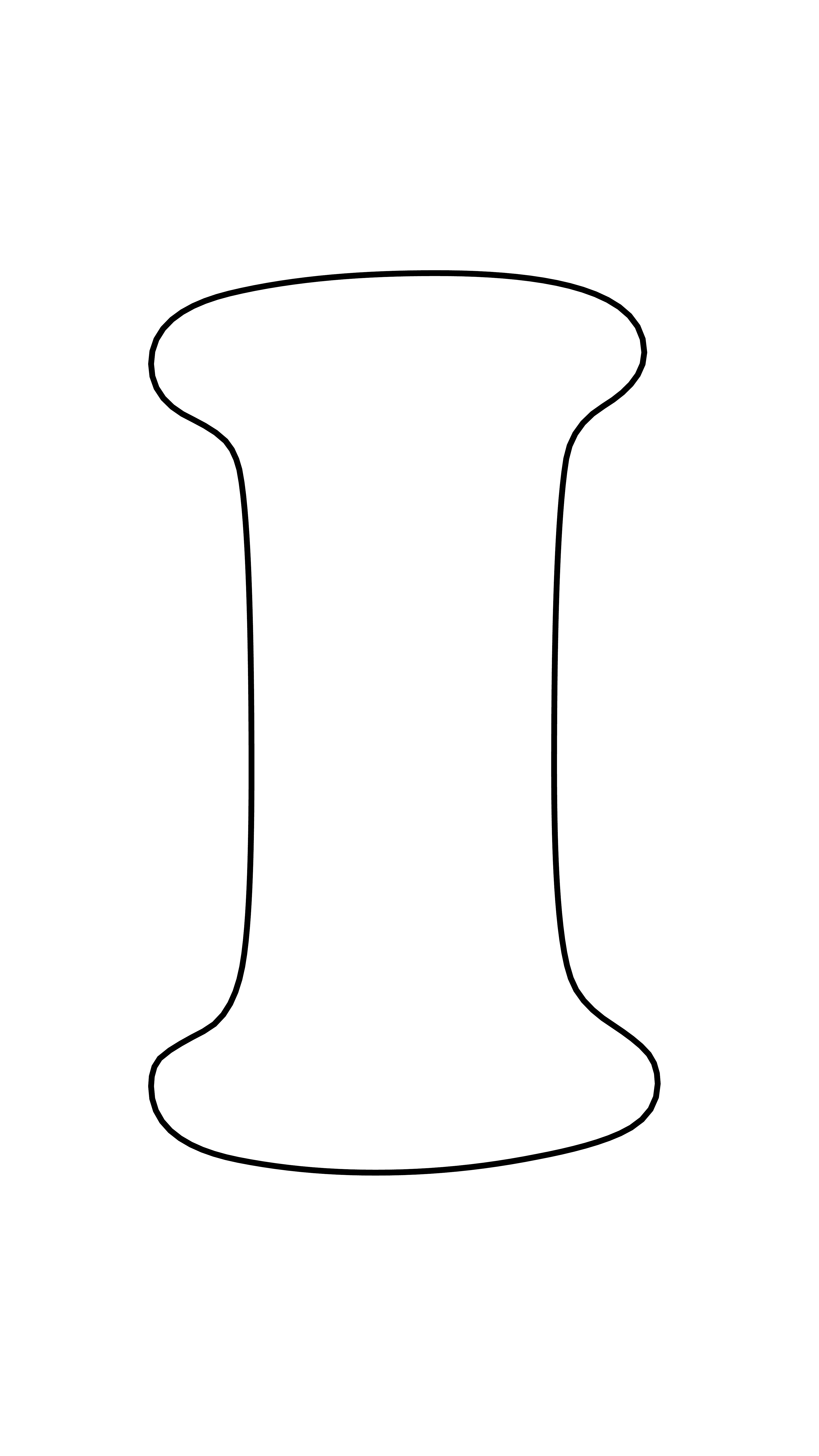}\\[-5pt]
				Reference
			\end{tabular}
		\end{adjustbox}
		&
		\begin{adjustbox}{valign=t}
			\begin{tabular}{c}
				\includegraphics[height=0.9in]{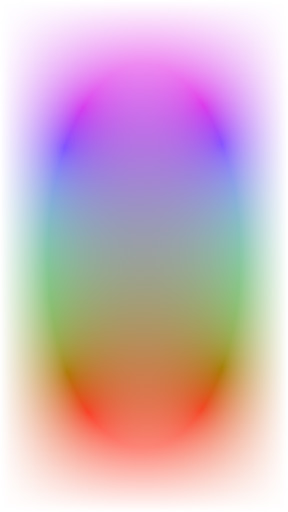}\\[-5pt]
				\includegraphics[height=0.9in]{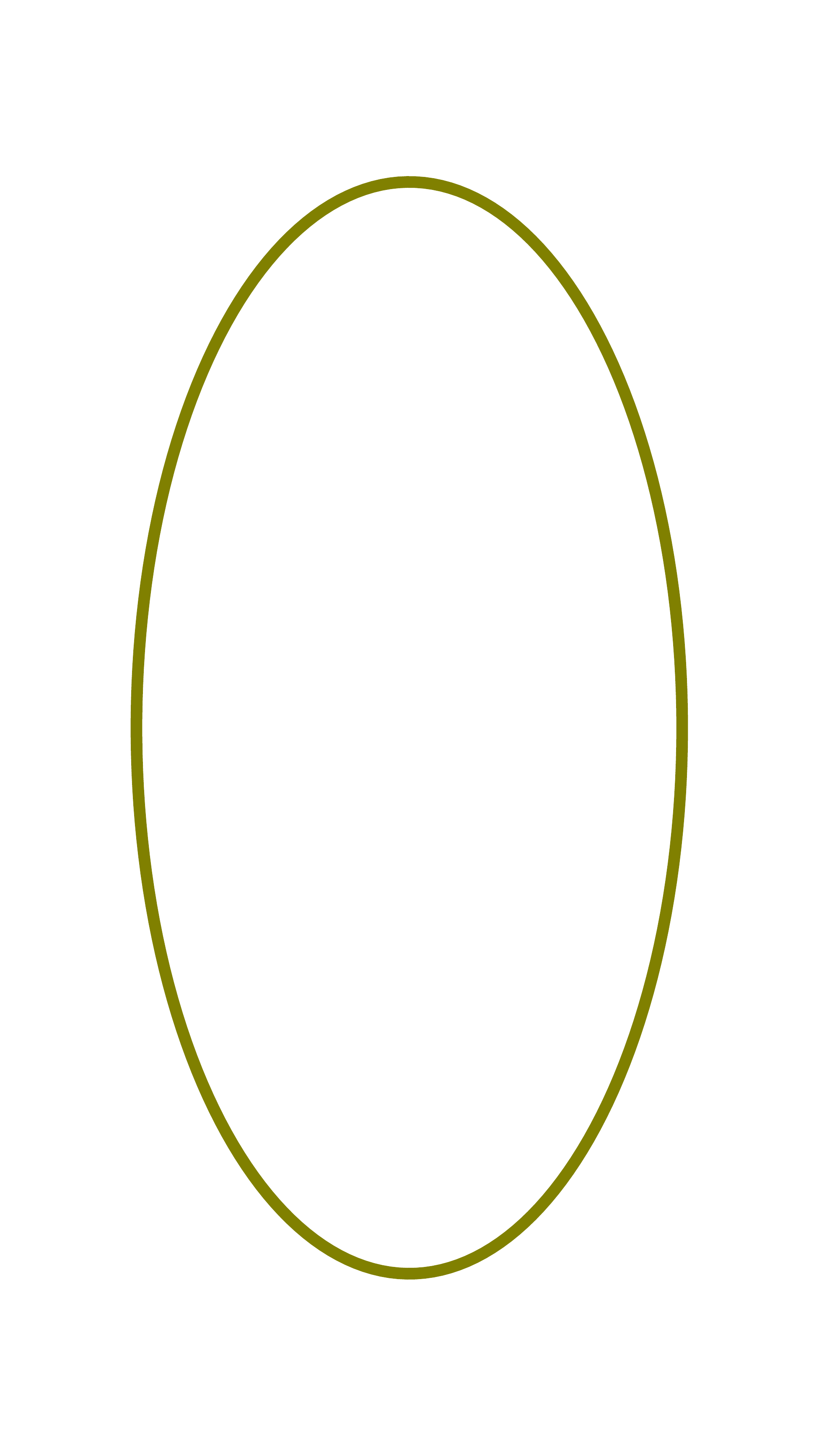}\\[-5pt]
				Initial
			\end{tabular}
		\end{adjustbox}
		&
		\begin{adjustbox}{valign=t}
			\begin{tabular}{c}
				\includegraphics[height=0.9in]{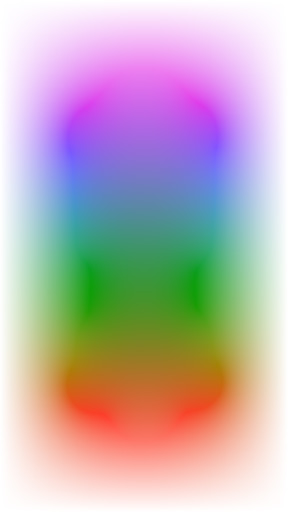}\\[-5pt]
				\includegraphics[height=0.9in]{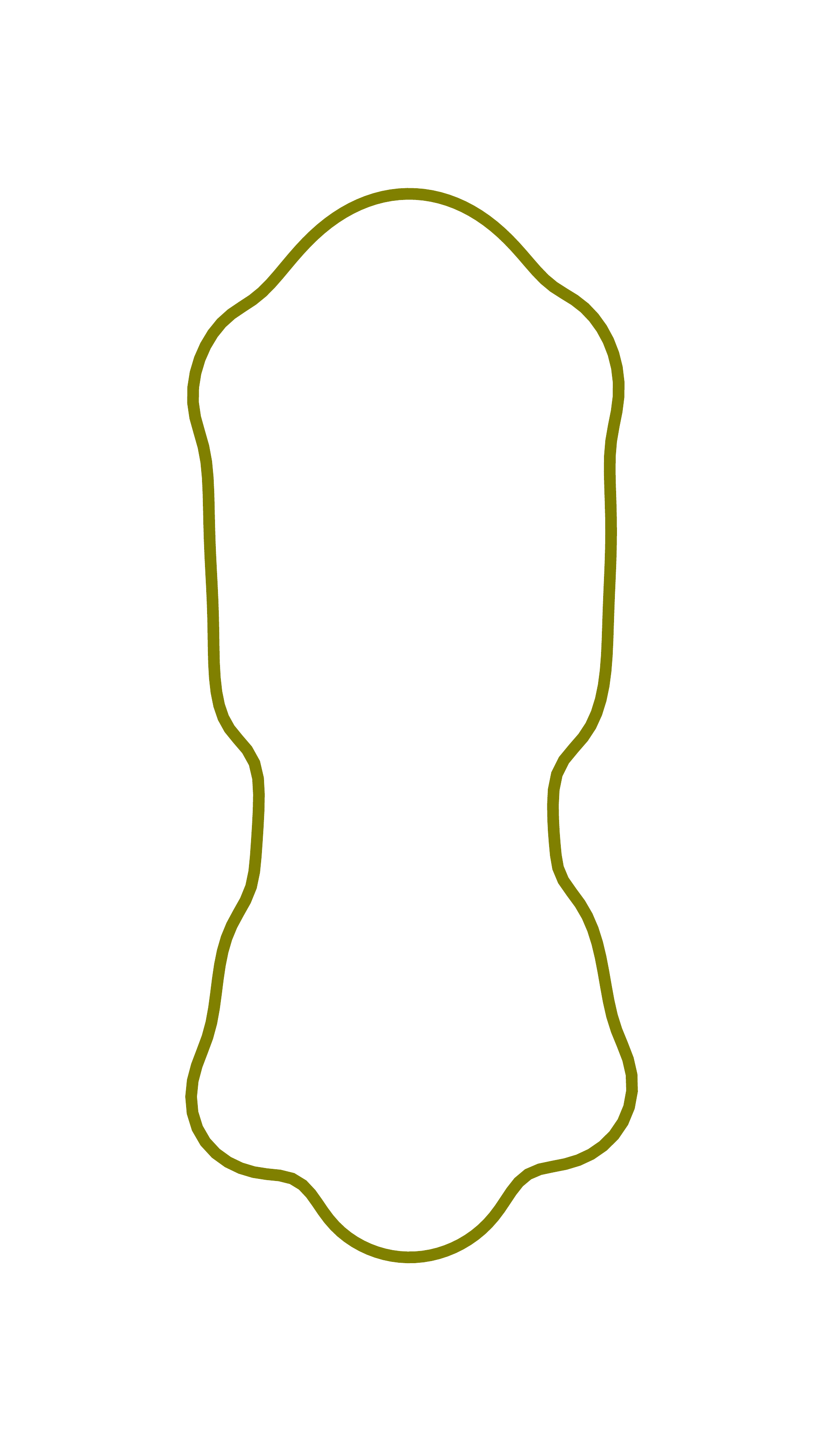}\\[-5pt]
				50 iter.
			\end{tabular}            
		\end{adjustbox}        
		&
		\begin{adjustbox}{valign=t}
			\begin{tabular}{c}
				\includegraphics[height=0.9in]{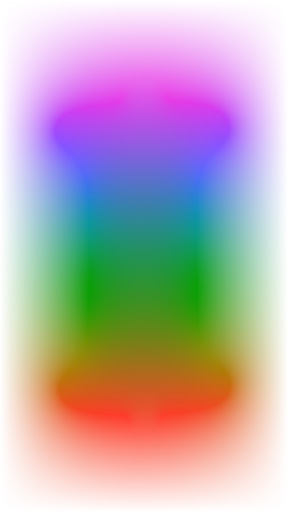}\\[-5pt]
				\includegraphics[height=0.9in]{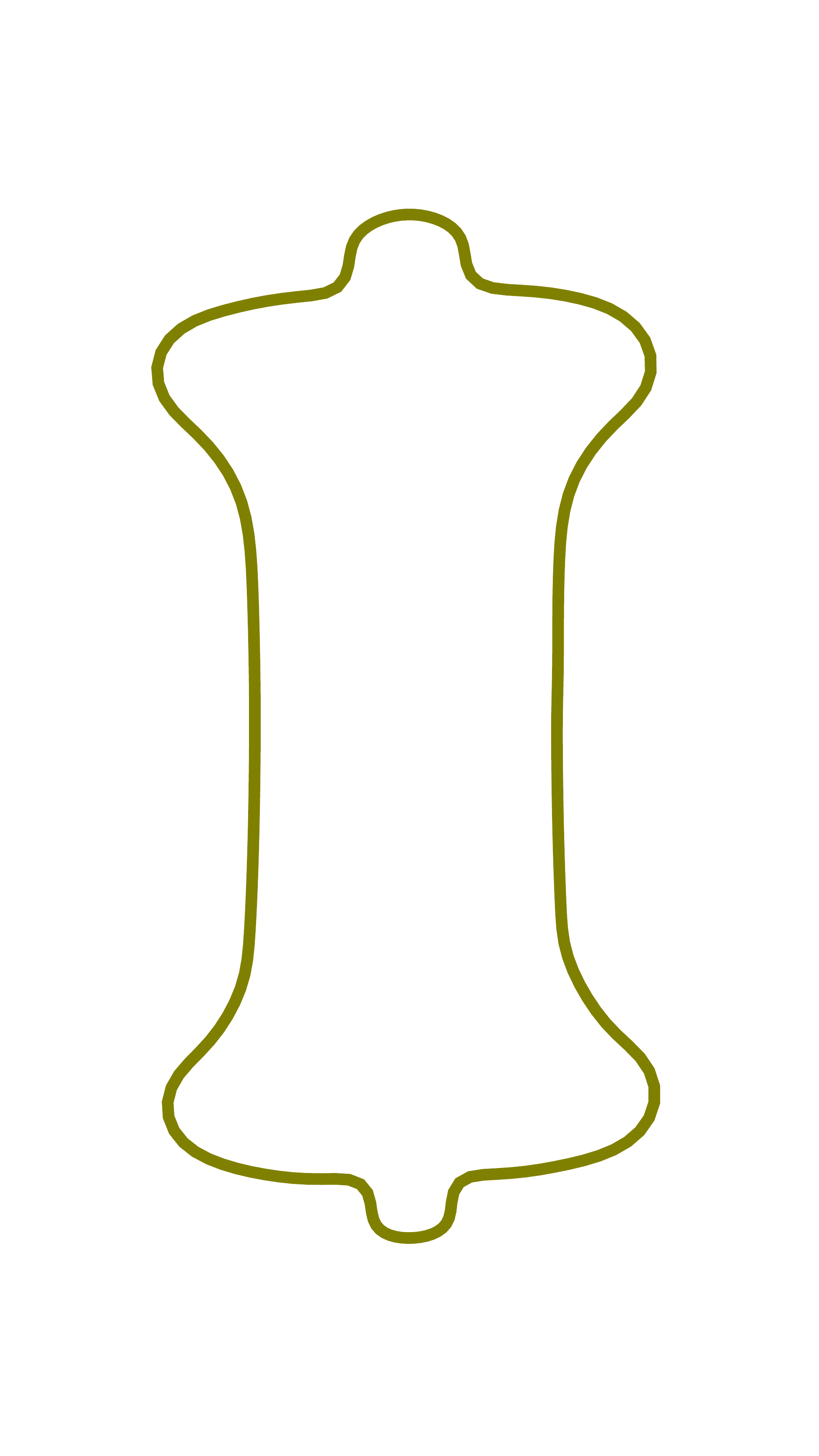}\\[-5pt]
				100 iter.
			\end{tabular}            
		\end{adjustbox}
		&
		\begin{adjustbox}{valign=t}
			\begin{tabular}{c}
				\includegraphics[height=0.9in]{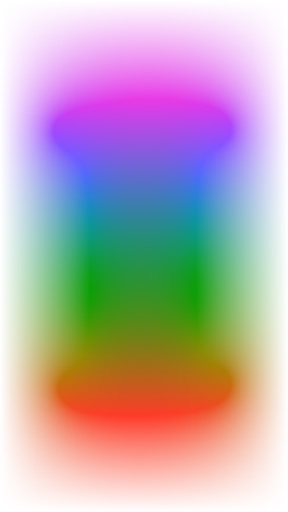}\\[-5pt]
				\includegraphics[height=0.9in]{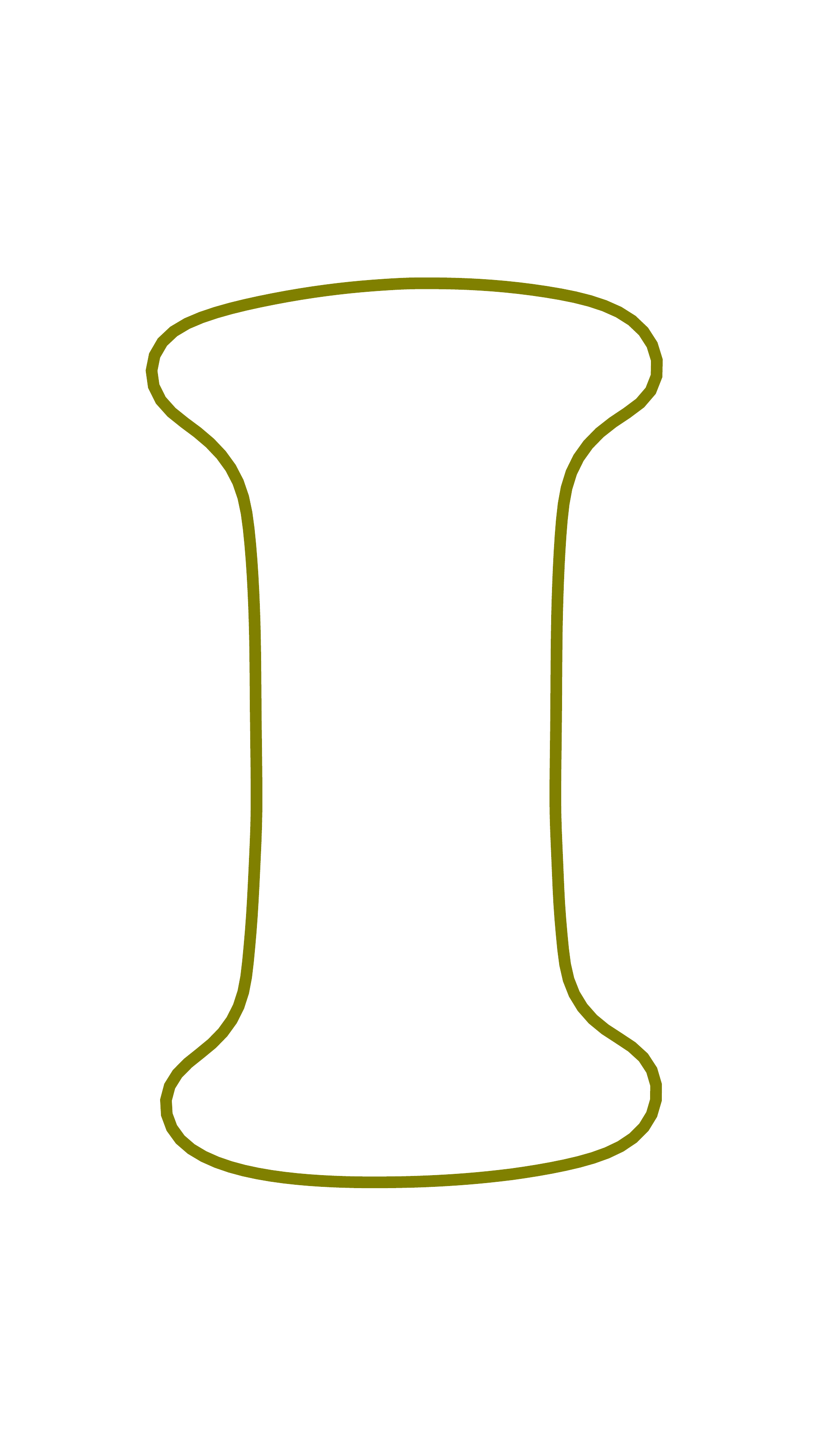}\\[-5pt]
				240 iter.
			\end{tabular}            
		\end{adjustbox}        
	\end{tabular}
	\endgroup
	\vspace{6pt}
	\begingroup
	\addtolength{\tabcolsep}{-6.5pt}
	\begin{tabular}{cccccccc}
		\multicolumn{8}{c}{
		\hspace{-10pt}
		$\overbrace{\hphantom{\hspace*{0.465\textwidth}}}^{\textbf{\normalsize Open Curve}}$}\\
		\includegraphics[height=0.33\columnwidth]{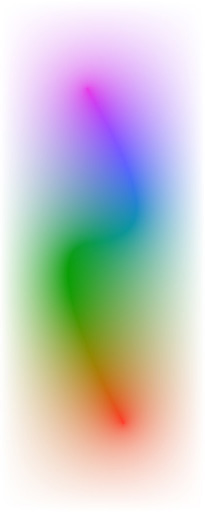} &
		\includegraphics[height=0.33\columnwidth]{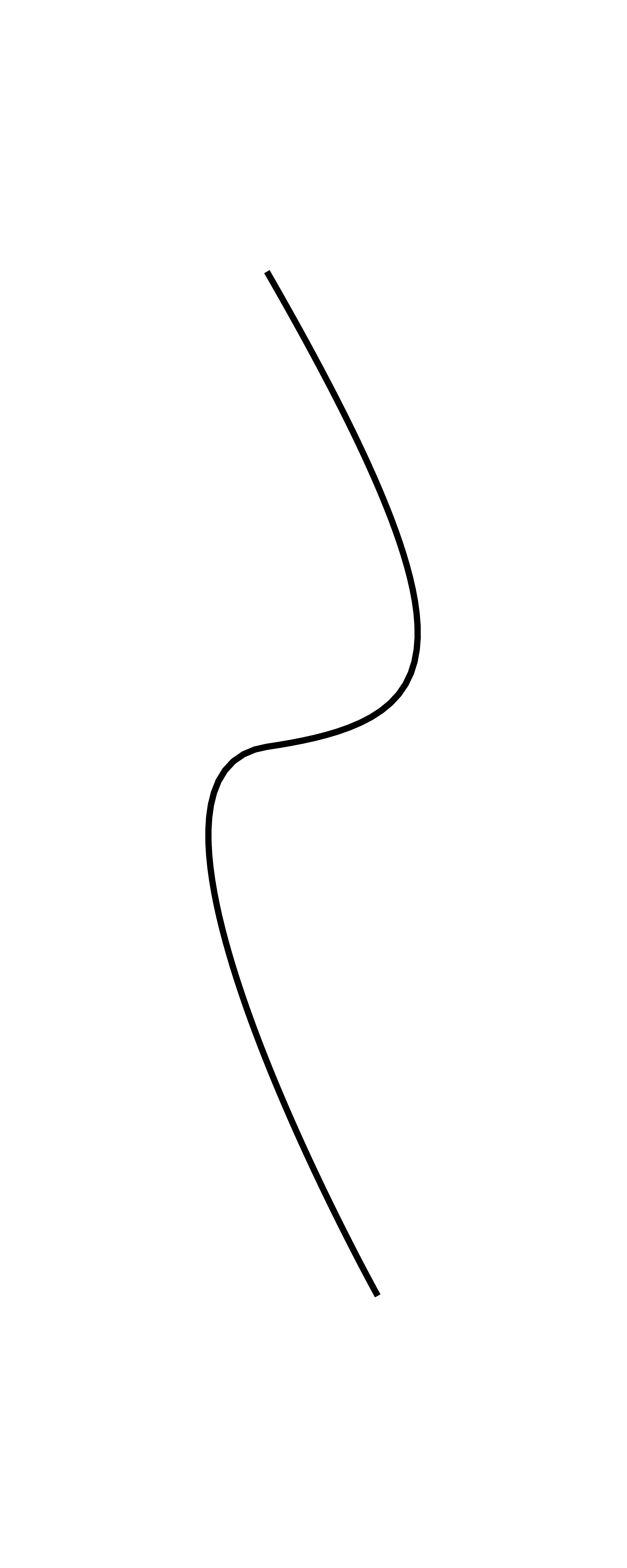} &
		\includegraphics[height=0.33\columnwidth]{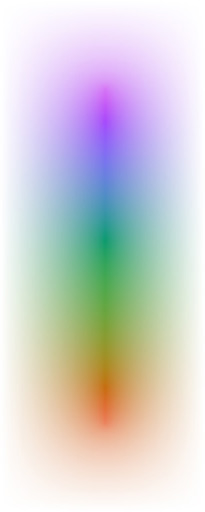} &
		\includegraphics[height=0.33\columnwidth]{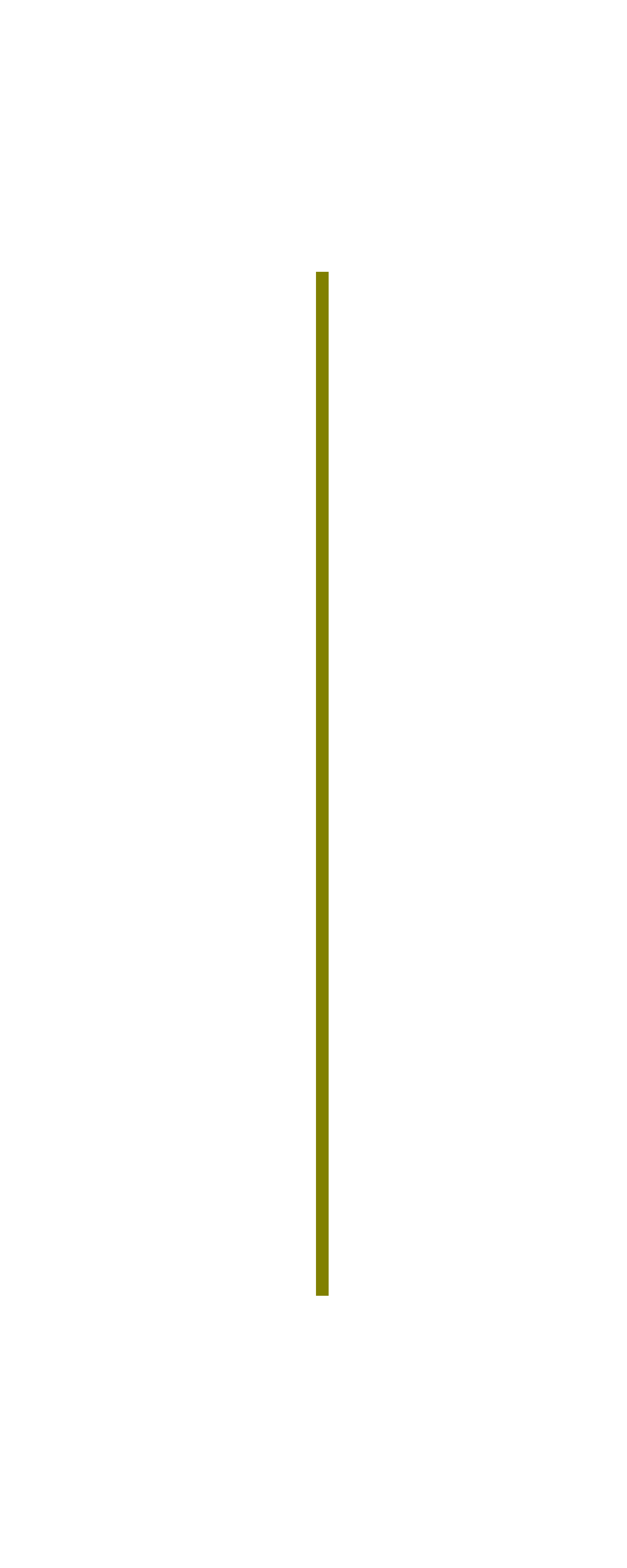} &
		\includegraphics[height=0.33\columnwidth]{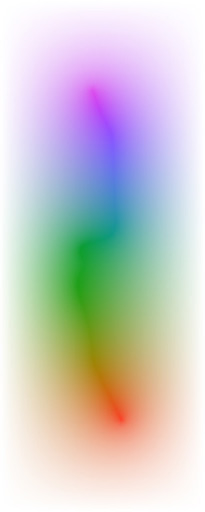} &
		\includegraphics[height=0.33\columnwidth]{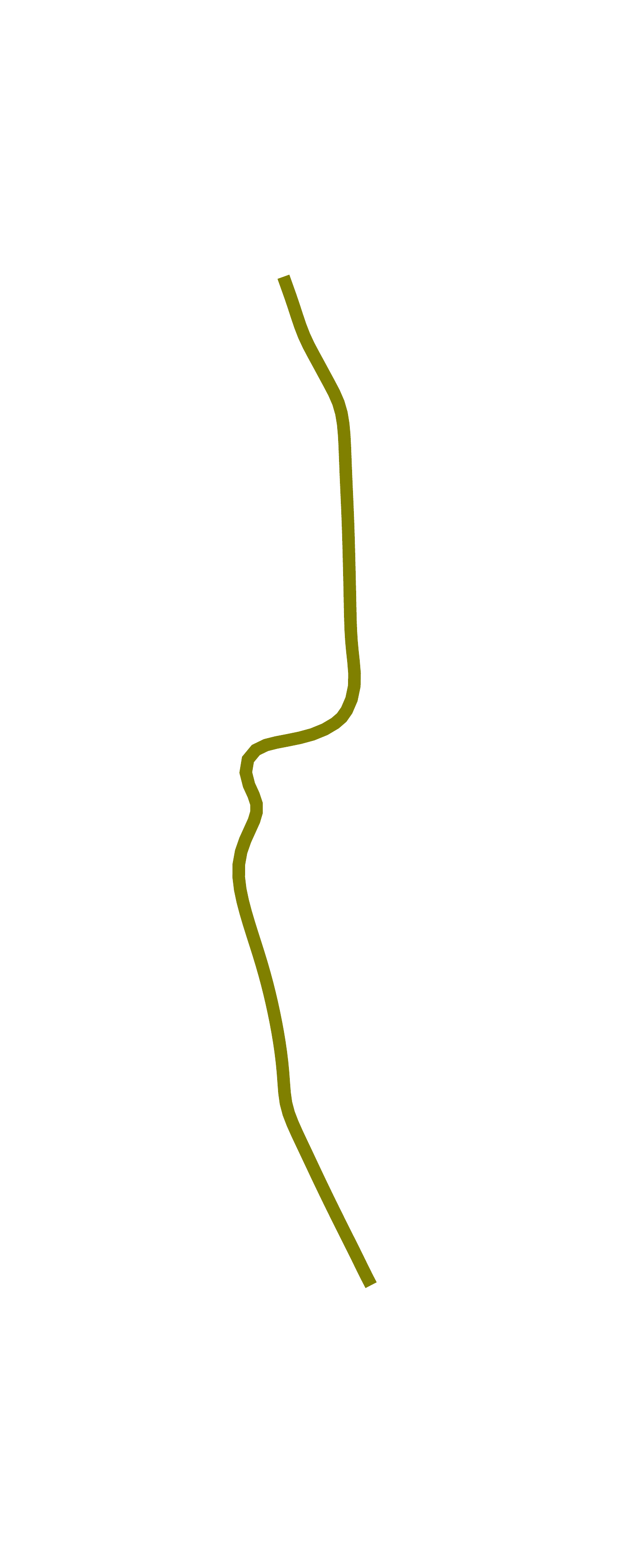} &		
		\includegraphics[height=0.33\columnwidth]{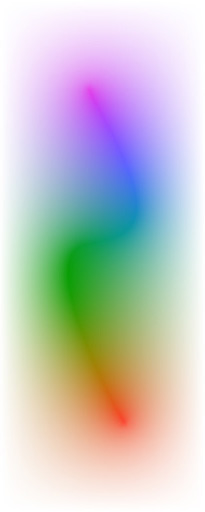} &
		\includegraphics[height=0.33\columnwidth]{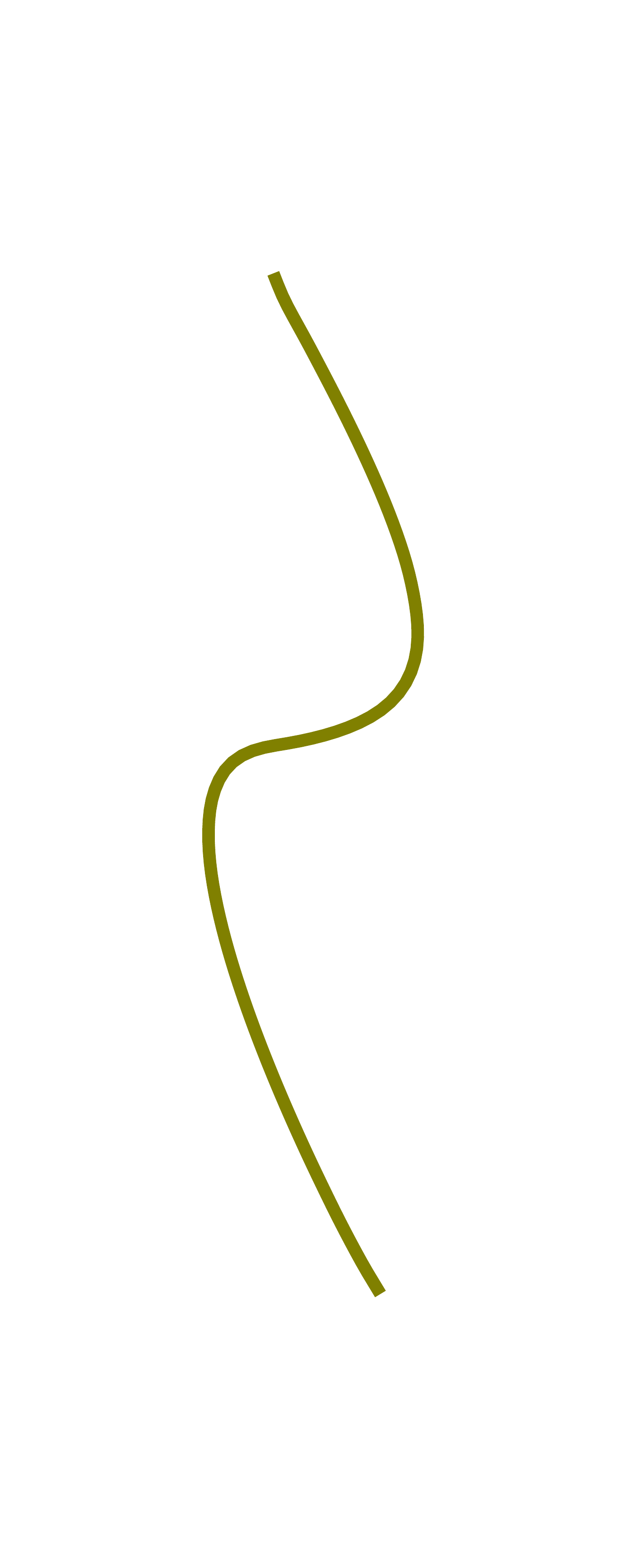}\\
		\multicolumn{2}{c}{Reference} &
		\multicolumn{2}{c}{Initial} &
		\multicolumn{2}{c}{25 iter.} &
		\multicolumn{2}{c}{100 iter.}
	\end{tabular}
	\endgroup
	\caption{\label{fig:syn_res} {\bf Synthetic validation} of our \emph{diffusion curve optimization} algorithm. The left-most column contains input color fields given by one closed curve (top) and one open curve (bottom).
	a varying number of iterations.
	Our resulting curves precisely match the original.}
\end{figure}

\section{Results}
\label{sec:results}

We first (\S\ref{sec:exp_res}) show experimental results demonstrating the validity of our curve optimization algorithm as well as how the regularization behaves in practice.
Then, in \S\ref{sec:main_res}, we show reconstructed diffusion curve images using input color fields represented in three forms: pixel image, 3D renderings, and gradient meshes.
In addition, we show preliminary results motivating possible future applications.

\subsection{Experimental Results}
\label{sec:exp_res}

\paragraph{Synthetic validations}
We design two synthetic tests (Figure~\ref{fig:syn_res}) to validate our diffusion curve optimization algorithm (Algorithm~\ref{alg:gd}).
In these tests, the input color fields $I$ are themselves diffusion curve images with continuous colors.
In this case, the optimal $\mathbb{B}$ is simply the set of diffusion curves used to generate $I$.
Although the shape optimization problem is in general non-convex, our method successfully finds the optimal solutions for both closed and open curves.
Please see the accompanying video for curve deformation animations.

\begingroup
\newlength{\lenDucky}
\setlength{\lenDucky}{1.05in}
\begin{figure}[t]
	\centering
	\addtolength{\tabcolsep}{-4pt}
	\begin{tabular}{ccc}
		(a) Reference &
		\multicolumn{2}{c}{(b) Stronger Regularization}\\
		\includegraphics[width=\lenDucky]{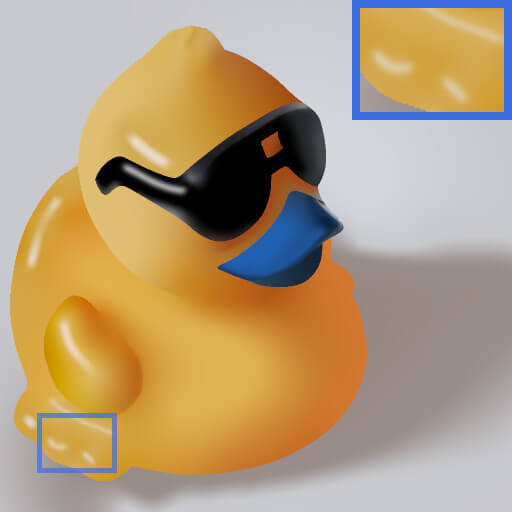} &
		\includegraphics[width=\lenDucky]{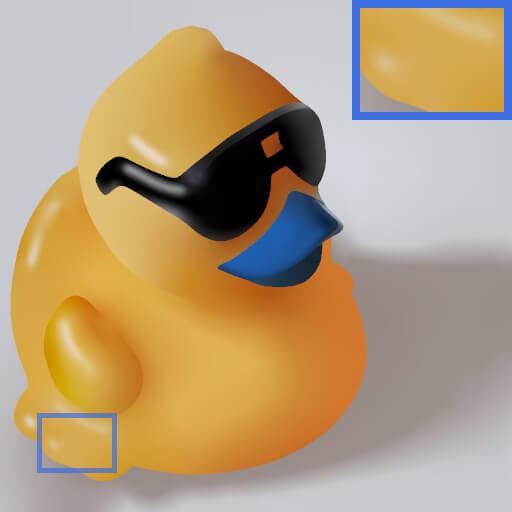} &
		\includegraphics[width=\lenDucky]{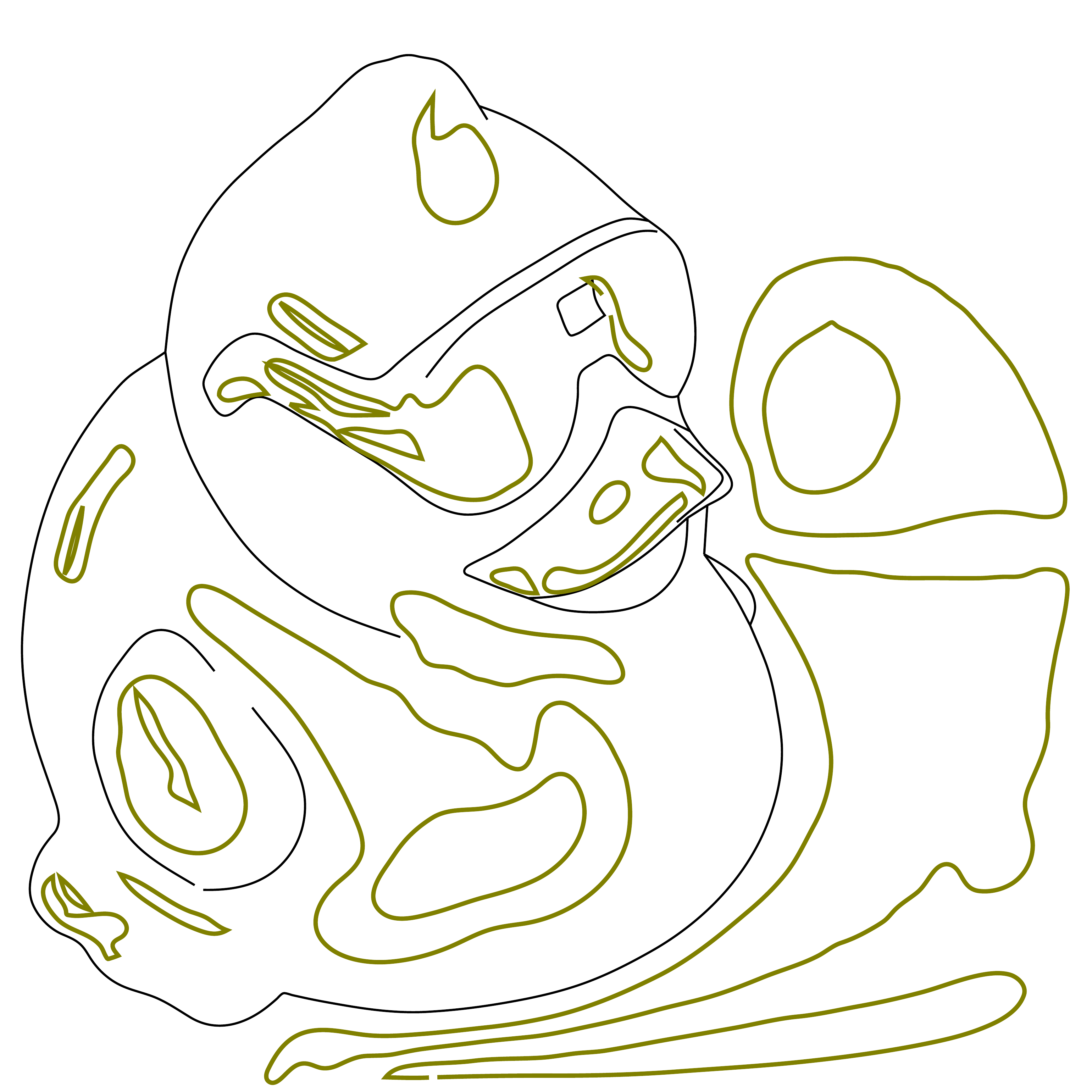}\\
		& RMSE: 0.0135 & Complexity: 17.56\\[5pt]
		&
		\multicolumn{2}{c}{(c) Weaker Regularization}\\
		&
		\includegraphics[width=\lenDucky]{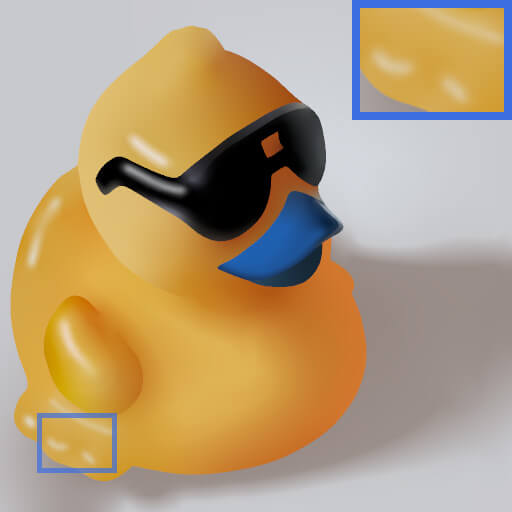} &
		\includegraphics[width=\lenDucky]{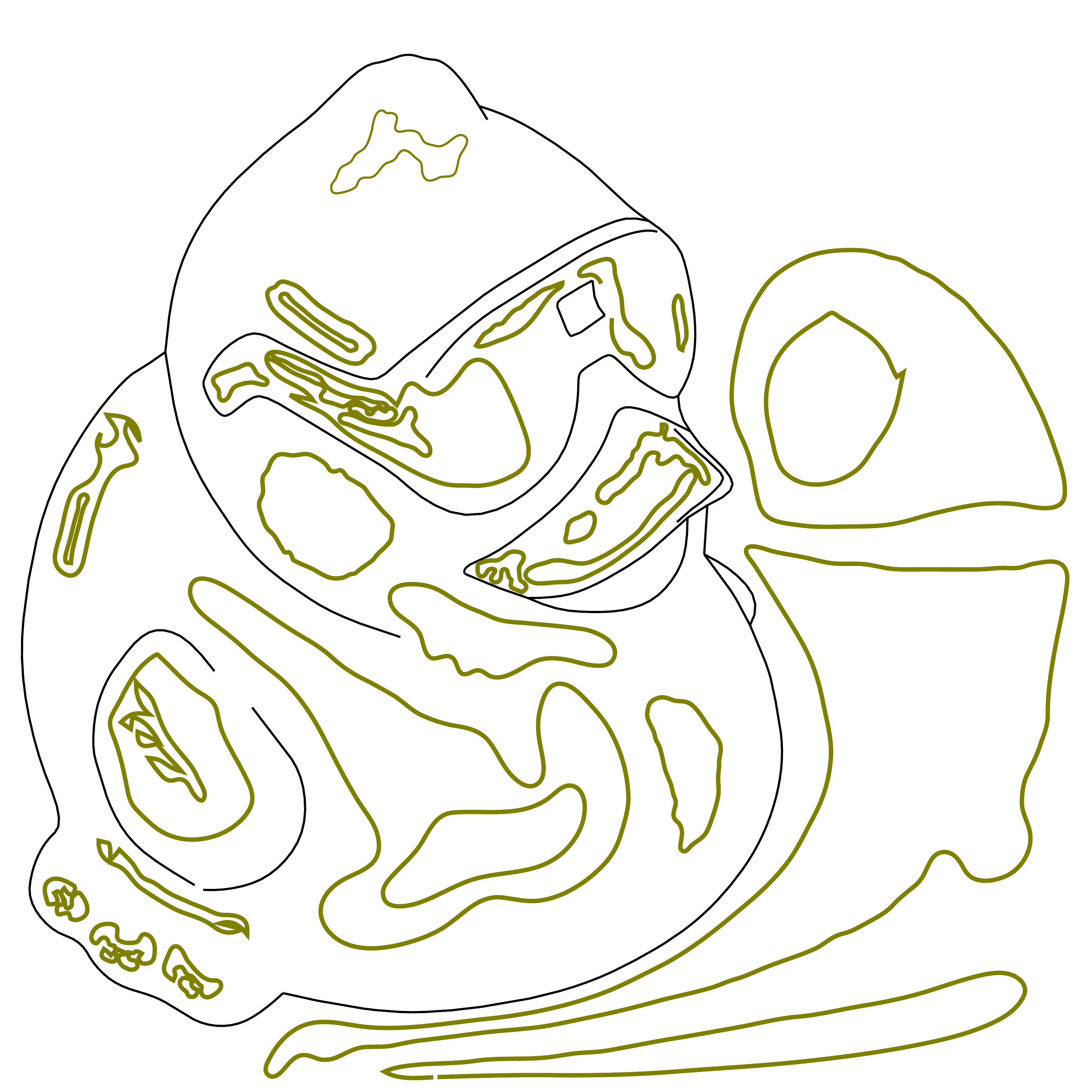}\\
		& RMSE: 0.0108 & Complexity: 20.22
	\end{tabular}
        \caption{\label{fig:ducky} Our method allows the user to {\bf balance} resulting accuracy with curve complexity by varying the strength of regularization.}
\end{figure}
\endgroup

\paragraph{Regularizing curve complexity}
As discussed in \S\ref{sec:balance}, our method is able to balance resulting accuracy and curve complexity by varying the strength $\alpha$ of regularization in (\ref{eq:objfunc2}).
Figure~\ref{fig:ducky} shows how $\alpha$ influences resulting curves generated with our full pipeline (Algorithm~\ref{alg:full_pipeline}).
Figure~\ref{fig:ducky}-a has simpler curves (due to greater $\alpha$), but some highlights at the bottom-left of the image are absent.
Figure~\ref{fig:ducky}-b, on the other hand, provides lower approximation error but at the cost of greater curve complexity (resulting from a lower $\alpha$).
In all our results, the complexity is numerically defined as total curve length normalized, so that the longest axis of each image's bounding box has unit length.

\subsection{Main Results}
\label{sec:main_res}

We now show diffusion curve images generated using our method (Algorithm~\ref{alg:full_pipeline}).
All our results utilize the per-pixel blurring described in \S\ref{subsec:curve_post}.
Please refer to the supplemental material for unblurred versions.

Theoretically, our approach does not require the input color field $I$ to have any particular representation or discretization.
In practice, we demonstrate such flexibility using three types of input: pixel images, 3D renderings, and gradient meshes.
The execution time for generating each of these results is summarized in Table~\ref{tab:perf}.
The supplemental video contains animations demonstrating the creation process for these results.

\begin{figure}
	\centering
	\includegraphics[width=\columnwidth]{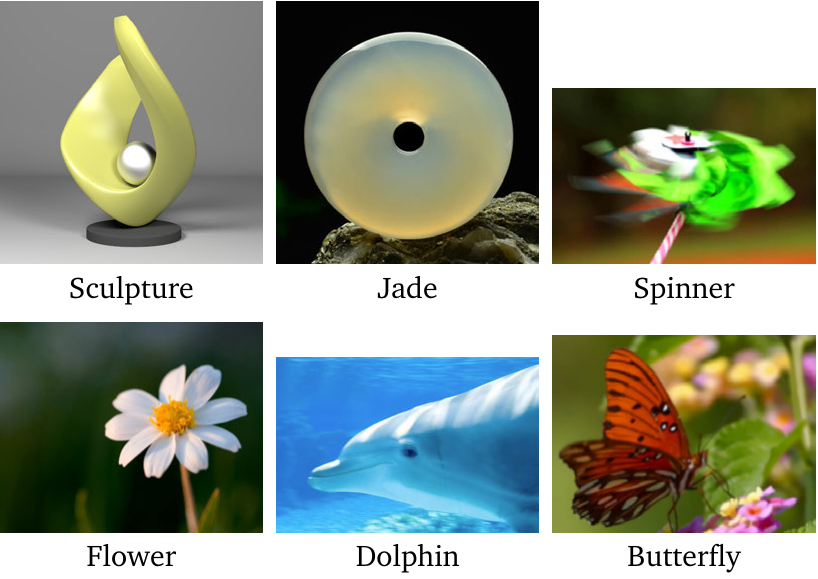}
	\caption{\label{fig:comparisons_ref} {\bf Input pixel images} for generating diffusion curve results in Figures~\protect\ref{fig:boundary_threshold} and \protect\ref{fig:comparisons}.}
\end{figure}

\begin{figure*}[t]
	\centering
	\includegraphics[width=\textwidth]{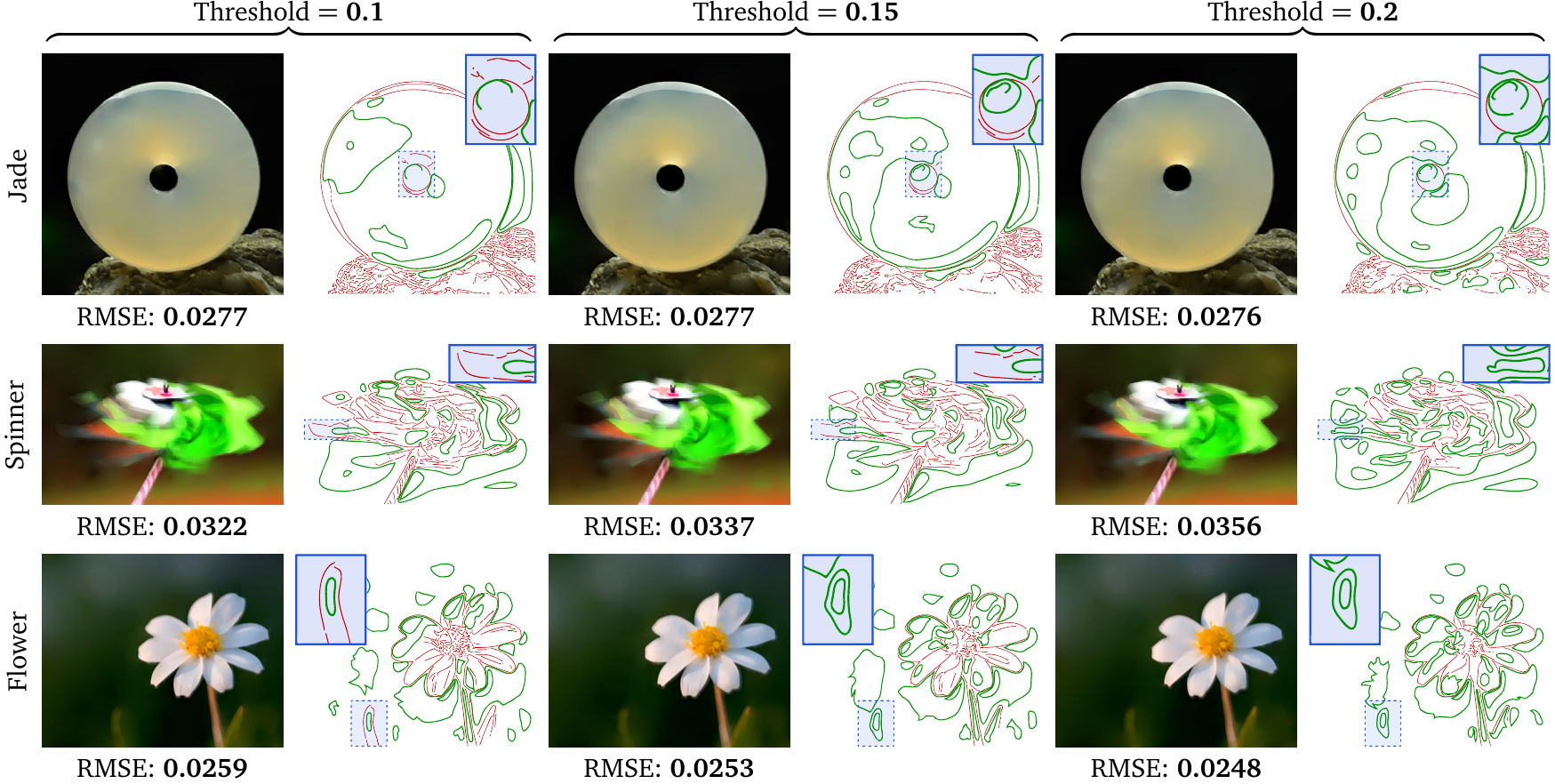}
	\caption{\label{fig:boundary_threshold}
		Our approach, when handling standard pixel images, is \textbf{insensitive to threshold values} used for detecting initial boundary edges.
		For the examples above, three different thresholds varying from 0.1 to 0.2 have been used.
		Boundary curves and additional ones generated by Algorithm~\ref{alg:full_pipeline} are shown in red and green, respectively.
		The resulting reconstructions under roughly identical curve complexities offer similar qualities.}
\end{figure*}

\begin{figure*}[t]
	\centering
	\includegraphics[width=\textwidth]{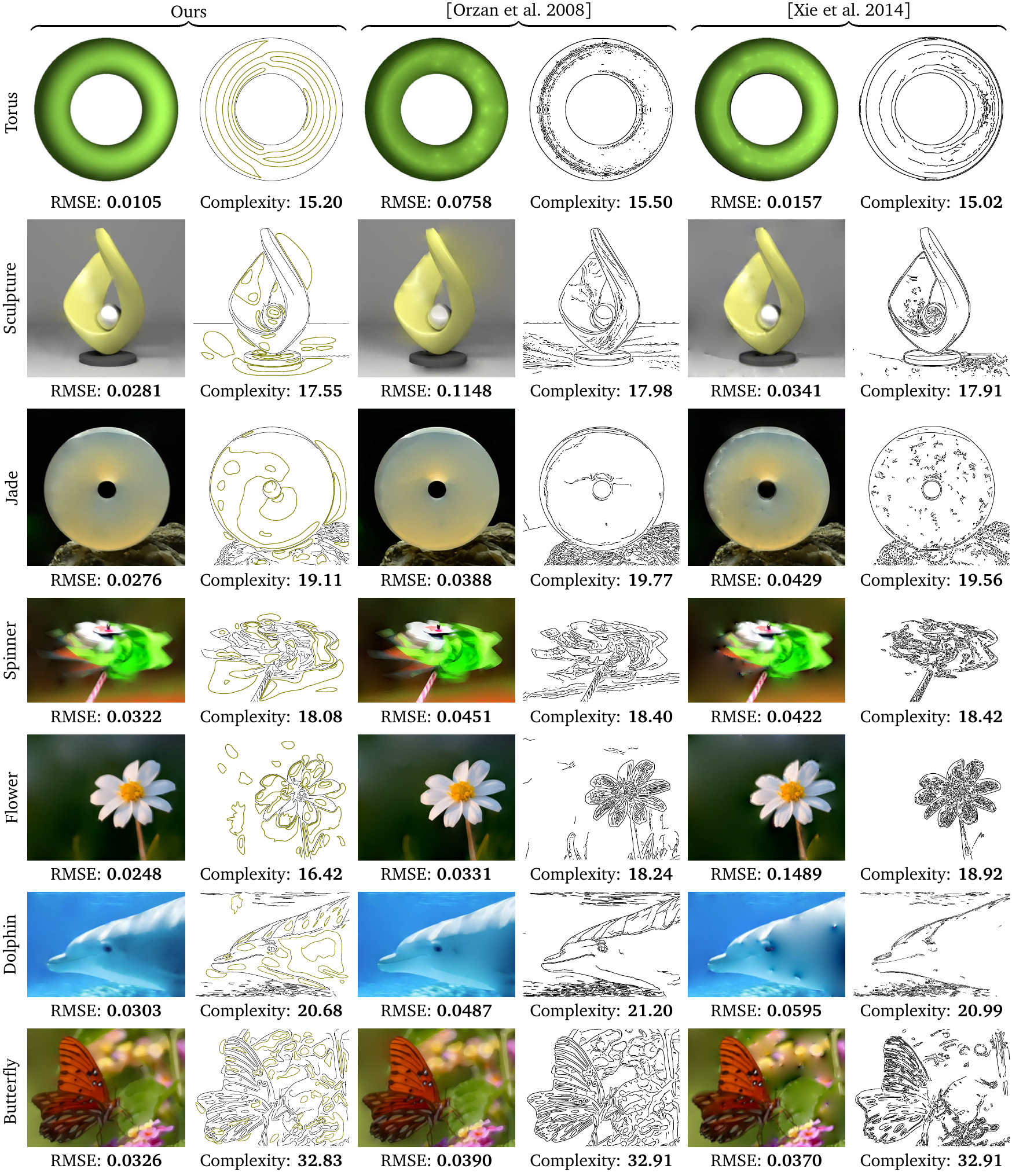}
	\caption{\label{fig:comparisons} {\bf Comparisons} between diffusion curve images generated by our approach and previous methods using {\bf pixel images} (top row) as input. The parameters are adjusted so that the resulting curves generated by each method have roughly \emph{identical complexities}. Our approach not only yields lower approximation error (measured in RMSE), but also generates better shaped and relatively simple curves.}
\end{figure*}

\paragraph{Pixel images}
One common way to represent a color field $I$ is to use standard \emph{pixel images}.
In this case, $I(\xx)$ is evaluated using bilinear interpolation, and $\nabla I(\xx)$ using finite difference.
As stated in \S\ref{subsec:curve_bound}, we perform edge detection to obtain the boundary curves $\partial\Omega$ required by Algorithm~\ref{alg:full_pipeline}.
Although color discontinuities are not well defined for standard pixel images, our method in practice is robust on the choice of boundary curves.
Figure~\ref{fig:boundary_threshold} shows three examples with boundary curves detected using Canny detector with three thresholds.
Notice how missing boundaries (when increasing the threshold) are handled by additional curves generated by Algorithm~\ref{alg:full_pipeline}.
All our results for standard pixel input used thresholds between 0.1 and 0.2.

Figure~\ref{fig:comparisons} contains diffusion curve images reconstructed from pixel input (Figures~\ref{fig:torusA_ref} and \ref{fig:comparisons_ref}) using Algorithm~\ref{alg:full_pipeline} as well as previous edge detection based methods~\cite{Orzan:2008,Xie:2014:HDC}.%
\footnote{We thank the authors of \cite{Xie:2014:HDC} for confirming the
correctness of results in Figure~\ref{fig:comparisons} generated with their
approach.}
The parameters for each method are selected such that the resulting curves have approximately \emph{identical complexities}.
Our method outperforms previous ones when handling smoothly varying color fields.
Notice that, in the bottom two rows (i.e., Flower and Jade), 
\cite{Xie:2014:HDC} has slightly higher approximation errors (measured in RMSE), because it requires higher curve complexities to work properly in these cases.
Please see the supplemental material for more comparisons.

\begin{figure*}[t]
	\centering
	\includegraphics[width=\textwidth]{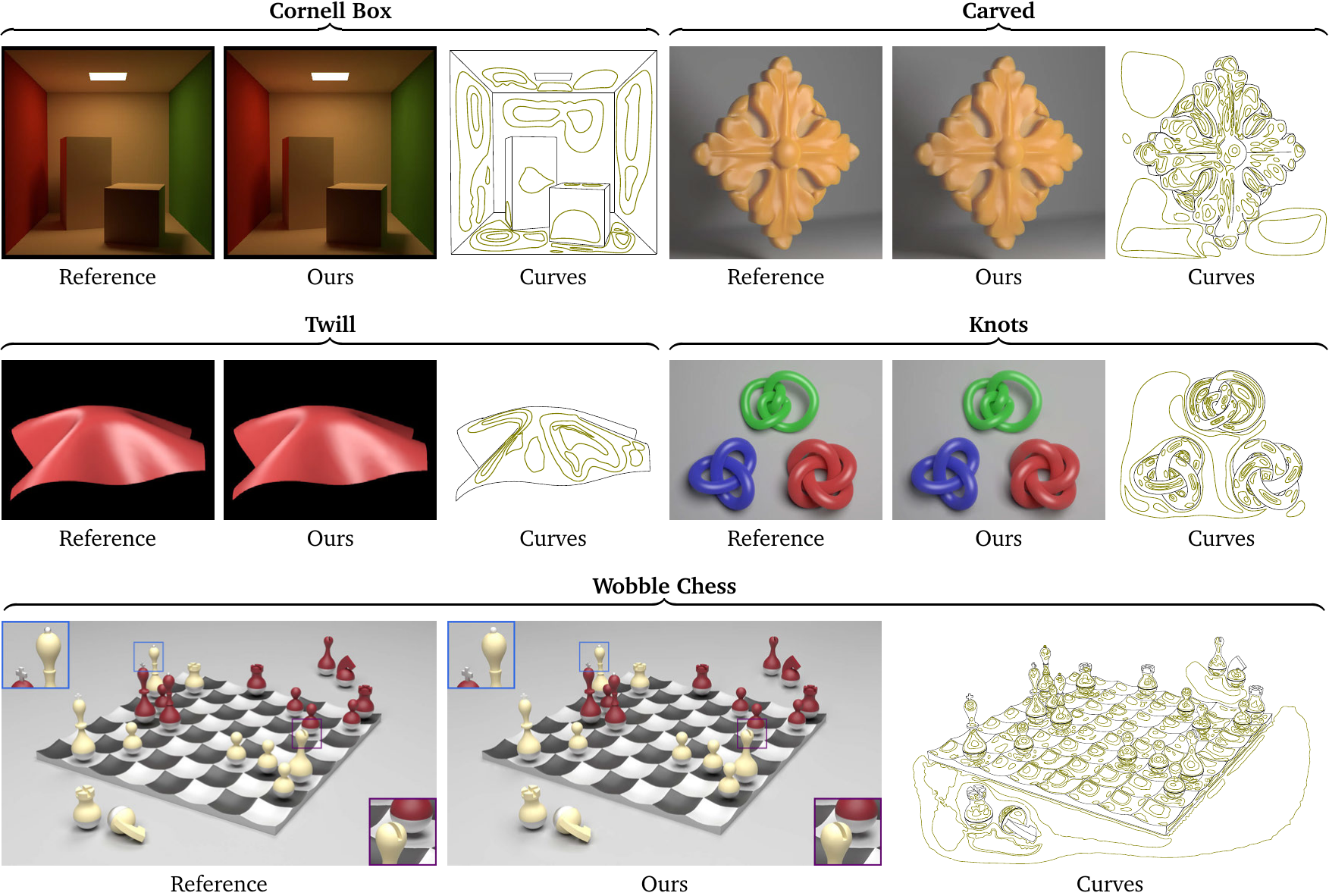}
	\caption{\label{fig:3d_render} Diffusion curve images generated from {\bf renderings of 3D scenes} using our approach. The boundary curves are obtained using mesh contours extracted from the scene geometries.}
\end{figure*}

\paragraph{3D renderings}
Another kind of color field common to computer graphics applications is renderings of 3D scenes. Our approach can be used to approximate these color fields with diffusion curve images. In this case, we represent $I$ as high-resolution pixel images and obtain the boundary curves $\partial \Omega$ directly using object contours.%
\footnote{In this example, we assume all objects to be homogeneous. To handle heterogeneity, $\partial\Omega$ needs to include color jumps across object surfaces.}
Figure~\ref{fig:3d_render} demonstrates diffusion curve images generated using our method from 3D renderings, where slightly higher curve complexities (compared to Figure~\ref{fig:comparisons}) are permitted to ensure low reconstruction errors. Our method successfully captures detailed appearances: from glossy surfaces to smooth shadow boundaries.

\begin{figure*}[t]
	\centering
	\includegraphics[width=\textwidth]{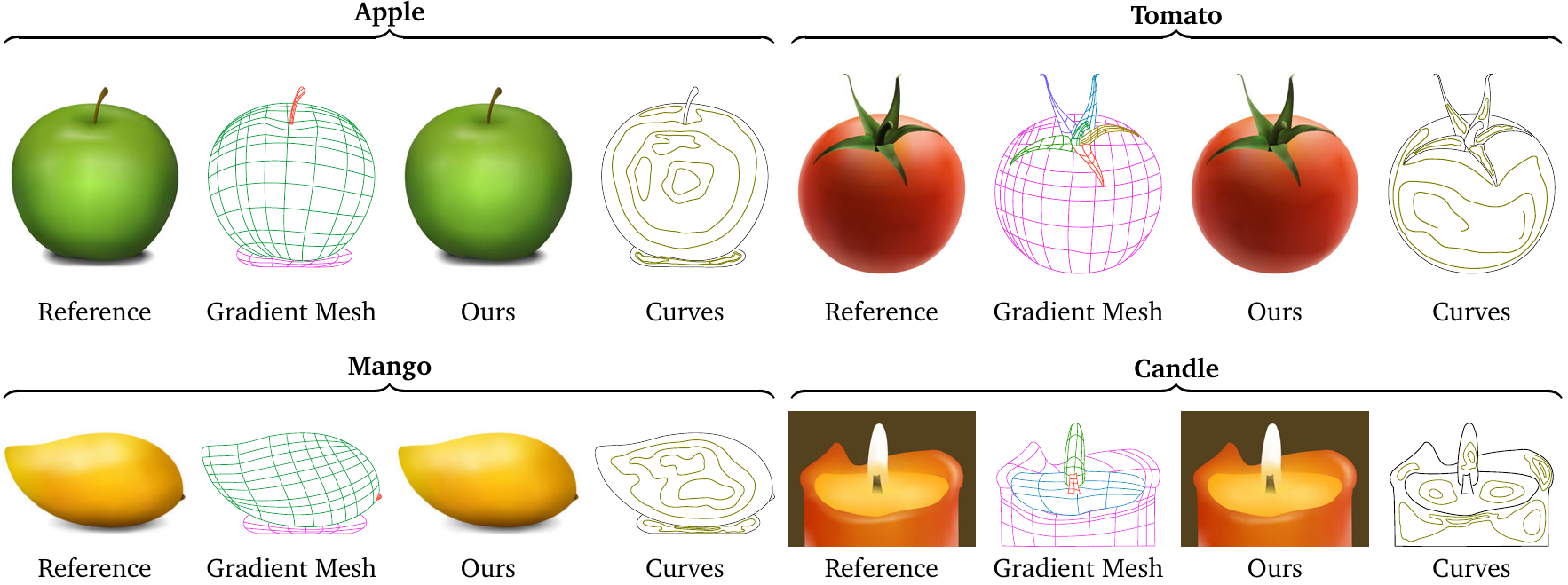}
	\caption{\label{fig:grad_mesh} Diffusion curve images generated from {\bf gradient meshes} directly (i.e., without rasterizing into pixel images) using our method. The boundary curves are given by the mesh boundaries.}
\end{figure*}

\paragraph{Gradient meshes}
Our approach can also generate diffusion curve images directly from input color fields $I$ represented using other vector formats.
We demonstrate this using input color fields represented as gradient meshes in SVG format~\cite{bah2011mesh} where each mesh grid is a Coons Patch~\cite{coons1967surfaces}.
In this case, the color $I(\xx)$ and gradient $\nabla I(\xx)$ can be evaluated analytically for any $\xx \in \Omega$, and the boundary curves $\partial\Omega$ are simply the mesh boundaries.
As shown in Figure~\ref{fig:grad_mesh}, our method directly creates diffusion curve images closely approximating the gradient meshes, without having to rasterize the input into pixel formats.

\begin{table}[t]
	\addtolength{\tabcolsep}{-1pt}
	\centering
	\begin{tabular}{llrlr}
                \whline{1.15pt}
		\textbf{Input Format}
		& \textbf{Scene} & \textbf{Time} & \textbf{Scene} & \textbf{Time}\\
                \whline{0.8pt}
		\multirow{4}{*}{\pbox{1in}{\em Pixel images\\(Figure~\ref{fig:comparisons})}}
		& Torus     & 14.5 & Sculpture & 48.3\\
		& Jade		& 37.6 & Spinner   & 62.7\\
		& Flower    & 64.2 & 
							 Dolphin   & 43.7\\
		& Butterfly	& 505.3\\
                \whline{0.8pt}
		\multirow{3}{*}{\pbox{1in}{\em 3D renderings\\(Figure~\ref{fig:3d_render})}}
		& Cornell Box  & 27.8  & Carved & 97.2\\
		& Twill	       & 13.5  & Knots  & 295.6\\
		& Wobble Chess & 460.3\\
                \whline{0.8pt}
		\multirow{2}{*}{\pbox{1in}{\em Grad. meshes\\(Figure~\ref{fig:grad_mesh})}}
		& Apple & 44.2 & Tomato & 16.7\\
		& Mango & 16.1 & Candle & 11.6\\
                \whline{1.15pt}
	\end{tabular}
	\caption{\label{tab:perf}
		Optimization time (in seconds) for generating results in Figures~\protect\ref{fig:comparisons}, \protect\ref{fig:3d_render}, and \protect\ref{fig:grad_mesh} using our approach on a Linux machine with an 8-core Intel Xeon E5 CPU.}
\end{table}

\subsection{Additional Results}
\label{sec:addtl_res}

\begingroup
\newlength{\lenPixel}
\setlength{\lenPixel}{1.02in}
\begin{figure}[t]
	\centering
	\addtolength{\tabcolsep}{-4pt}	
	\begin{tabular}{cccc}
		& \textbf{Reference} & \textbf{Ours} & \textbf{Curves}\\[2pt]
		\hspace{-5pt}\raisebox{20pt}{\rotatebox[origin=c]{90}{Leaves}} &
		\includegraphics[width=\lenPixel]{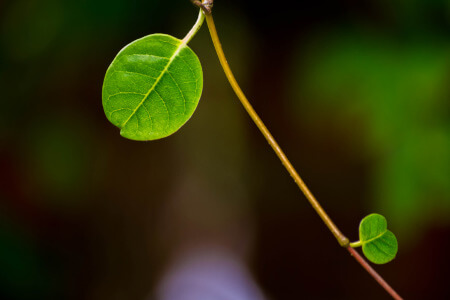} &
		\includegraphics[width=\lenPixel]{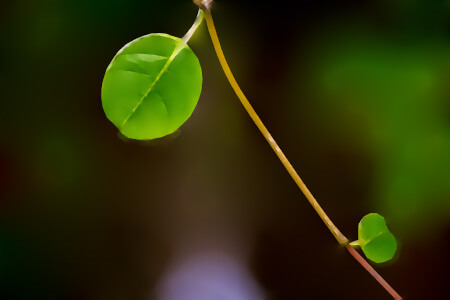} &		
		\includegraphics[width=\lenPixel]{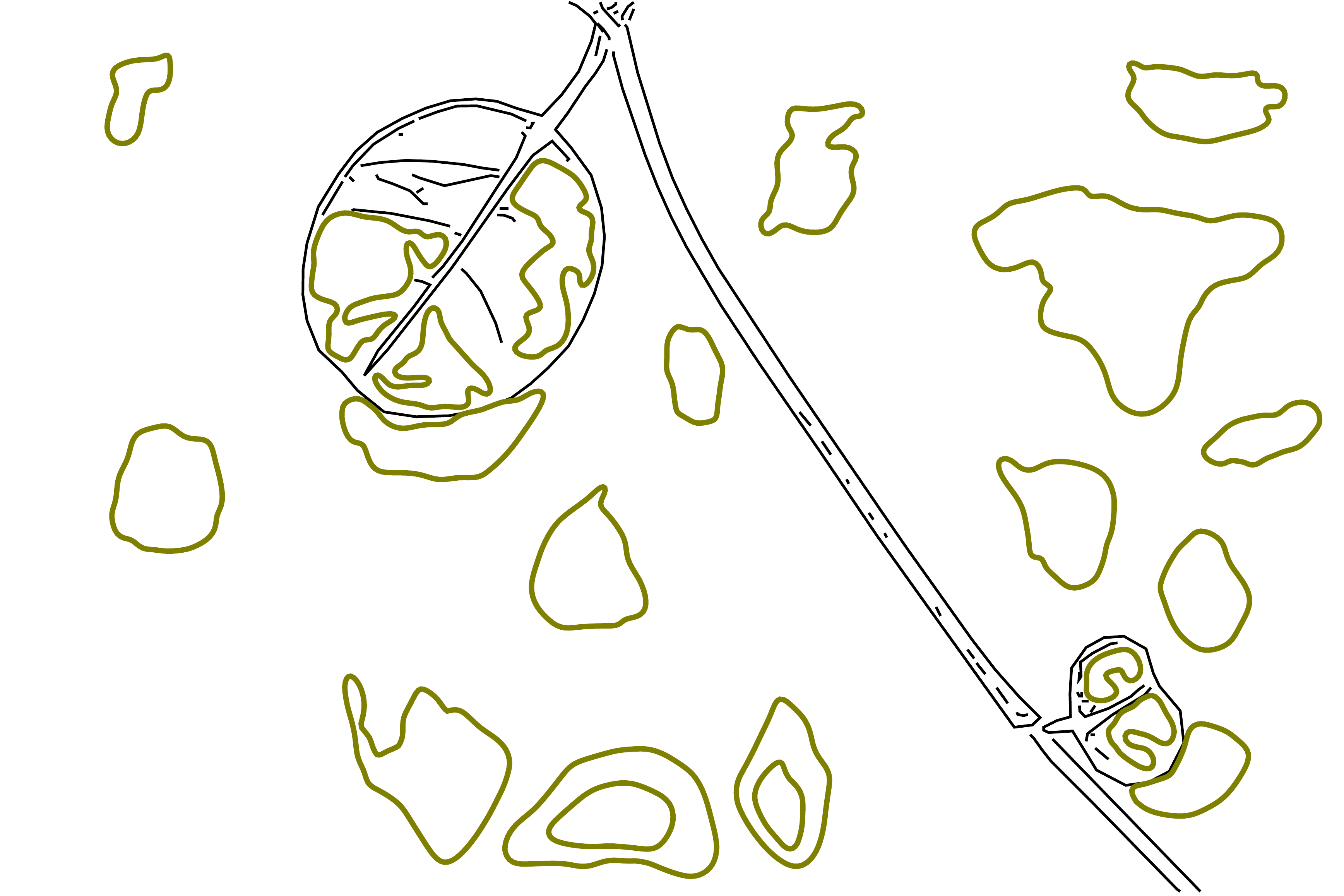}\\[2pt]
		\hspace{-5pt}\raisebox{20pt}{\rotatebox[origin=c]{90}{Apple}} &
		\includegraphics[width=\lenPixel]{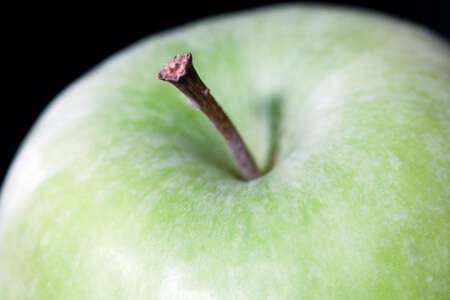} &
		\includegraphics[width=\lenPixel]{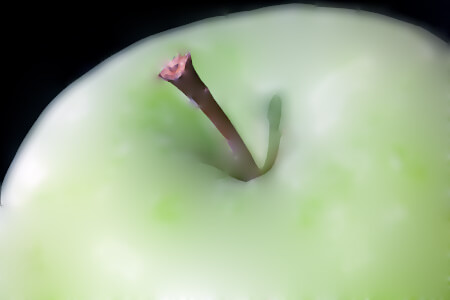} &		
		\includegraphics[width=\lenPixel]{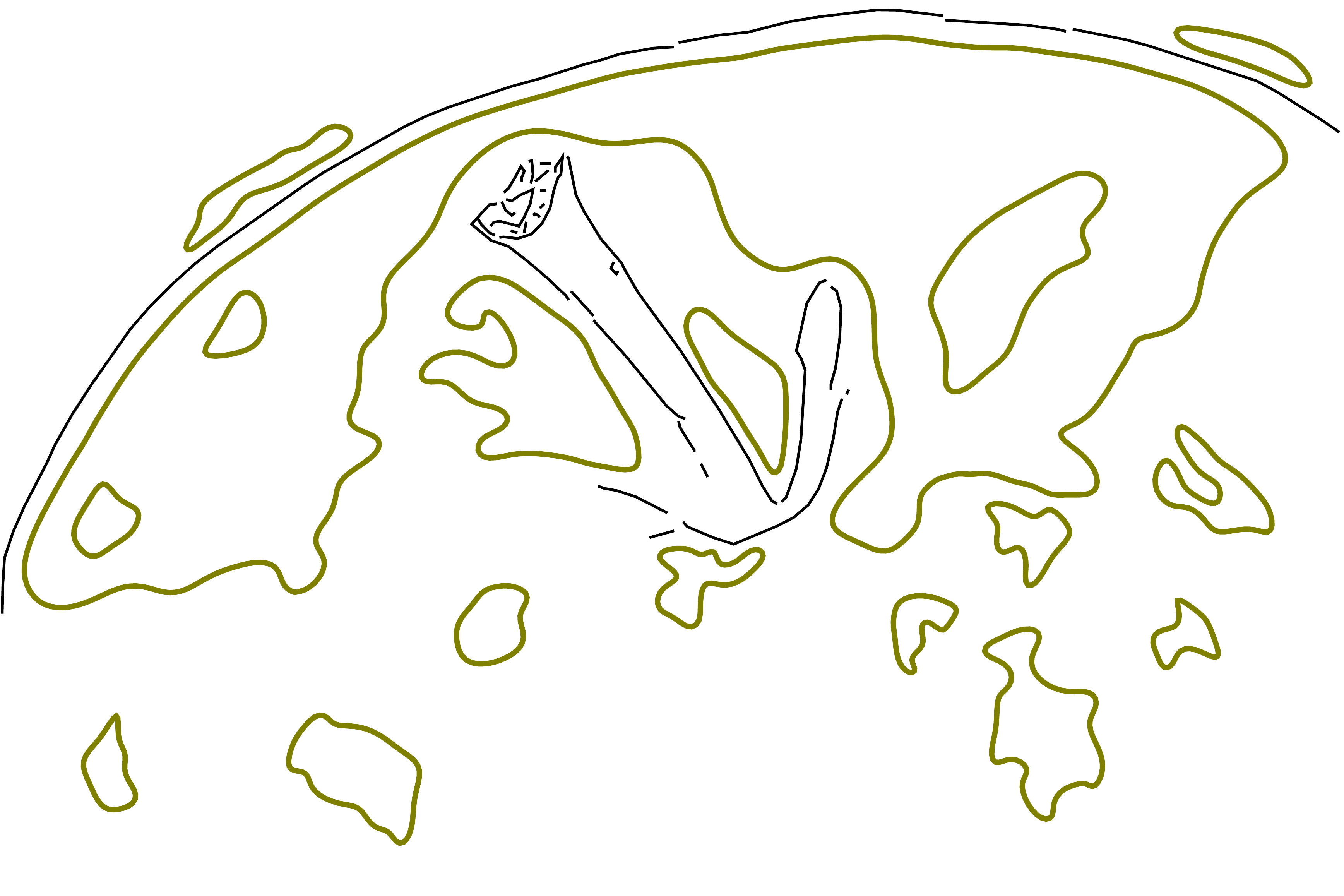}\\[2pt]
		\hspace{-5pt}\raisebox{20pt}{\rotatebox[origin=c]{90}{Eggplant}} &
		\includegraphics[width=\lenPixel]{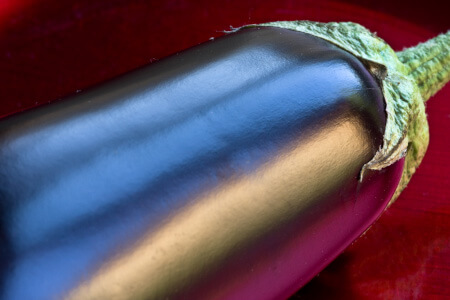} &
		\includegraphics[width=\lenPixel]{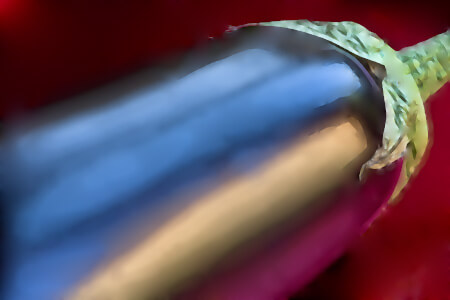} &		
		\includegraphics[width=\lenPixel]{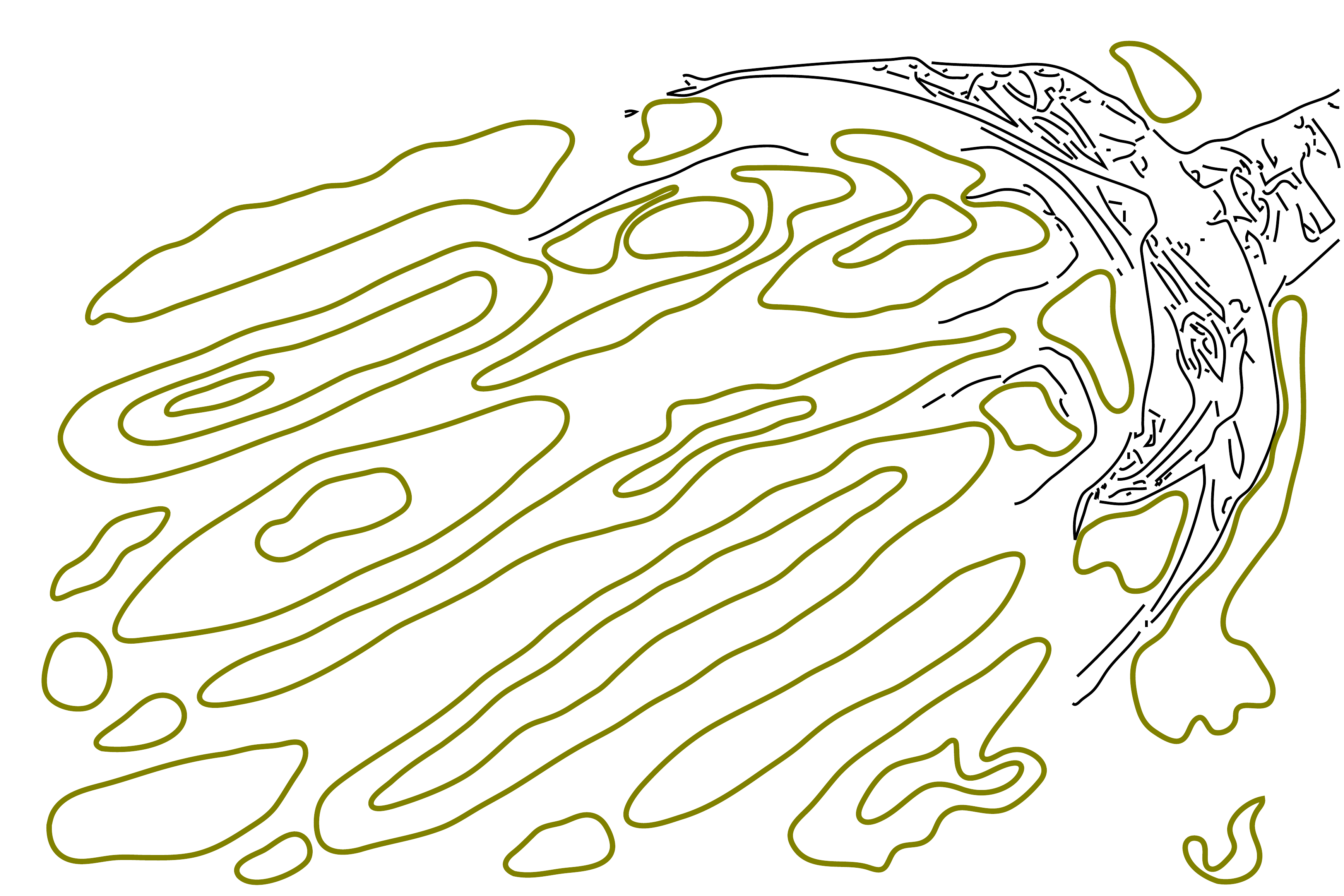}\\[2pt]		
		\hspace{-5pt}\raisebox{16pt}{\rotatebox[origin=c]{90}{Fruits}} &
		\includegraphics[width=\lenPixel]{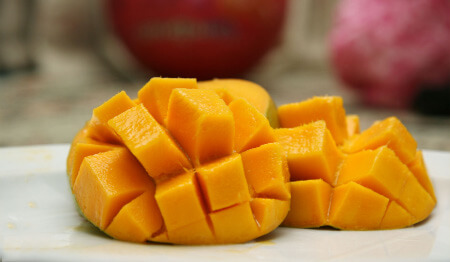} &
		\includegraphics[width=\lenPixel]{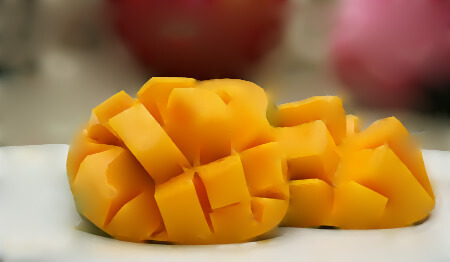} &		
		\includegraphics[width=\lenPixel]{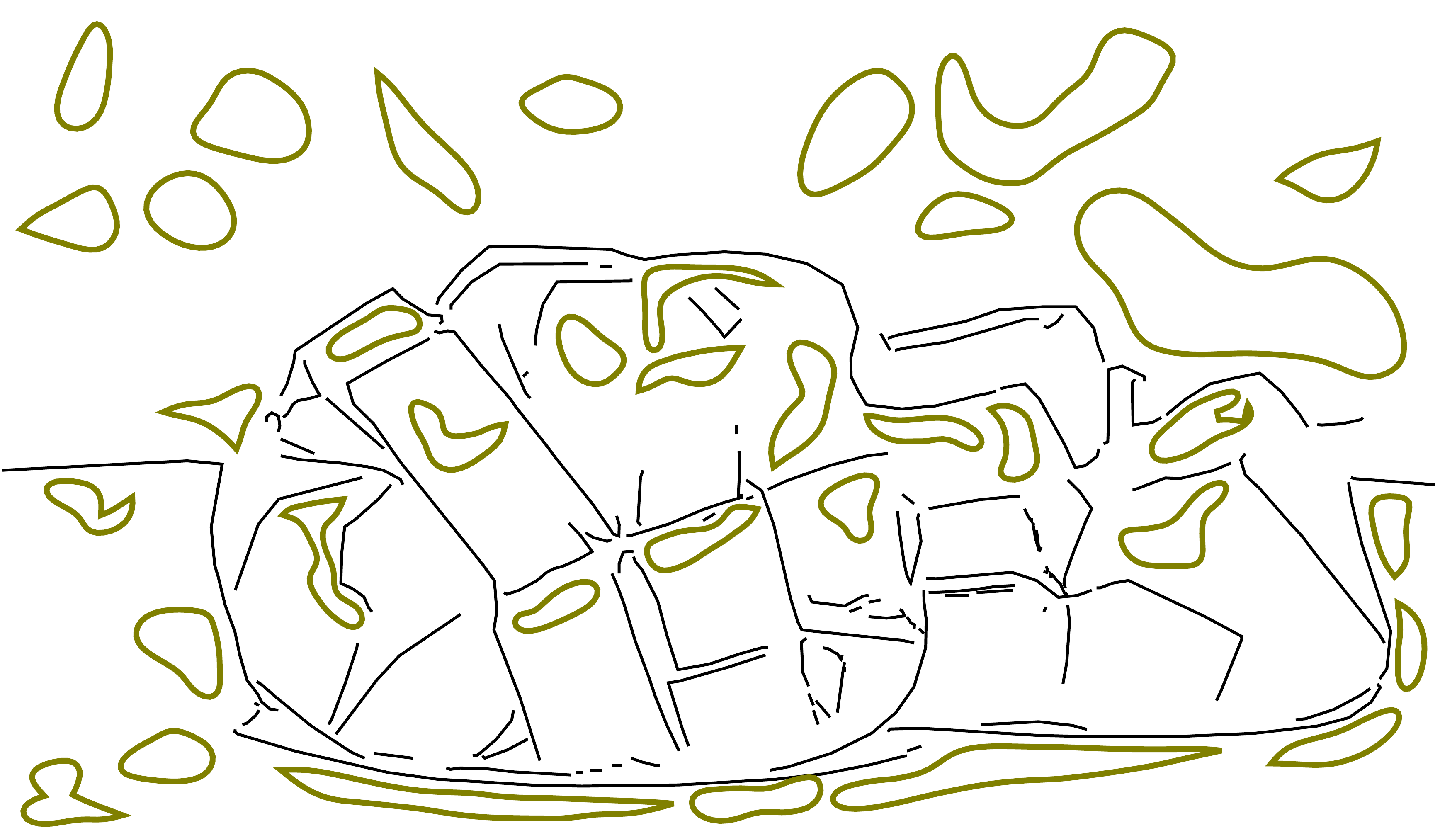}\\[2pt]
		\hspace{-5pt}\raisebox{21pt}{\rotatebox[origin=c]{90}{Flamingo}} &
		\includegraphics[width=\lenPixel]{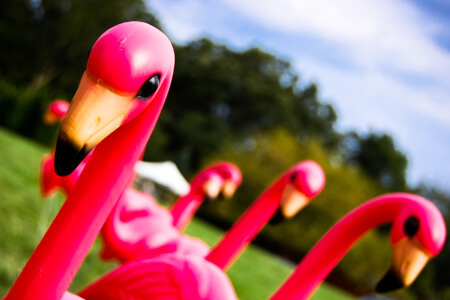} &
		\includegraphics[width=\lenPixel]{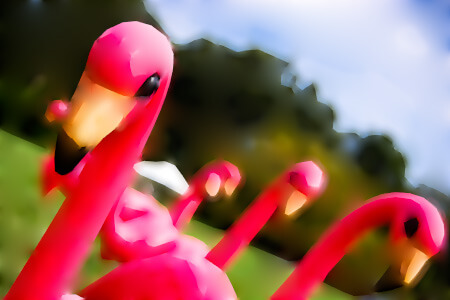} &		
		\includegraphics[width=\lenPixel]{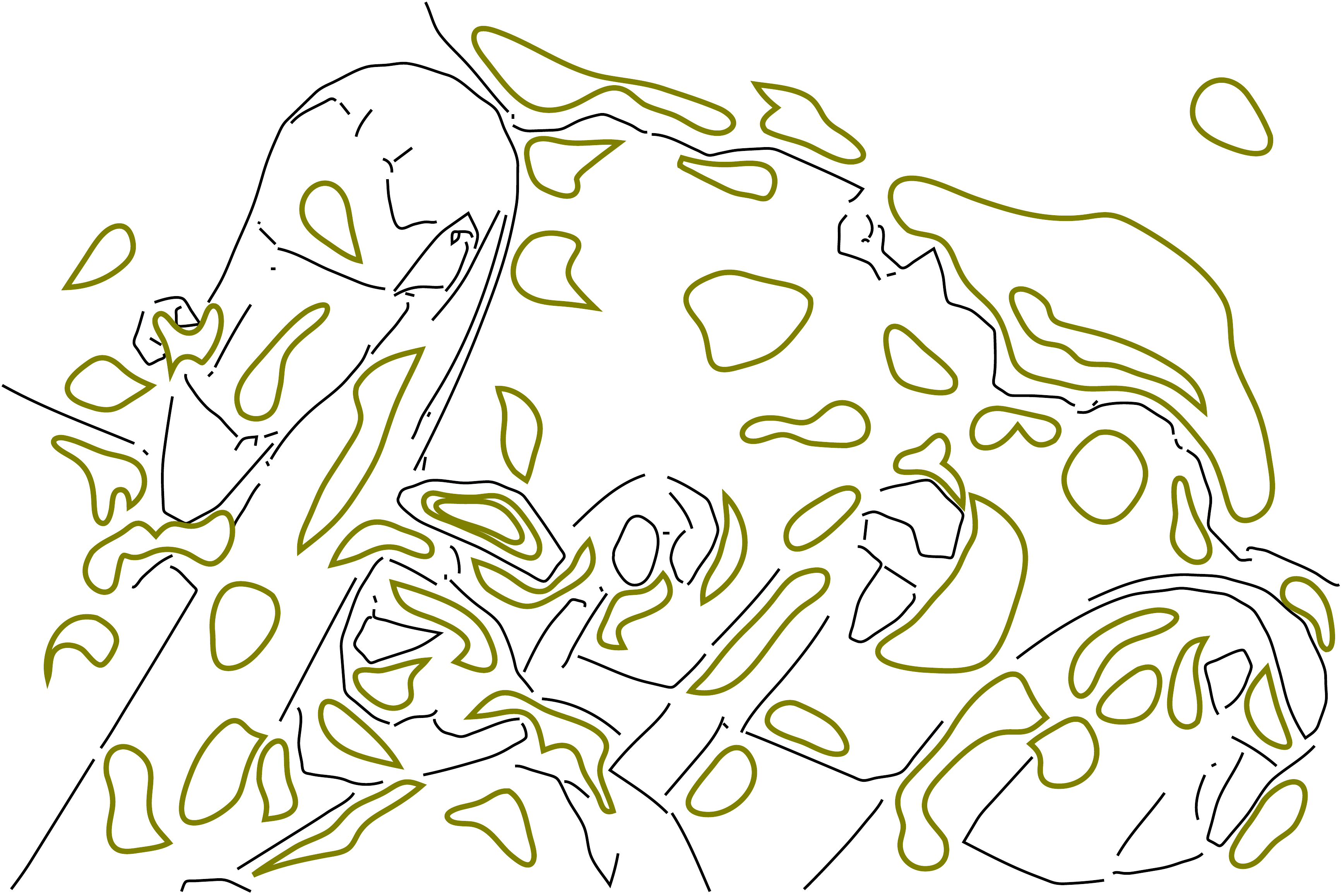}
	\end{tabular}
	\caption{\label{fig:pixel_extra}
		Additional results generated by our approach from {\bf pixel input}.
        Please see the supplementary materials for more results.
	}
\end{figure}
\endgroup

\paragraph{Coloring optimization}
As discussed in \S\ref{sec:curveopt_discuss}, our technique is orthogonal and complementary to coloring optimization techniques which take diffusion curve geometry as input and optimizes colors (and color gradients) at both sides of the curves.
These techniques can be used to replace our simple coloring scheme which samples color values directly (\S\ref{sec:opt_prob}) and performs per-pixel blurring (\S\ref{subsec:curve_post}).
Figure~\ref{fig:coloring_opt} demonstrates that our optimized curve geometry can be coupled with the coloring optimization scheme introduced by Xie~et~al.~\shortcite{Xie:2014:HDC} to further reduce reconstruction errors.
Notice that, since this technique explicitly optimizes color gradients across each curve, we did not perform per-pixel blurring (as stated in \S\ref{subsec:curve_post}).

\begingroup
\newlength{\lenPic}
\setlength{\lenPic}{1.05in}
\begin{figure}
	\centering
	\addtolength{\tabcolsep}{-4pt}	
	\begin{tabular}{ccc}
		\textbf{Our Optimized} & \textbf{Our Curves +} & \textbf{Our Curves +}\\
		\textbf{Curve Geometry} & \textbf{Our Coloring} & \textbf{\cite{Xie:2014:HDC}}\\[4pt]
		%
		\includegraphics[width=\lenPic]{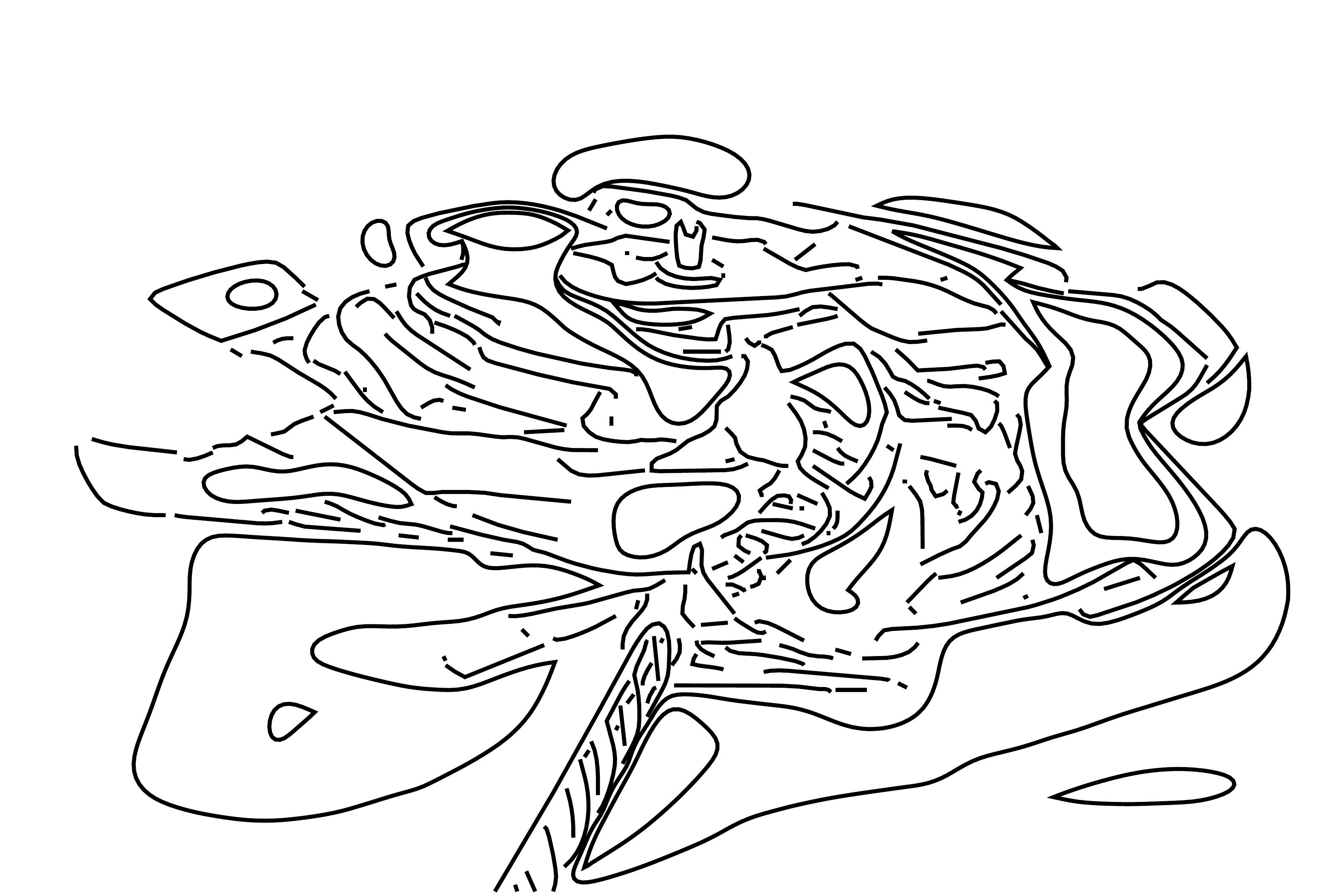} &		
		\includegraphics[width=\lenPic]{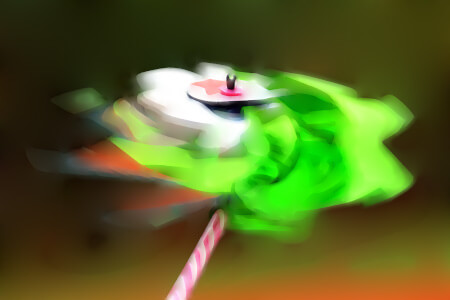} &
		\includegraphics[width=\lenPic]{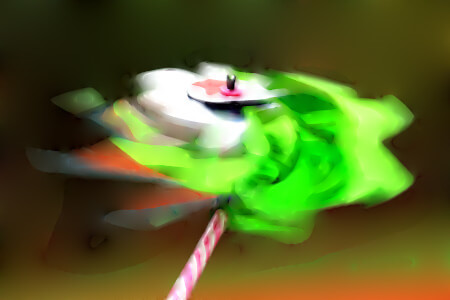}\\
		& RMSE: \textbf{0.0322} & RMSE: \textbf{0.0237}\\[3pt]
		\includegraphics[width=\lenPic]{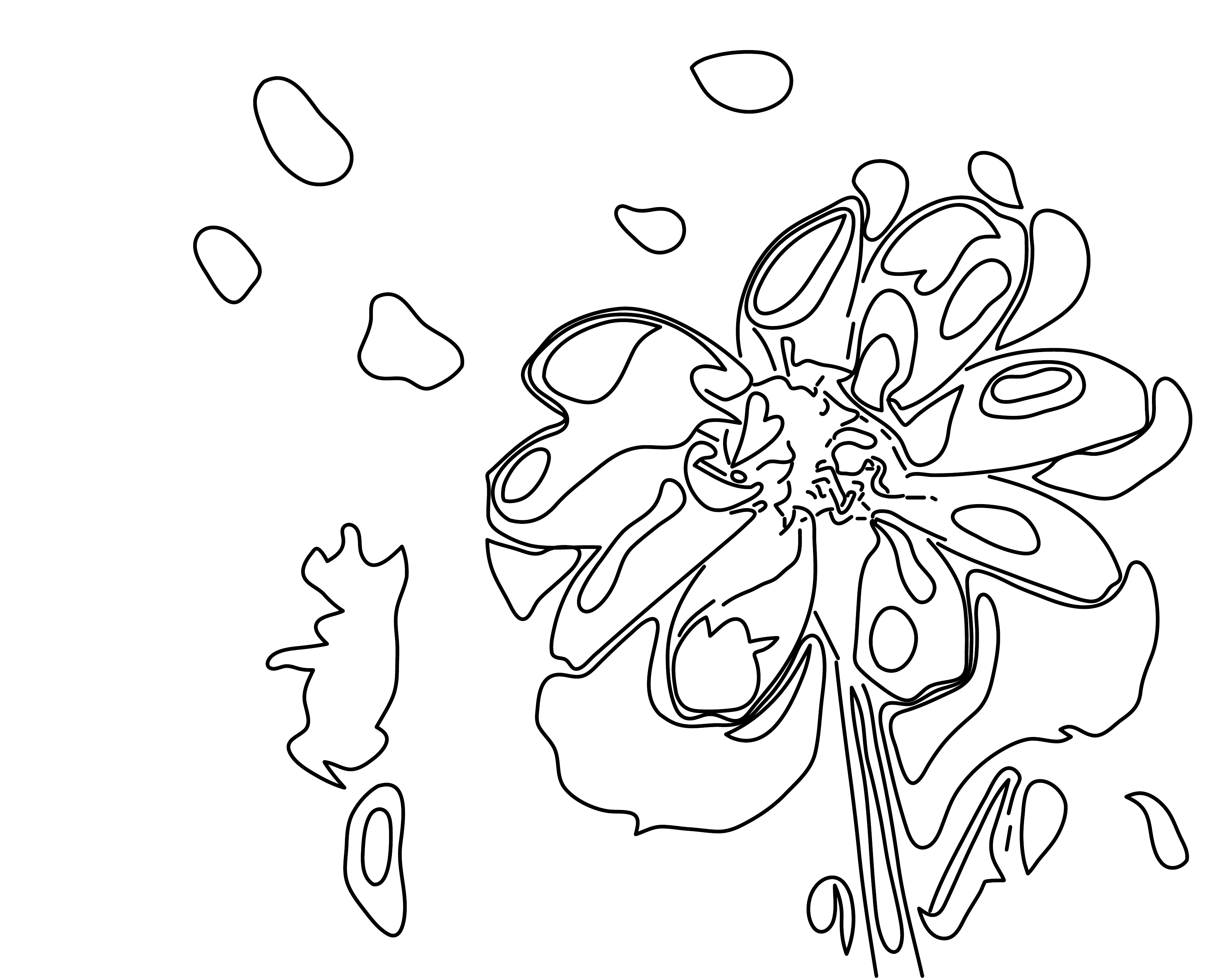} &		
		\includegraphics[width=\lenPic]{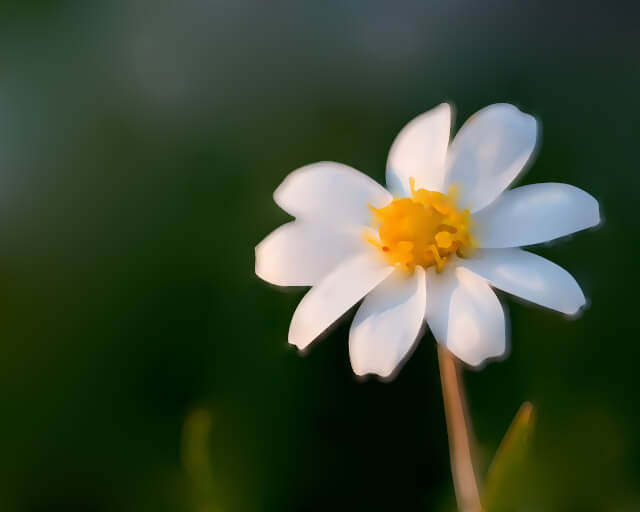} &
		\includegraphics[width=\lenPic]{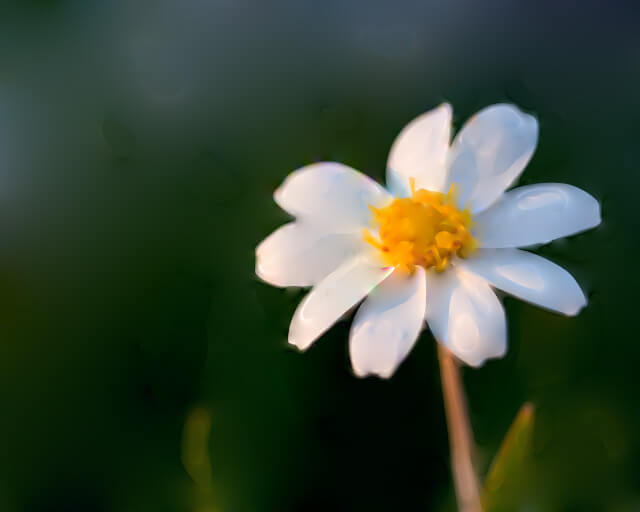}\\
		& RMSE: \textbf{0.0248} & RMSE: \textbf{0.0221}\\[3pt]
		\includegraphics[width=\lenPic]{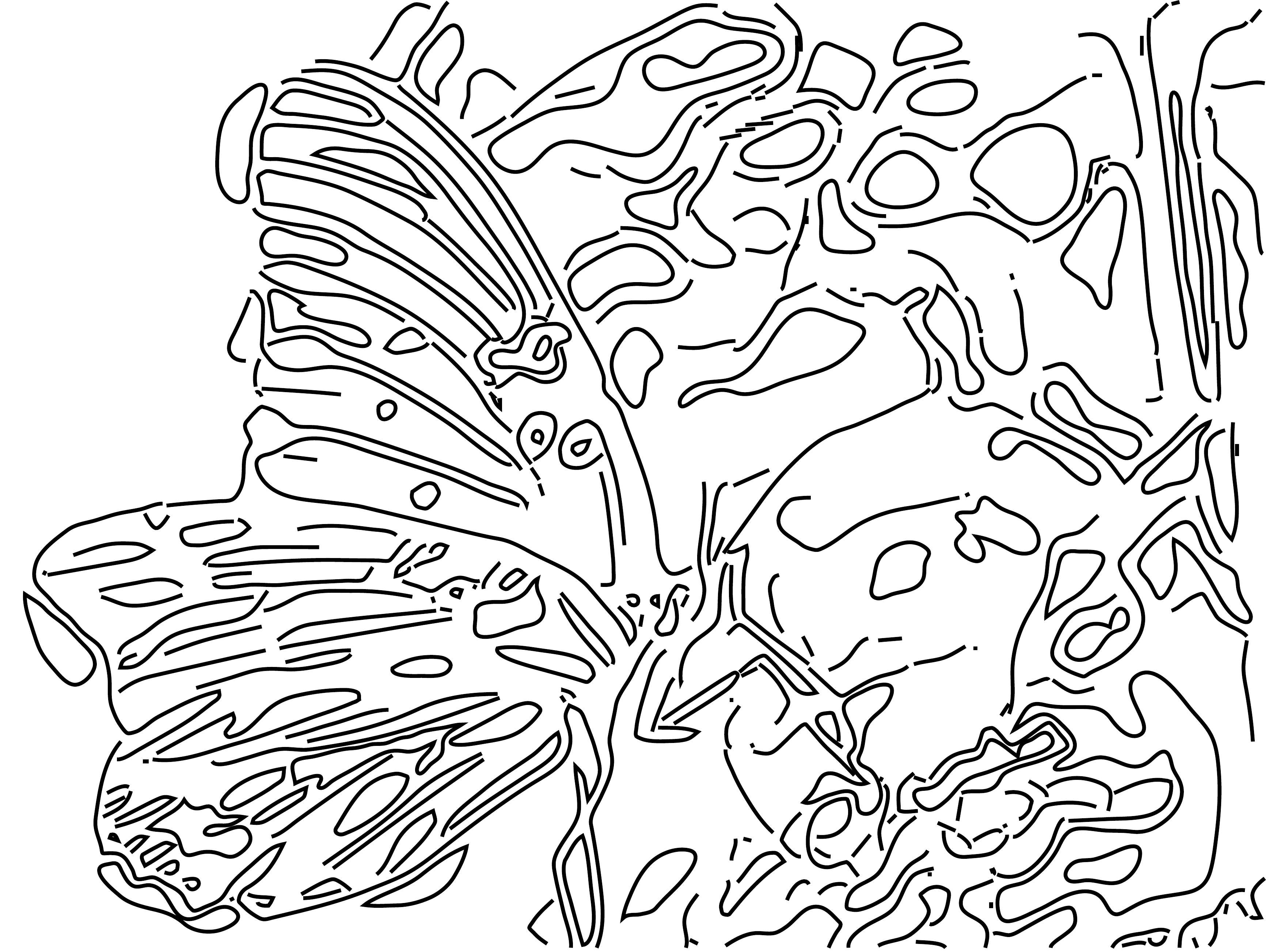} &		
		\includegraphics[width=\lenPic]{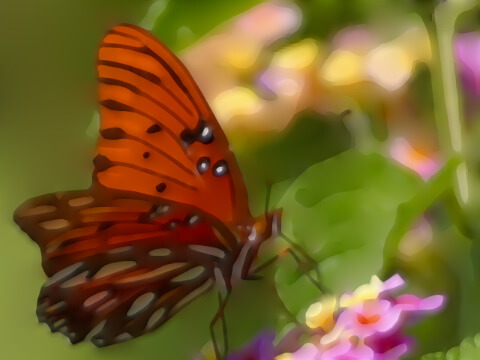} &
		\includegraphics[width=\lenPic]{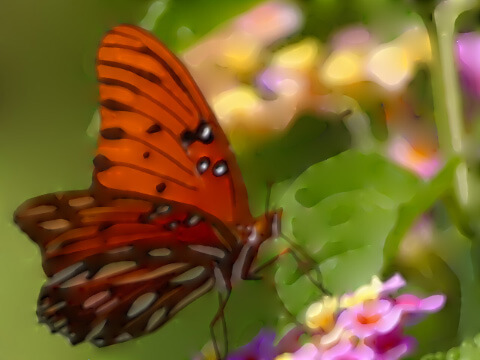}\\
		& RMSE: \textbf{0.0326} & RMSE: \textbf{0.0235}
	\end{tabular}
	\caption{\label{fig:coloring_opt}
		Our core approach is completely \textbf{orthogonal} and \textbf{complementary} to coloring optimization techniques.
		In particular, sophisticated coloring optimization schemes such as \protect\cite{Xie:2014:HDC} can be applied to our optimized curve geometry to further improve reconstruction accuracy.}
\end{figure}
\endgroup

\paragraph{Higher-order domain}
As discussed in \S\ref{sec:curveopt_discuss}, our approach can be applied to higher-order domains for generating curves with higher-order smoothness.
For instance, as shown in Figure~\ref{fig:higher_order}, our pipeline can be applied to color gradients rather than original color values.
In other words, given a RGB image $I$, we can use $\nabla I$, a six-channel image, as the input to Algorithm~\ref{alg:full_pipeline}.
Given our reconstructed gradient image, we solve an additional least square problem to recover the final image.
	
However, as observed by Xie~et~al.~\shortcite{Xie:2014:HDC}, we found that for natural images, solving the optimization at higher-order domains normally does not lead to better approximation accuracy under similar curve complexities.
This is because higher-order domains are generally filled with significantly more high-frequency contents that require complex (almost space-filling) curve geometry to accurately reconstruct.

\begingroup
%
\setlength{\lenPic}{1.02in}
\begin{figure}
	\centering
	\addtolength{\tabcolsep}{-4pt}	
	\begin{tabular}{cccc}
		\hspace{-5pt}\raisebox{21pt}{\rotatebox[origin=c]{90}{Reference}} &
		\includegraphics[width=\lenPic]{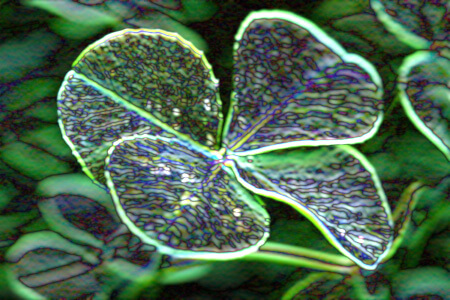} &		
		\includegraphics[width=\lenPic]{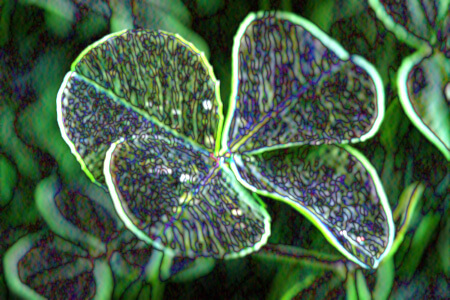} &
		\includegraphics[width=\lenPic]{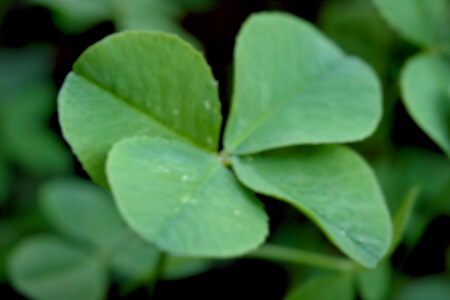}\\
		\hspace{-5pt}\raisebox{21pt}{\rotatebox[origin=c]{90}{Ours}} &
		\includegraphics[width=\lenPic]{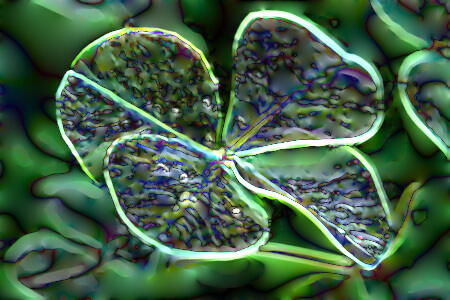} &		
		\includegraphics[width=\lenPic]{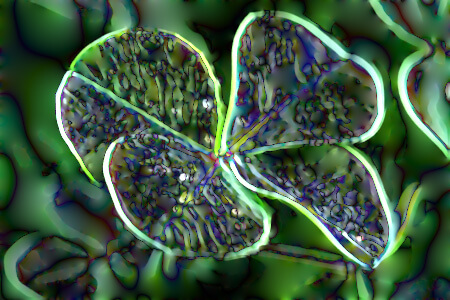} &
		\includegraphics[width=\lenPic]{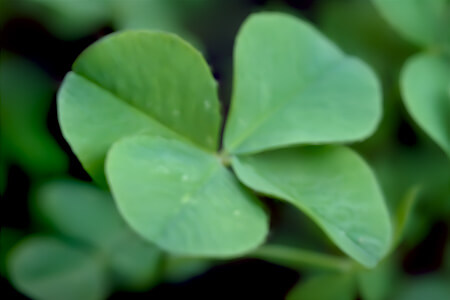}\\
		%
		\hspace{-5pt}\raisebox{21pt}{\rotatebox[origin=c]{90}{Reference}} &
		\includegraphics[width=\lenPic]{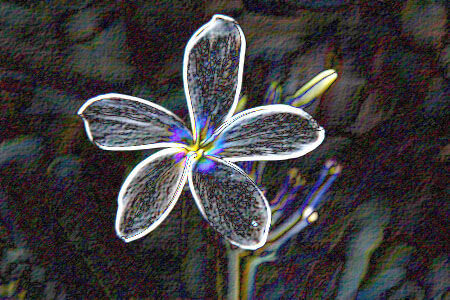} &		
		\includegraphics[width=\lenPic]{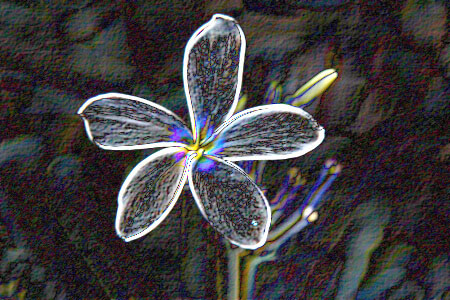} &
		\includegraphics[width=\lenPic]{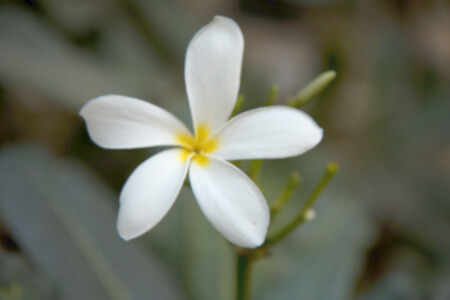}\\
		\hspace{-5pt}\raisebox{21pt}{\rotatebox[origin=c]{90}{Ours}} &
		\includegraphics[width=\lenPic]{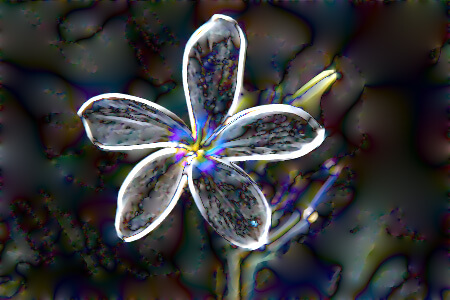} &		
		\includegraphics[width=\lenPic]{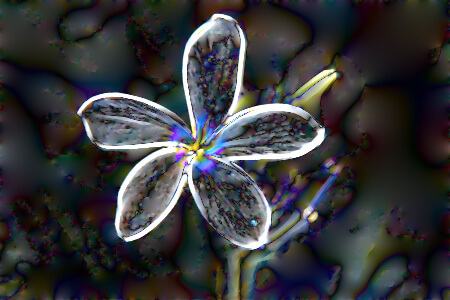} &
		\includegraphics[width=\lenPic]{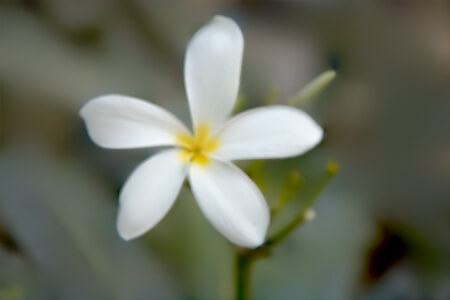}\\
		& Grad. Image X 
		& Grad. Image Y 
		& Original Image\\[-4pt]
		%
	\end{tabular}
	\addtolength{\tabcolsep}{2pt}
	\begin{tabular}{ccc}
		\hspace{-5pt}\raisebox{21pt}{\rotatebox[origin=c]{90}{Our Curves}} &
		\includegraphics[width=\lenPic]{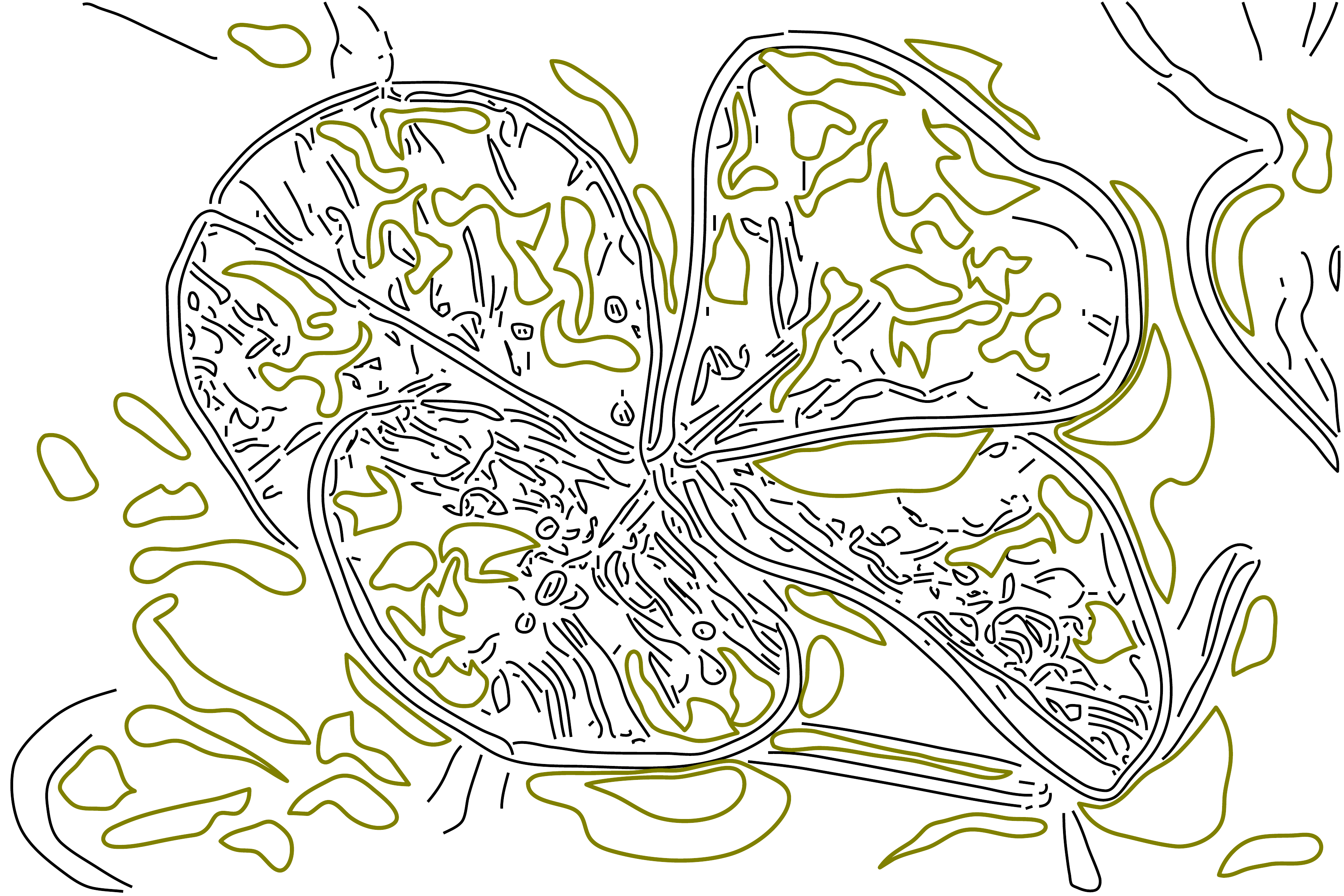} &
		\includegraphics[width=\lenPic]{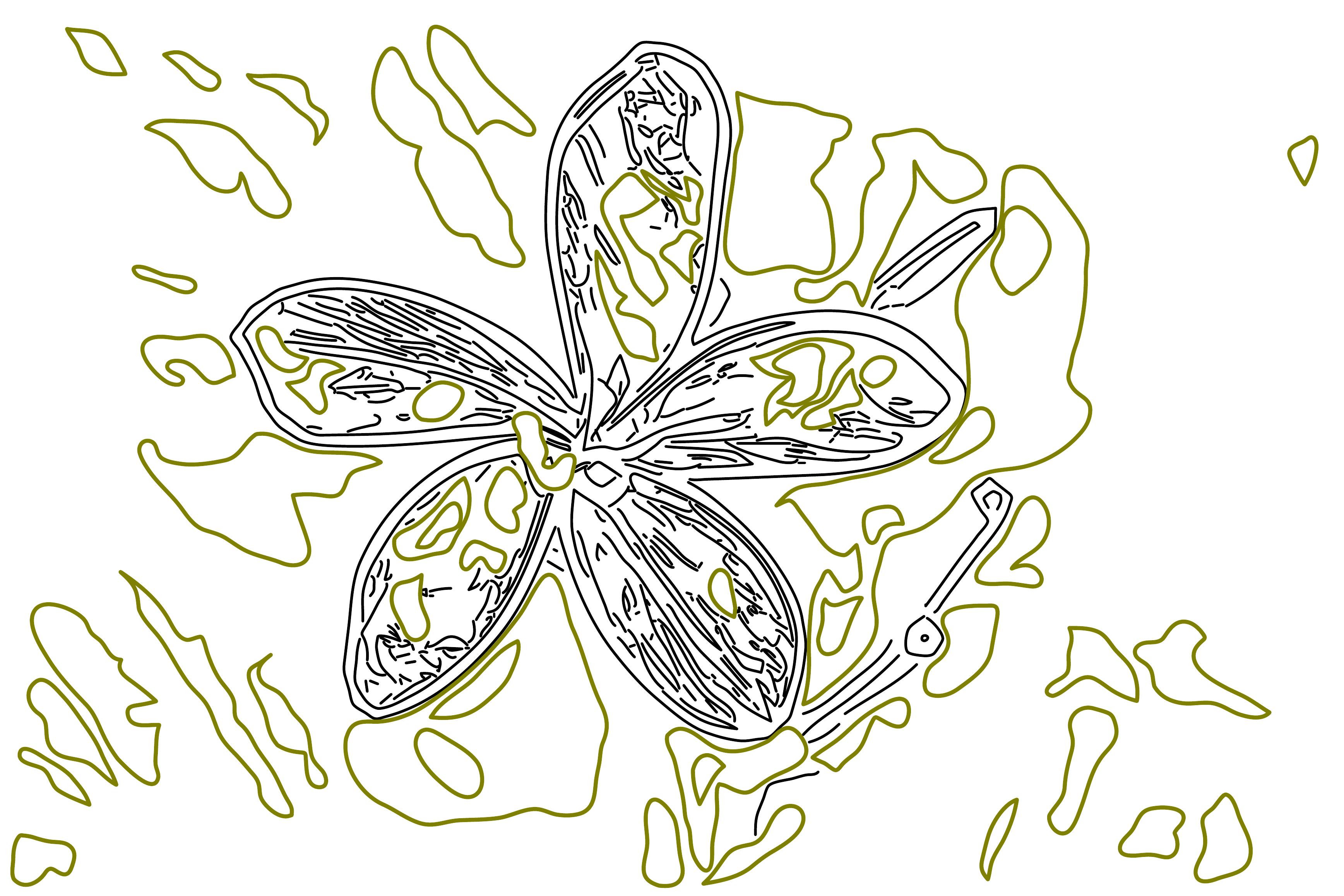}
	\end{tabular}
	\caption{\label{fig:higher_order}
		Application of our method on the \textbf{color gradient domain} instead of the original domain. In this case, the input color field to our approach (Algorithm~\protect\ref{alg:full_pipeline}) is a six-channel image representing color gradients in horizontal (X) and vertical (Y) directions of the original image.}
\end{figure}
\endgroup

\paragraph{Animated result}
Lastly, we show preliminary results to motivate future applications of our approach.
Since our method optimizes the shape of diffusion curves iteratively, it is
suitable for generating animated results from a sequence of gradually changing
input color fields.
The basic idea is \emph{curve reusing}: taking optimized curve geometry from one frame as the initial configuration to ``warm start'' the next one.

Figure~\ref{fig:reusing} and the accompanying video show a proof-of-concept example.
The input is the relighting (i.e., the object stays static while the light source moves) of a shiny torus knot.
In this case, the boundary curves keep unchanged throughout all frames,
and optimized curve geometry from one frame remains valid for all other frames.
Previous methods~\cite{Orzan:2008,Xie:2014:HDC} cannot easily enforce curve coherence across different frames, leading to temporally noisy animations.
By modifying the curve initialization step in Algorithm~\ref{alg:full_pipeline} to reuse optimized curve geometry, we are able to accelerate the optimization process by $2.3\times$, and the resulting animation has lower approximation error and little noise.
Please see the supplementary video for full animations.

\begingroup
\newlength{\lenAni}
\setlength{\lenAni}{0.85in}
\begin{figure}[t]
	\centering
	\addtolength{\tabcolsep}{-4pt}	
	\begin{tabular}{cccc}
		& \textbf{Frame 1} & \textbf{Frame 25} & \textbf{Frame 50}\\
		\hspace{-5pt}\raisebox{33pt}{\rotatebox[origin=c]{90}{Reference}} &
		\includegraphics[width=\lenAni]{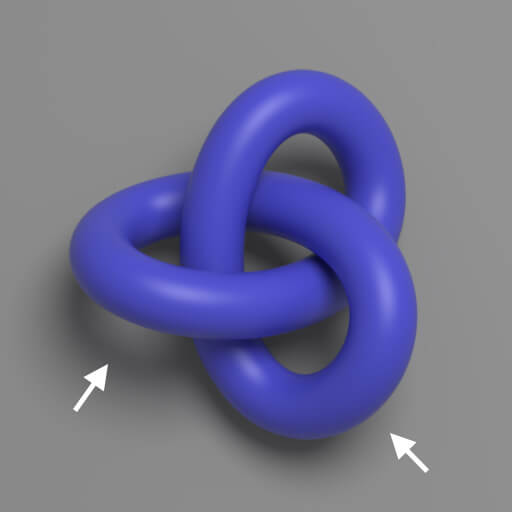} &
		\includegraphics[width=\lenAni]{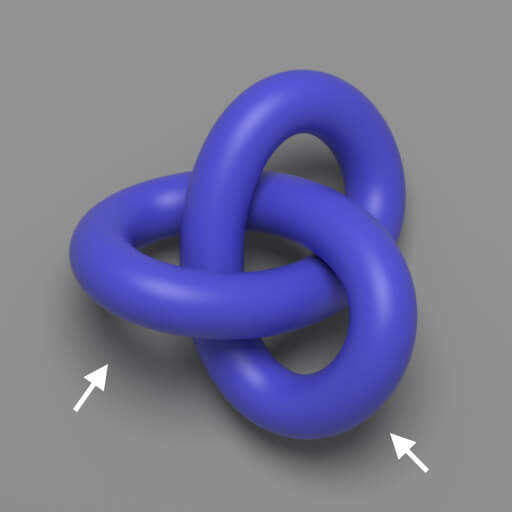} &		
		\includegraphics[width=\lenAni]{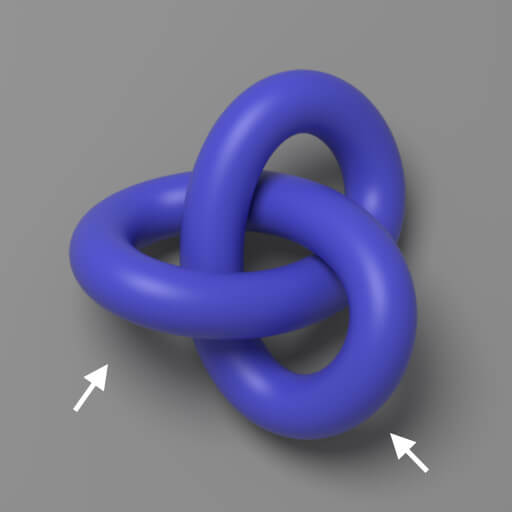}\\[-10pt]
	\end{tabular}
	\includegraphics[width=0.9\columnwidth]{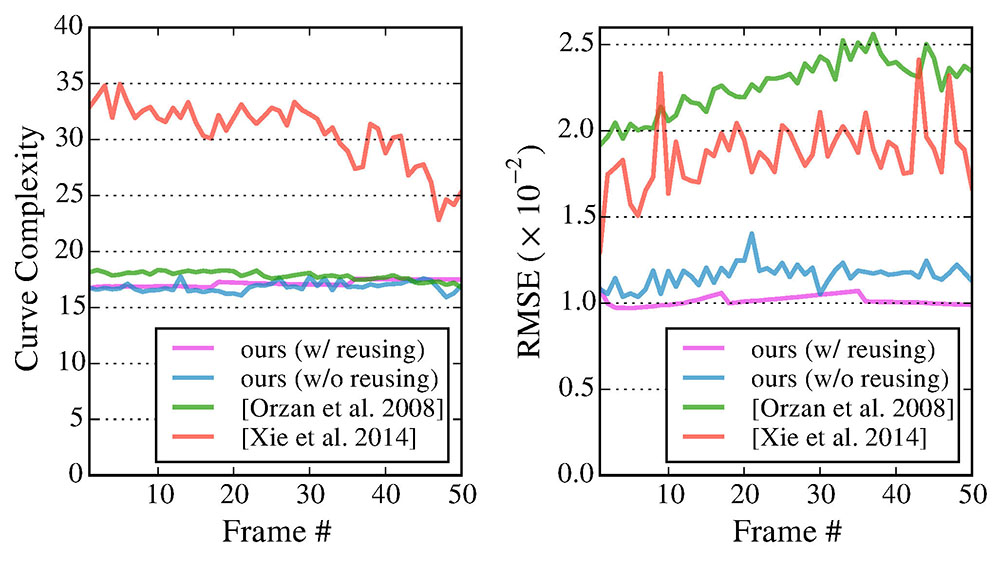}\\[-4pt]
	\caption{\label{fig:reusing}
		{\bf Animated results} consisting of 50 frames from relighting a torus knot.
		Three of these frames 
		are shown on the top (where white arrows indicate regions with moving shadows).
		Higher curve complexities (left plot) are used for \protect\cite{Xie:2014:HDC} 
		for fewer artifacts.
		{\bf Reusing} optimized curves from one frame as the starting point (i.e., initial curves) for the optimization of the next frame leads to temporally coherent curve geometry and better approximation accuracy (right plot).
		See the supplementary video for full animations.}
\end{figure}
\endgroup

%
\section{Limitation and Conclusion}

\paragraph{Limitation}
Our approach has a few limitations that can inspire future work.
First, it requires the color field to be $C^0$ continuous everywhere except at the given boundaries.
Robustly finding clean boundary curves, however, can be challenging.
Second, if the color field contains spatially high-frequency features, very fine triangulation may be needed to fully resolve them, slowing down our optimization process.

\paragraph{Conclusion}
This paper introduces a novel solution to the inverse diffusion curve problem.
The key component of our approach is a curve optimization algorithm that iteratively deforms a set of diffusion curves in a way that the reduction of approximation error is guaranteed.
Based upon the core algorithm, we develop a full pipeline that takes an input color field plus a set of boundary curves and produces an image with well-shaped and clean curves that closely matches the input. 
Our approach offers the generality to take input presented in different formats, which we demonstrate using three types: pixel images, 3D renderings, and gradient meshes.

%
\appendix
\section{Brief Derivation of Equation~\ref{eq:dJ1}}
\label{app:deriv}
We now briefly outline the derivation of~\eqref{eq:dJ1}.
This derivation has been developed in Shape Optimization Theory.
We therefore refer to the supplementary document and Chapter 10.6 of the book~\cite{sokolowski1992introduction} for a detailed exposition.
First, the \FC derivative of $C$ with a general integrand function $y$ can be expressed as
\begin{equation}\label{eq:dC}
\intd C(\Omega;\BO) = \int_{\Omega} y'(\xx;\BO)\,\intd\Omega + \int_{\Gamma_0}y(\xx;\BO)v_n(\xx)\,\intd\Gamma,
\end{equation}
where $y'$ is the so-called \emph{shape derivative} of $y$ under a given velocity field $\vv$, defined as
\begin{equation}
\label{eq:shape_deriv}
y'(\xx;\BO) = \dot{y}(\xx;\BO) - \nabla y(\xx;\BO)\cdot\vv(\xx).
\end{equation}
The concept of shape derivative is very much analogous to those in continuum mechanics~\cite{bonet97}, which has been widely adopted for creating computer animations.
In particular, $\dot{y}$ is equivalent to the material derivative, measuring the change of $y$ in the undeformed (material) space, while $y'$ indicates the derivative value in the deformed space, that is, the change rate of $y$ due to the boundary changes only.
See \S1.2 of the supplementary document for a rigorous mathematical definition of $\dot{y}$.

We notice that because of the Dirichlet boundary condition~\eqref{eq:bc}, the approximated color $u$ agrees with the input $I$ for all $\xx$ on the boundary (i.e. $\xx \in \Gamma_0$).
Therefore, when $y(\xx;\BO) = (u(\xx)-I(\xx))^2$, the second integral term in~\eqref{eq:dC} vanishes, leaving only the first domain integral term,
\begin{equation}\label{eq:dR}
\intd R(\Omega;\BO)=\int_{\Omega}\left(u(\xx)-I(\xx)\right) u'(\xx)\,\intd\Omega. 
\end{equation}
Here the shape derivative $u'$ follows the same definition as in~\eqref{eq:shape_deriv};
we use the fact that the shape derivatives, just like the conventional ones, satisfy the chain rule.

One can prove that the shape derivative $u'$ satisfies another Laplace equation (see \S1.5 of the supplementary document),
\begingroup
\footnotesize
\begin{equation}\label{eq:deriv_laplace}
\begin{aligned}
    \Delta u'(\xx) &= 0, \quad\\
    u'(\xx) &= \left(\frac{\partial I(\xx)}{\partial n}-\frac{\partial u(\xx)}{\partial n}\right)v_n(\xx), \;\; \forall\xx\in\Gamma_0, 
\end{aligned}
\end{equation}
\endgroup
where the Dirichlet boundary condition is determined by the normal derivative of both the provided (i.e., $I$) and approximated (i.e., $u$) color fields on the same boundaries.
Since the Laplacian operator is self-adjoint and $u'$ is used in an integral~\eqref{eq:dR}, instead of solving the Laplace equation~\eqref{eq:deriv_laplace}, we solve its adjoint problem~\eqref{eq:adj_laplace}, whose solution enables us to transform the domain integral~\eqref{eq:dR} into a desired boundary integral~\eqref{eq:surf_int}, because
\begingroup
\footnotesize
\sLNM
\begin{equation*}
\begin{split}
&\int_{\Omega_0} (u(\xx)-I(\xx)) u'(\xx)\,\intd\Omega = \int_{\Omega_0} \Delta p(\xx) u'(\xx)\,\intd\Omega \\
= & \int_{\Gamma_0}\frac{\partial p(\xx)}{\partial n} u'(\xx)\,\intd\Gamma 
-\int_{\Gamma_0}p(\xx)\frac{\partial u'(\xx)}{\partial n}\intd\Gamma + \int_{\Omega_0} p(\xx) \Delta u'(\xx)\,\intd\Omega,
\end{split}\end{equation*}
\tLNM
\endgroup
where the last equality follows the integration by parts and Green's formula. 
Further, the last two integral terms vanish, following the fact that both $p$ (according to the boundary condition of~\eqref{eq:adj_laplace}) and $\nabla u'$ (according to~\eqref{eq:deriv_laplace}) are zero on the boundary $\Gamma_0$.
Eventually, only the first integral term in the last expression remains, yielding the \FC derivative of $R$ as a boundary linear form of $v_n$, that is,
$\intd R(\Omega;\BO) = \int_{\Gamma_0} B(\xx)\ v_n(\xx)\,\intd\Gamma$ with $B(\xx)$ expressed as in~\eqref{eq:dJ1}.

%
%
\bibliographystyle{acmsiggraph}
\bibliography{ref}

\begin{thebibliography}{\protect\citename{Mohammadi et~al\mbox{.} }2001}

\bibitem[\protect\citename{Bah }2011]{bah2011mesh}
{\sc Bah, T.}, 2011.
\newblock Advanced gradients for {SVG} (online).

\bibitem[\protect\citename{Bonet and Wood }1997]{bonet97}
{\sc Bonet, J., and Wood, R.~D.}
\newblock 1997.
\newblock {\em Nonlinear Continuum Mechanics for Finite Element Analysis}.
\newblock Cambridge University Press.

\bibitem[\protect\citename{Boy{\'e} et~al\mbox{.} }2012]{Boye:2012:VSF}
{\sc Boy{\'e}, S., Barla, P., and Guennebaud, G.}
\newblock 2012.
\newblock A vectorial solver for free-form vector gradients.
\newblock {\em ACM Trans. Graph. 31}, 6, 173:1--173:9.

\bibitem[\protect\citename{Brakke }1992]{brakke1992surface}
{\sc Brakke, K.~A.}
\newblock 1992.
\newblock The surface evolver.
\newblock {\em Experimental mathematics 1}, 2, 141--165.

\bibitem[\protect\citename{Brochu and Bridson }2009]{brochu2009robust}
{\sc Brochu, T., and Bridson, R.}
\newblock 2009.
\newblock Robust topological operations for dynamic explicit surfaces.
\newblock {\em SIAM Journal on Scientific Computing 31}, 4, 2472--2493.

\bibitem[\protect\citename{Burger et~al\mbox{.} }2004]{burger2004inverse}
{\sc Burger, M., Osher, S.~J., and Yablonovitch, E.}
\newblock 2004.
\newblock Inverse problem techniques for the design of photonic crystals.
\newblock {\em IEICE transactions on electronics 87}, 3, 258--265.

\bibitem[\protect\citename{Canny }1986]{canny1986computational}
{\sc Canny, J.}
\newblock 1986.
\newblock A computational approach to edge detection.
\newblock {\em Pattern Analysis and Machine Intelligence, IEEE Transactions
  on}, 6, 679--698.

\bibitem[\protect\citename{Coleman }2012]{coleman2012calculus}
{\sc Coleman, R.}
\newblock 2012.
\newblock {\em Calculus on Normed Vector Spaces}.
\newblock Springer.

\bibitem[\protect\citename{Coons }1967]{coons1967surfaces}
{\sc Coons, S.~A.}
\newblock 1967.
\newblock Surfaces for computer-aided design of space forms.
\newblock Tech. rep., DTIC Document.

\bibitem[\protect\citename{Crane et~al\mbox{.} }2013]{Crane:2013:RFC}
{\sc Crane, K., Pinkall, U., and Schr\"{o}der, P.}
\newblock 2013.
\newblock Robust fairing via conformal curvature flow.
\newblock {\em ACM Trans. Graph. 32}, 4.

\bibitem[\protect\citename{Delfour and Zol{\'e}sio }2011]{delfour2011shapes}
{\sc Delfour, M.~C., and Zol{\'e}sio, J.-P.}
\newblock 2011.
\newblock {\em Shapes and geometries: metrics, analysis, differential calculus,
  and optimization}, vol.~22.
\newblock Siam.

\bibitem[\protect\citename{Finch et~al\mbox{.} }2011]{Finch:2011}
{\sc Finch, M., Snyder, J., and Hoppe, H.}
\newblock 2011.
\newblock Freeform vector graphics with controlled thin-plate splines.
\newblock {\em ACM Trans. Graph. 30}, 6, 166:1--166:10.

\bibitem[\protect\citename{Haslinger et~al\mbox{.}
  }2003]{haslinger2003introduction}
{\sc Haslinger, J., et~al.}
\newblock 2003.
\newblock {\em Introduction to shape optimization: theory, approximation, and
  computation}, vol.~7.
\newblock Siam.

\bibitem[\protect\citename{Herbulot et~al\mbox{.}
  }2006]{herbulot2006segmentation}
{\sc Herbulot, A., Jehan-Besson, S., Duffner, S., Barlaud, M., and Aubert, G.}
\newblock 2006.
\newblock Segmentation of vectorial image features using shape gradients and
  information measures.
\newblock {\em Journal of Mathematical Imaging and Vision 25}, 3, 365--386.

\bibitem[\protect\citename{Ilbery et~al\mbox{.} }2013]{Ilbery:2013:BDC}
{\sc Ilbery, P., Kendall, L., Concolato, C., and McCosker, M.}
\newblock 2013.
\newblock Biharmonic diffusion curve images from boundary elements.
\newblock {\em ACM Trans. Graph. 32}, 6.

\bibitem[\protect\citename{Jeschke et~al\mbox{.} }2009]{Jeschke:2009}
{\sc Jeschke, S., Cline, D., and Wonka, P.}
\newblock 2009.
\newblock A {GPU} {Laplacian} solver for diffusion curves and {Poisson} image
  editing.
\newblock {\em ACM Trans. Graph. 28}, 5.

\bibitem[\protect\citename{Jeschke et~al\mbox{.} }2011]{jeschke2011estimating}
{\sc Jeschke, S., Cline, D., and Wonka, P.}
\newblock 2011.
\newblock Estimating color and texture parameters for vector graphics.
\newblock In {\em Computer Graphics Forum}, vol.~30, 523--532.

\bibitem[\protect\citename{Jung et~al\mbox{.} }2012]{jung2012nonlocal}
{\sc Jung, M., Peyr{\'e}, G., and Cohen, L.~D.}
\newblock 2012.
\newblock Nonlocal active contours.
\newblock {\em SIAM Journal on Imaging Sciences 5}, 3, 1022--1054.

\bibitem[\protect\citename{Kass et~al\mbox{.} }1988]{kass1988snakes}
{\sc Kass, M., Witkin, A., and Terzopoulos, D.}
\newblock 1988.
\newblock Snakes: Active contour models.
\newblock {\em International journal of computer vision 1}, 4, 321--331.

\bibitem[\protect\citename{Mantegazza }2011]{mantegazza2011lecture}
{\sc Mantegazza, C.}
\newblock 2011.
\newblock {\em Lecture notes on mean curvature flow}, vol.~290.
\newblock Springer.

\bibitem[\protect\citename{Mohammadi et~al\mbox{.} }2001]{mohammadi2001applied}
{\sc Mohammadi, B., Pironneau, O., Mohammadi, B., and Pironneau, O.}
\newblock 2001.
\newblock {\em Applied shape optimization for fluids}, vol.~28.
\newblock Oxford University Press Oxford.

\bibitem[\protect\citename{Mumford and Shah }1989]{mumford1989optimal}
{\sc Mumford, D., and Shah, J.}
\newblock 1989.
\newblock Optimal approximations by piecewise smooth functions and associated
  variational problems.
\newblock {\em Communications on pure and applied mathematics 42,5\/}.

\bibitem[\protect\citename{Orzan et~al\mbox{.} }2008]{Orzan:2008}
{\sc Orzan, A., Bousseau, A., Winnem\"{o}ller, H., Barla, P., Thollot, J., and
  Salesin, D.}
\newblock 2008.
\newblock Diffusion curves: A vector representation for smooth-shaded images.
\newblock {\em ACM Trans. Graph. 27}, 3, 92:1--92:8.

\bibitem[\protect\citename{Pang et~al\mbox{.} }2012]{Pang:12}
{\sc Pang, W.-M., Qin, J., Cohen, M., Heng, P.-A., and Choi, K.-S.}
\newblock 2012.
\newblock Fast rendering of diffusion curves with triangles.
\newblock {\em IEEE Computer Graphics and Applications 32}, 4, 68--78.

\bibitem[\protect\citename{Pinnau and Ulbrich }2008]{pinnau2008optimization}
{\sc Pinnau, R., and Ulbrich, M.}
\newblock 2008.
\newblock {\em Optimization with PDE constraints}, vol.~23.
\newblock Springer.

\bibitem[\protect\citename{Pr{\'e}vost et~al\mbox{.}
  }2014]{prevost2014vectorial}
{\sc Pr{\'e}vost, R., Jarosz, W., and Sorkine-Hornung, O.}
\newblock 2014.
\newblock A vectorial framework for ray traced diffusion curves.
\newblock In {\em Computer Graphics Forum}.

\bibitem[\protect\citename{Ramer }1972]{ramer1972iterative}
{\sc Ramer, U.}
\newblock 1972.
\newblock An iterative procedure for the polygonal approximation of plane
  curves.
\newblock {\em Computer Graphics and Image Processing 1}, 3, 244--256.

\bibitem[\protect\citename{Schneider and Kobbelt }2001]{schneider2001geometric}
{\sc Schneider, R., and Kobbelt, L.}
\newblock 2001.
\newblock Geometric fairing of irregular meshes for free-form surface design.
\newblock {\em Computer aided geometric design 18}, 4, 359--379.

\bibitem[\protect\citename{Selinger }2003]{selinger2003potrace}
{\sc Selinger, P.}
\newblock 2003.
\newblock Potrace: a polygon-based tracing algorithm.
\newblock {\em Potrace (online)\/}.

\bibitem[\protect\citename{Sethian et~al\mbox{.} }2003]{sethian2003level}
{\sc Sethian, J.~A., et~al.}
\newblock 2003.
\newblock Level set methods and fast marching methods.
\newblock {\em Journal of Computing and Information Technology 11}, 1, 1--2.

\bibitem[\protect\citename{Shewchuk }1996]{shewchuk1996triangle}
{\sc Shewchuk, J.~R.}
\newblock 1996.
\newblock Triangle: Engineering a 2d quality mesh generator and delaunay
  triangulator.
\newblock In {\em Applied computational geometry towards geometric
  engineering}. Springer, 203--222.

\bibitem[\protect\citename{Sokolowski and Zol{\'e}sio
  }1992]{sokolowski1992introduction}
{\sc Sokolowski, J., and Zol{\'e}sio, J.-P.}
\newblock 1992.
\newblock {\em Introduction to shape optimization}.
\newblock Springer.

\bibitem[\protect\citename{Sun et~al\mbox{.} }2012]{Sun:2012:DCT}
{\sc Sun, X., Xie, G., Dong, Y., Lin, S., Xu, W., Wang, W., Tong, X., and Guo,
  B.}
\newblock 2012.
\newblock Diffusion curve textures for resolution independent texture mapping.
\newblock {\em ACM Trans. Graph. 31}, 4.

\bibitem[\protect\citename{Sun et~al\mbox{.} }2014]{Sun:2014:FMR}
{\sc Sun, T., Thamjaroenporn, P., and Zheng, C.}
\newblock 2014.
\newblock Fast multipole representation of diffusion curves and points.
\newblock {\em ACM Trans. Graph. 33}, 4, 53:1--53:12.

\bibitem[\protect\citename{Takayama et~al\mbox{.} }2010]{TSNI10}
{\sc Takayama, K., Sorkine, O., Nealen, A., and Igarashi, T.}
\newblock 2010.
\newblock Volumetric modeling with diffusion surfaces.
\newblock {\em ACM Trans. Graph. 29}, 6, 180:1--180:8.

\bibitem[\protect\citename{Xie et~al\mbox{.} }2014]{Xie:2014:HDC}
{\sc Xie, G., Sun, X., Tong, X., and Nowrouzezahrai, D.}
\newblock 2014.
\newblock Hierarchical diffusion curves for accurate automatic image
  vectorization.
\newblock {\em ACM Trans. Graph. 33}, 6, 230:1--230:11.

\bibitem[\protect\citename{Zienkiewicz and Morice }1971]{zienkiewicz1971finite}
{\sc Zienkiewicz, O.~C., and Morice, P.}
\newblock 1971.
\newblock {\em The finite element method in engineering science}, vol.~1977.
\newblock McGraw-hill London.

\end{thebibliography}
\end{document}